\documentclass[11pt]{article}
\usepackage{graphicx}
\usepackage{amsfonts}
\usepackage{amsmath}
\usepackage{amssymb}
\usepackage{url}
\usepackage{fancyhdr}
\usepackage{indentfirst}
\usepackage{enumerate}
\usepackage{dsfont,footmisc}
\usepackage{multirow}
\usepackage{float}
\usepackage{bm}
\usepackage{booktabs}
\usepackage{tablefootnote}
\usepackage{todonotes}
\usepackage{longtable}
\usepackage{mathrsfs}

\usepackage[misc]{ifsym}

\usepackage{setspace}

\usepackage{amsthm}
\usepackage{color}
\usepackage[round]{natbib}
\usepackage[british]{babel}
\usepackage[colorlinks=true,citecolor=blue]{hyperref}
\usepackage[right,pagewise,displaymath, mathlines]{lineno}

\defcitealias{BCBS2016}{BCBS\textcolor{black}{,} 2016}
\defcitealias{BCBS2013}{BCBS \textcolor{black}{(}2013\textcolor{black}{)}}

\usepackage{soul,xcolor}
\setstcolor{red}

\def\d{\,\mathrm{d}}

\def\lawis{\buildrel \d \over \sim}

\newcommand{\var}{\mathrm{var}}

\newcommand{\VaR}{\mathrm{VaR}}

\newcommand{\ES}{\mathrm{ES}}
\newcommand{\E}{\mathbb{E}}
\newcommand{\R}{\mathbb{R}}

\newcommand{\N}{\mathbb{N}}
\newcommand{\M}{\mathcal{M}}
\newcommand{\p}{\mathbb{P}}
\newcommand{\PP}{\mathcal{P}}

\newcommand{\id}{\mathds{1}}

\renewcommand{\ge}{\geqslant}
\renewcommand{\le}{\leqslant}

\renewcommand{\epsilon}{\varepsilon}

\theoremstyle{plain}
\newtheorem{theorem}{Theorem}%
\newtheorem{lemma}{Lemma}%
\newtheorem{proposition}{Proposition}%

\theoremstyle{definition}
\newtheorem{definition}{Definition}%
\newtheorem{example}{Example}%
\newtheorem{assumption}{Assumption}

\theoremstyle{remark}
\newtheorem{remark}{Remark}%

\theoremstyle{definition}

\renewcommand{\cite}{\citet}
\renewcommand{\cdots}{\dots}
\DeclareMathOperator*{\argmin}{arg\,min}
\DeclareMathOperator*{\argmax}{arg\,max}
\begin{document}

\title{E-backtesting
}

\author{Qiuqi Wang\thanks{Maurice R.~Greenberg School of Risk Science, Georgia State University, U.S.A. \Letter~\url{qwang30@gsu.edu}}\and Ruodu Wang\thanks{Department of Statistics and Actuarial Science, University of Waterloo, Canada. \Letter~\url{wang@uwaterloo.ca}}\and Johanna Ziegel\thanks{Seminar of Statistics, ETH Zurich, Switzerland. \Letter~\url{ziegel@stat.math.ethz.ch}}}

 \date{6th November 2026}

\maketitle

\begin{abstract} In the recent Basel Accords, the Expected Shortfall (ES) replaces the Value-at-Risk (VaR) as the standard risk measure for market risk in the banking sector, making it the most important risk measure in financial regulation. One of the most challenging tasks in risk modeling practice is to backtest ES forecasts provided by financial institutions. To design a model-free backtesting procedure for ES, we make use of the recently developed techniques of e-values and e-processes. Backtest e-statistics are introduced to formulate e-processes for risk measure forecasts, and unique forms of backtest e-statistics for VaR and ES are characterized using recent results on identification functions. For a given backtest e-statistic, a few criteria for optimally constructing the e-processes are studied. The proposed method can be naturally applied to many other risk measures and statistical quantities. We conduct extensive simulation studies and data analysis to illustrate the advantages of the  model-free backtesting method, and compare it with the ones in the literature.

\end{abstract}

\section{Introduction}

\label{sec:1}

Forecasting risk measures is important for financial institutions to calculate capital reserves for risk management purposes. Regulators are responsible to monitor whether risk forecasts are correctly reported by conducting hypothesis tests known as backtests \citep[see e.g.,][for general treatments]{Christoffersen2011, McNeilETAL2015}. Regulatory backtests have several features distinct from traditional testing problems; see \cite{AcerbiSzekely2014} and \cite{NoldeZiegel2017}. First, risk forecasts and realized losses arrive sequentially over time. 
Second, due to frequently changing portfolio positions and the complicated temporal nature of financial  data, the losses and risk predictions are neither independent nor identically distributed, and they do not follow any standard time-series models.
Third, the tester (e.g., a regulator)  is concerned about risk measure underestimation, which means high insolvency risk, whereas overestimation (i.e., being conservative) is  secondary or acceptable.
Fourth, the tester does not necessarily accurately know the underlying model used by a financial institution to produce risk predictions.

In financial practice, a well-adopted simple approach exists for backtesting the Value-at-Risk (VaR), which is the so-called three-zone approach based on binomial tests described in \citetalias{BCBS2013}; this approach is \emph{model-free} in the sense that one directly tests the risk forecast without testing any specific family of models. 
In other statistical context, model-free methods are often called ``nonparametric", but we use the term ``model-free" to emphasize our specific setting of   requiring no distributional  forecast.
More recently, the Basel Committee on Banking Supervision \citep{BCBS2016, BCBS2019} replaced  VaR by  the Expected Shortfall (ES)  as the standard regulatory measure for market risk, mostly due to the convenient  properties of ES, in particular,  being able to capture tail risk.\footnote{Quoting \cite[p.1]{BCBS2016}: \emph{Use of ES will help to ensure a more prudent capture of ``tail risk" and capital adequacy during periods of significant financial market stress.} See also \cite{WangZitikis2021} for an axiomatic justification of ES in financial regulation.} However, as discussed by \cite{Gneiting2011}, ES is not elicitable, and  backtesting ES is substantially more challenging than VaR.
Table \ref{tab:comp} summarizes the main features of existing methods backtesting ES. To the best of our knowledge, there is no model-free non-asymptotic backtesting method for ES. Moreover, except for \cite{HogaDemetrescu2022}, most of the backtesting methods in the existing literature only work for a fixed data size, and are thus not valid under optional stopping, or equivalently, not anytime valid (see e.g., \citealp{RamdasETAL2022}). This creates limitations to financial regulation practice where early rejections are highly desirable.

\begin{table}[t]
\def\arraystretch{1.5}

    \small
\begin{center}
\caption{Comparison of existing backtesting methods for ES}
\vspace{.1in}
\begin{tabular}{cccccc} 
\toprule
\small {Literature} 
& {\begin{minipage}[c]{0.13\textwidth}\begin{center} \small   Parametric or dependence assumptions  \end{center} \end{minipage}}
 & {\begin{minipage}[c]{0.13\textwidth}\begin{center} \small  Forecast  structural assumptions  \end{center}\end{minipage}}     
 &  {\begin{minipage}[c]{0.13\textwidth} \begin{center} \small   Fixed    sample size   \end{center}\end{minipage}}  
&  {\begin{minipage}[c]{0.13\textwidth} \begin{center} \small  Asymptotic test \end{center}\end{minipage}}
 & \small {\begin{minipage}[c]{0.16\textwidth} \begin{center} \small   Reliance on VaR or distributional  forecasts   \end{center}\end{minipage}}  \\ 
\midrule
 MF00  & yes   & yes   & yes   & yes   & yes \\
AS14  & yes   & yes & yes   & yes   & yes \\
DE17  & yes   & yes   & yes   & yes   & yes \\
NZ17  & yes   & yes   & yes   & yes   & yes \\
SQPQ21 & yes & yes & yes & yes & yes\\
BD22  & yes   & yes   & yes   & yes   & no \\
HD22  & yes   & yes   & no    & no    & yes \\
This paper & no    & no    & no    & no    & yes \\
\bottomrule
\end{tabular}
\end{center}
\vspace{0.2cm}
\emph{Notes:} We use shortcuts MF00 for \cite{McNeilFrey00}, AS14 for \cite{AcerbiSzekely2014}, DE17 for \cite{DuEscanciano2017}, NZ17 for \cite{NoldeZiegel2017}, SQPQ21 for \cite{SuETAL2021}, BD22 for \cite{BayerDimitriadis2022}, and HD22 for \cite{HogaDemetrescu2022}. Parametric or dependence assumptions refer to those on loss distributions, time series models, stationarity, or strong mixing. Forecast structural assumptions refer to requirements on the forms and properties of risk forecasts. \cite{AcerbiSzekely2014} proposed three methods of backtesting ES; The first two methods do not require specific forms of ES forecasts, but the third method requires ES to be estimated as realized ranks. 
\label{tab:comp}
\end{table}

In this paper, we develop a model-free backtesting method for risk measures, including ES, using the concepts of e-values and e-tests \citep{Shafer2021,VovkWang2021, GrunwaldHeideETAL2020}.
E-tests have important advantages over classical statistical tests (p-tests) based on p-values. 
\cite[Section 2]{WangRamdas22} collect many reasons for using e-values and e-tests, regarding high-dimensional asymptotics, composite models, sequential (any-time valid) inference, information accumulation, and robustness to model misspecification and dependence;  other advantages of e-values have been illustrated by \cite{VovkWang2021},  \cite{VovkWangWang22} and  \citet{GrunwaldHeideETAL2020}.
As a particularly relevant feature to our context,
our proposed e-tests allow regulators to get alerted early as the e-process accumulates to a reasonably large value. This is different from scientific discoveries (such as genome studies) where a scientist may not be entitled to reject a hypothesis based on merely ``substantial'' evidence. Noticing this, a multi-zone approach similar to the three-zone approach can be developed by setting different e-value thresholds in financial regulation. 
A simple illustrative example  of our methodology is presented in Section \ref{sec:R2-2}.

The main contribution of this paper is four-fold: First, we define backtest e-statistics and propose new backtesting methods. In particular, we obtain backtest e-statistics for ES in Section \ref{sec:2} (Theorem \ref{th:ep}), allowing us to construct e-processes to backtest ES, as well as other risk measures  %
in Section \ref{sec:etest} (Theorem \ref{lem:Ville}). The backtesting method for VaR and ES is discussed in Section \ref{sec:cons}. Second, with backtest e-statistics chosen, the next important step to construct an e-process is choosing a suitable betting process, which we address in Section \ref{sec:betting}. We propose three methods to calculate the betting processes based on data. It turns out that these methods are asymptotically optimal (equivalent to an oracle betting process) in different situations (Theorem \ref{th:opt}). Third, we characterize backtest e-statistics for the mean, the variance, VaR (Theorem \ref{prop:uniquantile}), and ES (Theorem \ref{th:es}) in Section \ref{sec:techresults}. All backtest e-statistics for these functionals take similar forms as mixtures between the constant function equal to $1$ and a simple backtest e-statistic. Finally, through the simulation study and data analysis in Sections \ref{sec:sim_stu} and \ref{sec:emp}, we demonstrate detailed procedures of backtesting VaR and ES using e-values for practical operations of financial regulations.

Five sections are put in the e-companion. %
Section \ref{app:vares} shows the betting processes calculated via Taylor approximation for VaR and ES.  Section \ref{app:iden} discusses the link between backtest e-statistics and identification functions, useful for the characterization results in Section \ref{sec:techresults}.
Except for a short proof of Theorem  \ref{th:ep}, proofs of all results are relegated to Section \ref{app:proofs}. 
Section \ref{app:num} contains additional simulation and data analyses and some necessary details of the ones in the paper. Section \ref{app:R1-EC-5} discusses a deliberate over-forecast strategy by the financial institution and how our methods can address this issue.

\subsection{Related literature}

Besides financial regulation, evaluating forecasting models and methods for major economic variables is also essential in the decision-making processes of government institutions and regulatory authorities. Earlier work on predictive ability tests and forecast selection includes \cite{DieboldMariano1995}, whose method was extended by \cite{West1996}, \cite{ClarkMcCracken2001}, and \cite{GiacominiWhite2006}. 
Unconditional backtests of VaR were considered by \cite{Kupiec1995} on testing Bernoulli distributions, which were extended by  \cite{Christoffersen1998} to include testing independence of the $\VaR$-violations.
\cite{EngleManganelli2004} tested conditional autoregressive VaR; \cite{EscancianoOlmo2010} further studied backtesting Value-at-Risk by considering the estimation risk; \cite{BerkowitzETAL2011} unified existing evaluation methods of VaR; and \cite{ZiggelETAL2014} proposed a Monte Carlo simulation-based backtesting method for VaR. 

Due to its increasing importance and challenging nature, there are ample studies in the more recent literature on backtesting ES with different approaches. \cite{McNeilFrey00} proposed bootstrap tests with iid innovations; \cite{AcerbiSzekely2014,AcerbiSzekely2017} studied three backtesting methods under independent losses; \cite{DuEscanciano2017} designed parametric tests using cumulative violations; their approach was extended by \cite{SuETAL2021} who adopted an efficient empirical likelihood method backtesting ES. \cite{NoldeZiegel2017} studied comparative backtests among forecasting methods, and considered conditional and unconditional calibration tests for risk measures. The backtests using e-processes that we propose in this paper are testing conditional calibration. Indeed, they can be seen as natural anytime-valid analogues of the conditional calibration tests in \cite{NoldeZiegel2017}. 
\cite{BayerDimitriadis2022} introduced backtesting through a linear regression model; and \cite{HogaDemetrescu2022} proposed sequential monitoring based on parametric distributions. The main features of these approaches are summarized in Table \ref{tab:comp}. See also \cite{MoldenhauerPitera2017} for a method of backtesting empirical ES forecasts.

The literature on e-values has also been growing fast recently. 
E-values were used in  the early literature in different disguises, although the term ``e-value" was proposed by \cite{VovkWang2021}. For instance,  e-values and e-tests were essentially used in the work of \cite{Wald1945} and \cite{DarlingRobbins1967}, and they are central to the ideas of testing by betting and martingales  \citep{ShaferETAL2011,ShaferVovk2019} and universal inference \citep{WassermanETAL2020}. E-values are shown to be useful in multiple hypothesis testing with dependence \citep{VovkWang2021},  Bayesian models with optional sampling \citep{GrunwaldHeideETAL2020},   false discovery rate control \citep{WangRamdas22},
high-dimensional regression \citep{ren2022derandomized}, and many other statistical applications. 
Confidence sequences for tail risk measures including ES are studied by \cite{agrawal2021optimal}, see also \cite{CasgrainLarssonETAL2024}. 
The recent survey paper of \cite{RamdasETAL2022} contains a review of this topic. 
 Compared to p-values, e-values have their own interpretations, such as wealth levels by betting against the null hypothesis, generalized likelihood ratios, and  post-hoc valid tests.
We refer to the recent monograph \cite{ramdas2024hypothesis} for a general treatment of e-values, including their statistical interpretations.

\section{The e-backtesting procedure: An illustrative example} 
\label{sec:R2-2}

To prepare for the general methodology of e-backtesting, we first present a simple example of testing the mean and the variance.
Denote by $[T]=\{1,\dots,T\}$ for any positive integer $T$. 
Suppose that we observe a sequence of iid data $(L_t)_{t\in [T]}$ from the random variable $L$.
For $r>0$ and $z\in \R$, we would like to test the nonparametric null hypothesis 
 \[
H_0: \mbox{$\var(L) \le r$ and $\E[L]=z$}
 \]
against  the alternative hypothesis $H_1$ that $\var(L) > r$, 
where $\E[L]$ is the expectation of $L$ and $\var(L)$ is the variance of $L$. We are concerned about the variance being large, but not the mean.
We consider sequential testing; that is, for each time point $t$ we decide whether we reject $H_0$ based on $(L_s)_{s\in[t]}$ or continue, until all $T$ data points are observed. 
Our approach based on an e-process can be described in the following steps.
\begin{enumerate}[(i)]
\item Pick a value $\alpha \in (0,1)$, which is a significance level. 
    \item Choose a function $e:\R^3 \to \R$ such that 
    $\E[e(L,r,z)]\le 1$ under the null hypothesis,
    and $\E[e(L,r,z)]> 1$ under the alternative hypothesis. 
    Define $X_t=e(L_t,r,z)$ for each $t\in [T]$. 
In this particular setting,
the function $e$ can be chosen as $e(x,r,z)= (x-z)^2/r$.
    \item For each $t\in [T]$, choose a value $\lambda_t \in [0,1] $ that can depend on $(L_s)_{s\in [t-1]}$.
    \item  Construct a stochastic process: $M_0=1$ and
 $$
M_t    = (1-\lambda_t +\lambda_t X_t)M_{t-1}  = \prod_{s=1}^t  (1-\lambda_s+\lambda_s X_s),~~t\in[T].
 $$
\item Reject  $H_0$ at time $t$ if we observe $M_t \ge 1/\alpha$.
\end{enumerate}

Under the null hypothesis, regardless of how $(\lambda_t)_{t\in [T]}$ is chosen, $(M_t)_{t\in [T]}$ is a non-negative supermartingale with respect to  the $\sigma$-algebra $\sigma(L_1,\dots,L_t)$, which can be 
easily checked by $$\E[M_t|\sigma(L_1,\dots,L_{t-1})] =(1-\lambda_t+\lambda_t\E[X_t] )M_{t-1} \le  M_{t-1}.$$
By   Ville's inequality, we have $\p(\sup_{t\in [T]} M_t\ge 1/\alpha)\le \alpha \E[M_0] =\alpha$ \citep{Ville1939}; see Theorem \ref{lem:Ville}.
Therefore, we have  type-I error control at level $\alpha$. 
Moreover,  under the alternative hypothesis $H_1$, because $\E[X_t]>1$, $(M_t)_{t\in [T]}$ has a growing mean (it is in fact a submartingale), and we may hope that it grows with high probability as $t$ increases, and thus it is able to reject the null hypothesis.
Indeed, with properly chosen $(\lambda_t)_{t\in [T]}$, this process grows exponentially in the iid setting under $H_1$.

The rest of the paper formalizes the above idea via the theory of e-values. We will prove the claims above in a much more general framework of risk measures and sequential forecasts. Some key ingredients are explained below.
\begin{enumerate}
    \item Variance and mean are replaced by general functionals $\rho$ (risk measure to backtest, e.g., capital requirement for the portfolio) and $\phi$ (auxiliary information, sometimes not needed), respectively.
    \item The random variable $X_t=e(L_t,r,z)$ in (ii)  is called an e-variable, and its realization an e-value. 
  The functions $e$ for general risk measures will be formally defined  in Section \ref{sec:2} with examples for several important cases.  The  functions $e$ do not depend on the specific distributions of the loss $L$, thus they are model-free.
  Their  characterizations  are studied in Section \ref{sec:techresults}.
    \item In our general setting, the data are not iid, and a bank provides sequential risk forecasts $(r_t,z_t)$  each day for their portfolio loss $L_t$ in the next day. As formally studied in Section \ref{sec:etest}, the null hypothesis is
 $$
H_0: ~~ r_t\ge \rho(L_t|\mathcal F_{t-1})~\mbox{and}~ z_t = \phi(L_t|\mathcal F_{t-1})~~
 \mbox{ for $t\in [T]$},
$$
and the alternative is $r_t < \rho(L_t|\mathcal F_{t-1})$ for some $t$ (i.e., the bank is under-forecasting their risk),
where $\mathcal F_t$ is the $\sigma$-algebra of available information at time $t$.
\item The choice of $(\lambda_t)_{t\in [T]}$ in (iii) is essential for the power of the above procedure. We offer several ways to compute  $(\lambda_t)_{t\in [T]}$ and study their optimality in Section \ref{sec:betting}.
\item  We allow $T$ to be infinite or data-dependent, which is  practically useful as we usually do not know a priori how many data points we will sample in sequential testing. The test can stop at any time. 
The flexibility of $T$ is a feature of e-tests, which allows us to address situations more general than the ones considered in the literature with fixed $T$.

\item The level $\alpha$ can be chosen differently from traditional tests, because tests using e-values have a different type of guarantee in addition to a type-I error control. Values of $ \alpha$ like $1/5$ and $1/10$ are already useful in the e-value literature (although their type-I error guarantee is not very small). This is discussed in Section \ref{sec:etest}.
\end{enumerate}

\section{E-values and backtest e-statistics}
\label{sec:2}

The approach in Section \ref{sec:R2-2} builds on the theory of e-values.
This section first reviews e-values and risk measures, and then introduces backtest e-statistics, which are further illustrated with  examples, including the important case of ES.

\subsection{E-values and risk measures}
Let $(\Omega,\mathcal F,\p)$ be any probability space.
A (composite) hypothesis $ H$ is a set of probability measures on $(\Omega,\mathcal F)$.
A hypothesis $  H$ is simple if it is a singleton. 
Following the terminology of \cite{VovkWang2021}, an \emph{e-variable} for  $  H$ is a random variable $E:\Omega \to [0,\infty]$ such that $\E^P[ E] \le 1$ for each $P\in H$. We denote by $\mathcal E_{H}$ the set of e-variables for a hypothesis $ H$ and by $\mathcal E_P$ the set of e-variables for the simple hypothesis $\{P\}$. An \emph{e-test} rejects the hypothesis $  H$ if a realized e-variable, called an \emph{e-value}, is larger than a given threshold. 
A common rule of thumb is that an e-value of $10$ represents strong evidence, see Remark \ref{rem:R1-threshold} for more discussion on  thresholds for e-values.
Markov's inequality  guarantees that the reciprocal of an e-value   is a conservative p-value.
 A non-negative stochastic process $(E_t)_{t\in K}$, $K\subseteq \N$, adapted to a given filtration, is an \emph{e-process} for $H$ if $\E^P[E_{\tau}]\le 1$ for all stopping times $\tau$ taking values in $K$ and each $P\in {H}$.

Let $d$ be a  positive integer. The \emph{model space} $\M$ is a set of distributions on $\R$.  
The value of the functional $\psi=(\rho,\phi_1,\dots,\phi_{d-1}): \M\to \R^{d}$ represents the collection of available statistical information, where $\rho$ is the risk prediction to be tested, and $\phi=(\phi_1,\dots,\phi_{d-1})$ contains auxiliary information.  A simple example to keep in mind for now is that $\rho$ is the variance (which we aim to backtest) and $\phi$ is the mean (which helps to backtest the variance) as in Section \ref{sec:R2-2}. In this case, $d=2$.
If $d=1$, then the only available information is the predicted value of $\rho$. We also write $\psi(X) = \psi(F)$ for a random variable $X \lawis F\in \M$, that is, when $F$ is the distribution of $X$.

Let $\M_q$, $\M_\infty$, and $\M_0$ represent the set of distributions on $\R$ with finite $q$-th moment for $q\in(0,\infty)$, that of compactly supported distributions on $\R$, and that of all distributions on $\R$, respectively.  
For level $p\in(0,1)$, the \emph{Value-at-Risk (VaR)} is defined as the lower $p$-quantile:
$$\VaR_p(F)=Q_p(F)=\inf\{x\in \R: F(x)\ge p\},~~F\in\M_0,$$
and the \emph{Expected Shortfall (ES)} is defined as
$$
\ES_p(F)=\frac{1}{1-p}\int_p^1 \VaR_\alpha(F) \d \alpha,~~F\in\M_1.
$$
VaR and ES belong to the class of dual utilities in \cite{Yaari1987} and \cite{Schmeidler1989}. 
The pair $(\rho,\phi)=(\ES_p,\VaR_p)$ is a main object of interest in our paper.
For  $\psi =(\rho,\phi_1,\dots,\phi_{d-1})$,
the natural domains of $\rho,\phi_1,\dots,\phi_{d-1}$  are not necessarily identical, and the domain of $\psi$ is  their intersection.

\subsection{Backtest e-statistics}

Next, we introduce the key tool we use for our e-tests.

\begin{definition}[Point and one-sided e-statistics]\label{def:1}
Let $\PP\subseteq\M$. A \emph{$\PP$-point e-statistic for $\psi:\M\to \R^d$} is a measurable function $e:\R^{d+1} \to [0,\infty]$ satisfying $\int_\R e(x,  \psi(F)) \d F(x)\le 1$ for each $F\in \PP$. 
 A measurable function $e:\R^{d+1} \to [0,\infty]$ is a \emph{$\PP$-one-sided e-statistic for $\psi = (\rho,\phi)$} if $$\int_\R e(x, r,\phi(F)) \d F(x)\le 1 \mbox{~for each $F\in \PP$ and $r \ge \rho(F)$.}$$ If $\PP=\M $, we speak of point or one-sided e-statistics for $\psi$, respectively, without mentioning $\mathcal P$.
\end{definition}

We have seen e-statistics used in the example in Section \ref{sec:R2-2}.
Using the language of e-values, a $\PP$-point e-statistic is a function $e$ such that $e(X,  r, z )$ is an e-variable for the hypothesis  
$$
H_0: \mbox{$X $ follows any $F\in \mathcal P$ with $\rho(F) = r$ and $\phi(F)=z$}
$$
for each $(r, z)$ in the range $\psi(\M)$ of $\psi=(\rho,\phi)$.
 Here, {the conditions in the null hypothesis are equalities.} 
 In case $d=1$,  $e(X,  r )$ is an e-variable for the hypothesis 
 $$
H_0: \mbox{$X $ follows any $F\in \mathcal P$ with $\rho(F) = r$.}
$$
A $\PP$-one-sided e-statistic for $\psi$ is a function $e$ such that $e(X,r,z)$ is an e-variable for the hypothesis
$$
H_0: \mbox{$X $ follows any $F\in \mathcal P$ with $\rho(F) \le r$ and $\phi(F)=z$}
$$
for each $(r, z)$ in the range of $\psi$ (see Section \ref{sec:R2-2}).
It is clear that one-sided e-statistics are always point e-statistics. The next definition states desirable properties of e-statistics under the alternative.

\begin{definition}[Backtest e-statistics]\label{def:2}
Let $\PP\subseteq\M$.  A  $\PP$-one-sided e-statistic $e:\R^{d+1}\to [0,\infty]$ for $\psi = (\rho,\phi)$ is a \emph{$\PP$-backtest e-statistic for $\psi$}   if  
$\int_\R e(x, r,z) \d F(x)> 1$ for all $(r,z) \in \psi(\PP)$ and $F\in \PP$ with $\rho(F)> r$. 
The backtest e-statistic  $e$ is called \emph{monotone} if, in addition, $r\mapsto e(x,r,z)$ is decreasing for each $(x,z)$.
\end{definition}

The property of having a mean larger than $1$ under the alternative is crucial for the e-statistic. If the risk $\rho$ is underestimated, 
then a backtest e-statistic will have a mean that is larger than $1$, regardless of whether the prediction of the auxiliary functional $\phi$ is truthful.  The function $e$  in Section \ref{sec:R2-2} is a backtest e-statistic for $\psi=(\mathrm{var},\E)$, with details explained in  Example \ref{ex:2}.

In our backtesting procedure,  detailed in Section \ref{sec:etest}, we will use e-variables of the form $(1-\lambda)+ \lambda e(X,r,z)$ for some $\lambda \in [0,1)$ and loss random variable $X$, such that $e(X,r,z)$ is an e-variable for the null hypothesis that $(r,z)$ is correctly specified. The following lemma justifies that using a backtest e-statistic for $\psi$ yields positive e-power for some $\lambda$.
The e-power of an e-variable $E$ for a probability measure $Q$ is defined as  $\E^Q[\log E]$ (\citealp{vovk2024nonparametric}). 
As we see in Section \ref{sec:betting}, the e-power is closely related to the Kelly criterion and growth-optimal portfolios, a central concept for e-values (e.g., \citealp{ShaferVovk2019,GrunwaldHeideETAL2020}).
\begin{lemma}\label{lem:R1}
 For any random variable $E\ge 0$, we have
$$
\E[E]>1 \iff \E[\log ((1-\lambda) +\lambda E)]>0 \mbox{~for some $\lambda \in [0,1]$}.
$$
\end{lemma}

By Lemma \ref{lem:R1}, if $e$ is a backtest e-statistic and $r<\rho(F)$, then for any $z$  there exists $\lambda\in (0,1)$ such that $(1-\lambda) +\lambda e(X,r,z)$ has positive e-power. This condition is sufficient for establishing consistency of our backtesting procedures in some settings; see Theorem \ref{th:opt} in Section  \ref{sec:betting}.

We are only interested in forecast values $(r,z)$ in the set $\psi(\mathcal P)$.
Any forecast values outside $\psi(\mathcal P)$ can be rejected automatically.  
Our idea of backtest e-statistics specifically addresses the underestimation of $\rho$, which is consistent with the motivation in regulatory backtesting. 
If a backtest e-statistic is monotone, then an overestimation of the risk is rewarded: An institution being scrutinized by the regulator can deliberately report a higher risk value (which typically means higher capital reserve) to pass to the regulatory test, thus rewarding prudence.   

\subsection{Examples}
Below, we give a few examples of backtest e-statistics for some common risk measures. Throughout, we use the convention 
that $0/0=1$ and $1/0=\infty$, 
and let $\R_+=[0,\infty)$. 

\begin{example}[Backtest e-statistic for the mean]
\label{ex:1}
Let $\PP$ be the set of distributions on $\R_+$ in $\M_1$.
Define the function
$e(x,r) = x/r$ for $x,r\ge 0$.
In this case, we have $\E[e(X,r)] \le 1$ for all random variables $X$ with distribution in $\PP$ and $ r \ge \E[X]$. Moreover, for any such $X$, $\E[X]>r\ge 0$ implies
$\E[e(X,r)]>1$. 
Therefore,  $e $ is a monotone $\PP$-backtest e-statistic for the mean. 
\end{example}

\begin{example}[Backtest e-statistic for $(\var,\E)$]
\label{ex:2} 
Consider $(\var,\E):\M_2\to\R^2$ as in Section \ref{sec:R2-2}.
The function
$e(x,r,z) = (x-z)^2/r$ for $x,z\in \R$ and $r\ge 0$ is a monotone backtest e-statistic  for $(\var,\E)$. To see this, for all random variables $X$ with distribution in $\M_2$, we have
$$\E[e(X, r, \E[X])] =\frac{\E[(X-\E[X])^2]}{r}\le 1$$
for $r \ge \var(X)$.
Moreover, since $z=\E[X]$ minimizes $\E[(X-z)^2]$ over $z\in \R$ and $\mathrm{var}(X)=\E[(X-\E[X])^2]$,
$\var(X)>r \ge 0$ implies
 $$\E[e(X, r,z)] =\frac{\E[(X-z)^2]}{r} \ge \frac{\var(X)}{r}  >1.$$ 

\end{example}

\begin{example}[Backtest e-statistic for a quantile]
\label{ex:5} 
Take $p\in (0,1)$.
Define the function
\begin{equation}\label{eq:eqp}
e_{p}^Q(x,r) = \frac{1}{1-p}\id_{\{x> r\}},~~x,r \in \R.
\end{equation}
We have $\E[e^Q_p(X,r)] \le 1$ for any random variable $X$ with distribution $F$ and $r \ge Q_p(F)$. Moreover, $Q_p(F)>r$ implies $\p(X>r)>1-p$, and hence
$\E[e^Q_p(X,r)]>1$. 
Therefore,  $e^Q_p $ is a monotone backtest e-statistic forthe $p$-quantile. 
\end{example}

\begin{example}[Backtest e-statistic for an expected loss]
\label{ex:3}
For some $a\in \R$, let $\ell :\R\to [a,\infty)$ be a function that is interpreted as a loss.
Define the function
$e(x,r) = (\ell(x)-a)/{(r-a)}$ for $x\in \R$ and $r\ge a$.
Analogously to Example \ref{ex:1},   $e $ is a monotone backtest e-statistic for the expected loss $F\mapsto \int \ell \d F$ on its natural domain.  
\end{example}

The choice of a backtest e-statistic $e$ for $\psi$ is not necessarily unique. 
For instance, a linear combination of $e$ with $1$ with the weight between $0$ and $1$ is also a backtest e-statistic for $\psi$. 
Depending on the specific situation, either e-statistic may be useful in practice. 

\begin{remark}
The functional $\psi =(\rho,\phi) = (\var,\E)$ in Example \ref{ex:2} is an example of a Bayes pair; that is, there exists a measurable function $L:\R^{d+1}\to \R$, called the \emph{loss function}, such that
\begin{equation}\label{eq:bayes}
 \phi(F)\in \argmin_{z \in \R^d } \int L(z,x) \d F(x)  \text{ and } \rho(F)=\min_{z \in \R^d}  \int L(z,x) \d F(x),\quad F\in \M,
 \end{equation}
where $\int L(z,x) \d F(x)$ is assumed to be well-defined for each $z\in \R^d$, $F \in \M$  \citep{FisslerZiegel2016,FR21, EMWW21}. 
The function $L$ is the square loss function in the case of $ (\var,\E)$.
Bayes pairs often admit backtest e-statistics. A typical example commonly used in risk management practice is $(\ES,\VaR)$ treated below,\footnote{We sometimes omit the probability level  $p$ in $\VaR_p$, $\ES_p$ and $(\VaR_p,\ES_p)$ in the text (but never in equations).} which is our main focus.
\end{remark}

We will see that, for $p\in (0,1)$, the function 
\begin{equation}\label{eq:ep}
e^{\mathrm{ES}}_p (x,r,z) = \frac{(x-z)_+}{(1-p)(r-z)},~~~x\in \R,~ z \le r
\end{equation}
defines a backtest e-statistic for $(\ES_p,\VaR_p)$, where $y_+$ denotes $\max\{y,0\}$ for any real number $y$.  Recall the convention that $0/0=1$ and $1/0=\infty$, and set $e^{\mathrm{ES}}_p(x,r,z)=\infty$ if $r<z$, which is a case of no relevance since $\ES_p(F)\ge \VaR_p(F)$ for any $F \in \M_1$. %

\begin{theorem}\label{th:ep}
The function $e^{\mathrm{ES}}_p$ is a monotone backtest e-statistic for $( \ES_p,\VaR_p)$.
\end{theorem}
\begin{proof}
By the VaR-ES relation of \cite{RockafellarUryasev2002},   for any random variable $X$ with finite mean,
\begin{equation}  \VaR _p(X)  \in   \argmin_{z\in \R} \left\{z + \frac{1}{1-p}\E[(X-z)_+]\right\}, \label{eq:var1}
 \end{equation}
 and \begin{equation} 
  \ES_p(X) =     \min_{z\in \R} \left\{z + \frac{1}{1-p}\E[(X-z)_+]\right\} .  \label{eq:es1}
 \end{equation}
 This indicates that $(\ES_p,\VaR_p)$ is a Bayes pair by \eqref{eq:bayes} with loss function $L:(z,x)\mapsto z + (x-z)_+/(1-p)$. Since $L(z,x) \ge z$, for $r \ge z$, we have that $e^{\mathrm{ES}}_p(x,r,z)=(L(z,x)-z)/(r-z) \ge 0$, and it is decreasing in $r$. We have for $r \ge \ES_p(X)$ that 
 \[
 \E\left[\frac{L(\VaR_p(X),X)-\VaR_p(X)}{r-\VaR_p(X)}\right] = \frac{\ES_p(X)-\VaR_p(X)}{r - \VaR_p(X)} \le 1.
 \]
 Furthermore, for $z < r \le \ES_p(X)$, 
\[
\E\left[\frac{L(z,X)-z}{r-z}\right] \ge \frac{\ES_p(X)-z}{r-z} \ge 1
\] 
with equality if and only if $r = \ES_p(X)$.
\end{proof}

While Examples \ref{ex:1}-\ref{ex:3}  and Theorem \ref{th:ep} show that interesting backtest e-statistics exist, much more can be said about their general structure; see Section \ref{sec:techresults}.

\section{E-backtesting risk measures}\label{sec:etest}

We next present general methodology for backtesting risk measures via e-statistics in a sequential setting. 
We will use one-sided e-statistics in Definition \ref{def:1}.
Backtest e-statistics in Definition \ref{def:2} are relevant to have powerful backtests (see Section \ref{sec:betting}) but we first consider validity in this section.

\subsection{General risk measures}
Let $T$ be any time horizon, which may be fixed, infinite, or adaptive, i.e., depending on the data observed.
Let the $\sigma$-algebra $\mathcal F_t$ represent all available information up to time $t\in [T]$, such that $\mathcal F_m \subseteq\mathcal F_n$ for all $m\le n$. Let  $(L_t)_{t\in  [T] }$ be a sequence of realized losses that are adapted to the filtration $(\mathcal F_t)_{t\in  [T] }$.
Denote by  $\rho(L_t|\mathcal F_{t-1})$ and $\phi(L_t|\mathcal F_{t-1})$, respectively, the values of $\rho$ and $\phi$ applied to the conditional distribution of $L_t$ given $\mathcal{F}_{t-1}$.
Let $r_t$ and $z_t$ be forecasts for $\rho (L_t|\mathcal F_{t-1})$ and $\phi (L_t|\mathcal F_{t-1})$ made at time $t-1$, respectively.
Note that $\rho (L_t|\mathcal F_{t-1})$ and $\phi (L_t|\mathcal F_{t-1})$ are  random variables and $\mathcal F_{t-1}$-measurable for all functionals of interest (see e.g., \citealp{FisslerHolzmann2022}).

We assume that the risk forecasts $r_t$ and $z_t$ are obtained based on past market information and all other possible factors that may affect the decisions of risk predictors in financial institutions. For instance, the information may even include throwing a die or random events such as coffee spilling; all these events up to time $t-1$ are included in $\mathcal F_{t-1}$.

We test the following null hypothesis:
\begin{equation}\label{eq:H0b}
H_0: ~~ r_t\ge \rho(L_t|\mathcal F_{t-1})~\mbox{and}~ z_t = \phi(L_t|\mathcal F_{t-1})~~
 \mbox{ for $t\in [T]$}.
 \end{equation} 
 Rejecting \eqref{eq:H0b} implies, in particular, rejecting 
 $r_t= \rho(L_t|\mathcal F_{t-1})$ and $z_t = \phi(L_t|\mathcal F_{t-1})$ for all $t\in [T]$.
 In the special case that $d=1$ (i.e.~we do not need auxiliary information from $\phi$), \eqref{eq:H0b} becomes  
\begin{equation*}
H_0: ~~ r_t\ge \rho(L_t|\mathcal F_{t-1})~~
 \mbox{ for $t\in [T]$}.
 \end{equation*} 
 
 \begin{remark}
 Since $\rho$ is the regulatory risk measure of interest, 
 over-predicting $\rho$  is conservative. 
 On the other hand, $\phi$ represents some additional statistical information and it  may not relate to measuring financial risk. Hence, over-predicting $\phi$  is not necessarily conservative. See Example \ref{ex:over_var} below for a sanity check.
Therefore, it is more natural to test an equality of the auxiliary information $z_t$ in \eqref{eq:H0b} instead of an inequality; note also that this hypothesis is still more lenient than testing a specified loss distribution.
For the case where a financial institution is conservative for both the risk measures $\rho$ and $\phi$, see Section \ref{sec:cons}.
 \end{remark}

For a non-negative function $e:\R^{d+1}\to[0,\infty]$, let $X_t=e(L_t,r_t,z_t)$ for each $t$. We construct the following stochastic process: $M_0=1$ and
\begin{equation}\label{eq:mtg}
M_t(\bm{\lambda})  = (1-\lambda_t +\lambda_t X_t)M_{t-1}(\bm{\lambda}) = \prod_{s=1}^t  (1-\lambda_s+\lambda_s X_s),~~t\in [T],
\end{equation}
where the process $\bm{\lambda}=(\lambda_t)_{t\in [T]}$ is chosen  such that $\lambda_t$ is a function of $(L_{s-1},r_s,z_s)_{s\in[t]}$ and takes values in $[0,1]$ for $t\in [T]$.\footnote{More generally, we may allow $\lambda_t$ to be  $\mathcal{F}_{t-1}$-measurable instead of $\sigma((L_{s-1},r_s,z_s)_{s\in [t]})$-measurable, but this adds no further methodological value.}
Suppose that $e$ is a  {one-sided e-statistic}   for $(\rho,\phi):\M\to \R\times \R^{d-1}$.
In the construction of \eqref{eq:mtg}, the user-chosen inputs are the process $\bm{\lambda}$ and the e-statistic $e$.

We have by definition that $X_t$ is an e-variable conditional on $\mathcal F_{t-1}$ under $H_0$ for all $t\in [T]$.
As suggested by \cite{VovkWang2020}, the only admissible (or unwasteful) way to   combine these e-variables is through the martingale function \eqref{eq:mtg}; see also the universal representation result in Proposition 3 of \cite{WaudbySmithRamdas2020}.
The e-process in \eqref{eq:mtg} may be interpreted as the payoffs of a betting strategy against the null hypothesis $H_0$ as in \cite{ShaferVovk2019}. In this betting game, the initial capital is $M_0=1$ and all the money is invested at each step, split between a random payoff $X_t$ and a fixed payoff $1$. The payoff per capital at each step is $1-\lambda_t+\lambda_t X_t$ for $t\in [T]$. As a result, the player earns money at step $t$ if $X_t>1$, meaning that there is some evidence against the null hypothesis, which is the interpretation of an e-value larger than $1$. In this sense, we call the process $\bm{\lambda}$ in \eqref{eq:mtg} a \emph{betting process} (see step (iii) in Section \ref{sec:R2-2}).
In particular, 
$\lambda_t<1$ can be interpreted as keeping part of one's money in the pocket, and $\lambda_t=0$ means one does not bet at all in this step.
The following theorem follows from Ville's well-known inequality \citep{Ville1939}, indicating that $(M_t(\bm{\lambda}))_{t\in \{0,\dots,T\} }$ in \eqref{eq:mtg} is a non-negative supermartingale, so in particular an e-process under the null hypothesis in \eqref{eq:H0b}. 
\begin{theorem}\label{lem:Ville}
Suppose that $e$ is a  {one-sided e-statistic}   for $(\rho,\phi):\M\to \R\times \R^{d-1}$. Under $H_0$ in \eqref{eq:H0b}, $(M_t(\bm{\lambda}))_{t\in \{0,\dots,T\}}$ in \eqref{eq:mtg} is a  non-negative supermartingale with $M_0=1$, 
 and  for each $\alpha \in (0,1)$,
 $$
 \p\left(\sup_{t\in \{0,\dots,T\}} M_{t}(\bm{\lambda}) \ge \frac 1{\alpha} \right)\le \alpha.
 $$
 \end{theorem} 

Based on Theorem \ref{lem:Ville}, we will use the e-test that arises from  the e-variable $M_\tau(\bm{\lambda})$ where $M(\bm{\lambda})$ is the e-process given by \eqref{eq:mtg} and $\tau$ is the stopping time $\min\{T,\inf\{t\ge 0: M_t(\bm{\lambda}) \ge  1/\alpha\}\}$ with $\inf\varnothing =\infty$. This is common practice in testing with e-values.

\begin{remark}\label{rem:R1-threshold}
The choice of $\alpha$ in the threshold $1/\alpha$ for e-processes has been extensively discussed in the literature; see \cite{Shafer2021, VovkWang2021, GrunwaldHeideETAL2020} and \cite{WangRamdas22}.
 Theorem \ref{lem:Ville} only gives a minimal guarantee of type-I error $\alpha$, but the guarantee is actually much stronger, although not in terms of a provable smaller type-I error. 
In particular, it gives an \emph{anytime-valid} type-I error control, that is, validity under arbitrary stopping times  and optional continuation. 
The empirical type-I error is usually much smaller than $\alpha$ for a moderate sample size (see Section \ref{sec:sim_stu}), and it can be close to $\alpha$ for a very large sample size (see Section \ref{app:type1}).
Choosing the threshold also depends on the objective of the backtest. E-value and p-value tests require different types of guarantee in general. If the goal is for early warning and not a decisive rejection, as in financial regulation, then a threshold $2$ or $5$ is useful. If the goal is to reject the null with a provable type-I error $\alpha$, as in the traditional setting,   then a   threshold $1/\alpha$ should be chosen. In our simulation and empirical studies, we will use the thresholds $2$, $5$ and $10$, which (although only with type-I error guarantee of 50\%, 20\% and
10\% by Theorem \ref{lem:Ville})
roughly correspond to minor, substantial and strong evidence for e-values according to \cite{VovkWang2021}.
\end{remark}

\subsection{E-backtesting Value-at-Risk and Expected Shortfall}
\label{sec:cons}

To put our general ideas in the context of financial regulation, we focus on backtesting VaR and ES in this section. Let $L_t$ be the random loss at time $t$. For the case of backtesting VaR, $(r_t)_{t\in [T]}$ are the forecasts for $\VaR_p(L_t|\mathcal F_{t-1})$, $p\in (0,1)$. As we see in Example \ref{ex:5}, the function $e^Q_p(L_t,r_t)$ defined in \eqref{eq:eqp} is an e-variable under the following null hypothesis that we are testing:
\begin{equation}
H_{0}: ~~\mbox{$r_t\ge \VaR_p(L_t|\mathcal F_{t-1} )$}, ~~~~t\in [T].
\label{eq:R2-H0VaR}
 \end{equation} 

For backtesting  ES,
$(r_t)_{t\in [T]}$ and $(z_t)_{t\in [T]}$ are the forecasts for $\ES_p(L_t|\mathcal F_{t-1})$ and $\VaR_p(L_t|\mathcal F_{t-1})$, $p\in (0,1)$, respectively.
By Theorem \ref{th:ep},  $e^{\mathrm{ES}}_p(L_t,r_t,z_t)$ is an e-variable under the null hypothesis
\begin{equation}\label{eq:H0t}
H_{0}: ~~\mbox{$r_t\ge \ES_p(L_t|\mathcal F_{t-1} )$ and $z_t=\VaR_p(L_t |\mathcal F_{t-1})$}, ~~~~t\in [T].
 \end{equation}

 In practice, a financial institution may use a conservative model for risk management purposes, which leads to underestimation of both VaR and ES. 
 In the following proposition, we illustrate that $e^{\mathrm{ES}}_p(L_t,r_t,z_t)$ is a valid e-variable in case both $\VaR_p(L_t|\mathcal{F}_{t-1})$ and $\ES_p(L_t|\mathcal{F}_{t-1})$ are over-predicted, together with their difference.  
  \begin{proposition}\label{prop:conservative}
  For $p\in (0,1)$,
  $(e^{\mathrm{ES}}_p(L_t,r_t,z_t))_{t\in [T]}$ are e-variables for
  \begin{equation}\label{eq:H0c}
H_{0}: ~z_t\ge \VaR_p(L_t|\mathcal F_{t-1})~~\mbox{and}~~ r_t-z_t\ge \ES_p(L_t|\mathcal F_{t-1}) - \VaR_p(L_t|\mathcal F_{t-1}),~t\in [T].
 \end{equation}
  \end{proposition}

In practice, the equality 
$z_t=\VaR_p(L_t|\mathcal F_{t-1})$ in \eqref{eq:H0t} is unlikely to hold exactly. For example, even with correctly specified forecasting models, there will still be estimation error in parameters. 
The obtained e-values should be seen as a quantification of how strong the evidence is against $H_0$; this is a useful feature of e-values, as discussed by e.g., \cite{Shafer2021} and \cite{GrunwaldHeideETAL2020}. On the other hand,
 $H_0$ in \eqref{eq:H0c} can hold in practice as it is formulated using inequalities.

The hypothesis $H_0$ in \eqref{eq:H0c}
is stronger than 
   \begin{equation}\label{eq:both-conser}
H_{0}: z_t \ge \VaR_p(L_t|\mathcal{F}_{t-1}) ~~\mbox{and}~~  r_t \ge \ES_p(L_t|\mathcal{F}_{t-1}).
 \end{equation}
In contrast to \eqref{eq:H0c}, $e^{\mathrm{ES}}_p(L_t,r_t,z_t)$ is not necessarily an e-variable for \eqref{eq:both-conser}.
For instance, $\E[e^{\mathrm{ES}}_p(L_t,r_t,z_t)]=\infty$ if $r_t=z_t=\ES_p(L_t|\mathcal{F}_{t-1})$ and $\p(L_t>z_t)>0$.
It implies that over-predicting $\VaR_p(L_t|\mathcal{F}_{t-1})$ does not always lead to a smaller e-value $e^{\mathrm{ES}}_p(L_t,r_t,z_t)$; in contrast, over-predicting $\ES_p(L_t|\mathcal{F}_{t-1})$ always reduces the resulting e-value. The following example shows that a poor VaR forecast could result in large e-values although it is obtained by over-prediction and satisfies \eqref{eq:both-conser}.

\begin{example}\label{ex:over_var}
For $p\in(0,1)$, a continuously distributed random variable $X$ with $a=\VaR_p(X)<\ES_p(X)=1$ (this implies $\p(X\le 1)<1$) has a heavier tail than $Y$ with $\VaR_p(Y)=\ES_p(Y)=1$ (this could happen if $Y$ has a point-mass at $1$, and it implies $\p(Y\le 1)=1$). Thus, intuitively, a forecaster producing the random loss $X$ is more conservative than that producing $Y$. This shows that over-predicting both $\VaR_p$ and $\ES_p$ does not always mean that the forecaster is more conservative about the risk. Our  backtest e-statistic $e^{\ES}_p$ can detect this, because  $\E[e^{\ES}_p(X,1,a)]\le 1$ while $\E[e^{\ES}_p(X,1,1)]=\infty$, thus correctly rejecting the forecast $(1,1)$ of $(\ES_p(X),\VaR_p(X))$ but not rejecting the truthful forecast $(1,a)$, although $a<1.$
\end{example}
 
   The following example collects some practical situations of conservative forecasts. In each case,  $e^{\mathrm{ES}}_p(L_t,r_t,z_t)$ yields a valid e-variable. 
 \begin{example}
  \begin{enumerate}[(i)]
  \item $r_t= \ES_q(L_t|\mathcal{F}_{t-1})$ and $z_t=\VaR_q(L_t|\mathcal{F}_{t-1})$ for $q> p$:
    $$\E[e^{\mathrm{ES}}_p(L_t,r_t,z_t)]=\frac{1-q}{1-p}  \E[e^{\ES}_q (L_t, \ES_q(L_t|\mathcal{F}_{t-1}),\VaR_q(L_t|\mathcal{F}_{t-1}))]=\frac{1-q}{1-p}<1.$$   In this situation, $\VaR_p$ and $\ES_p$ are over-predicted by lifting the confidence level $p$ to $q$. For instance, this may represent the output of a stress-testing scenario which amplifies the probability of extreme losses. 
  \item  $r_t=c_1 \ES_p(L_t|\mathcal{F}_{t-1})$ and $z_t=c_2 \VaR_p(L_t|\mathcal{F}_{t-1})\ge 0$ for $c_1\ge c_2 \ge 1$, as justified by Proposition \ref{prop:conservative}. 
In particular, $\VaR_p$ and $\ES_p$ can  be over-predicted by the same  multiplicative factor.
  \item $r_t=\ES_p(L_t|\mathcal{F}_{t-1}) + b_1$ and $z_t= \VaR_p(L_t|\mathcal{F}_{t-1})+b_2$ for $b_1\ge b_2 \ge 0$, as justified by Proposition \ref{prop:conservative}. 
In particular, $\VaR_p$ and $\ES_p$ can  be over-predicted by the same  absolute amount.
  \end{enumerate} 
  \end{example}

\begin{remark}
Our e-backtesting method of ES and the method based on cumulative violations introduced in \cite{DuEscanciano2017} (we call it the cumulative violation method) have several different features. First, the cumulative violation method requires distributional forecasts $\hat{u}_t(\hat{\theta})$ as input based on some parametric distribution; our e-backtesting method needs ES and VaR forecasts that can be arbitrarily reported. This provides more flexibility in practice in the sense that our method does not require special treatment of estimation effects as in \cite{DuEscanciano2017} and \cite{HogaDemetrescu2022}. Second, the cumulative violation method is a two-sided test and focuses on detecting model misspecification; our method is a one-sided test focusing only on the underestimation of ES. This means that we do not reject the null as long as ES is not underestimated even though the forecasts are obtained based on a wrong model or no specific model is assumed. Third, the cumulative violation method relies on a fixed sample size $T$ and an asymptotic model, which means its statistical validity requires it to be only evaluated at the end of the sampling period $T$ that is large enough; our method is sequential and is valid at any stopping time, where detections can be achieved much earlier. This is desirable in risk management applications in timely detecting insufficient risk predictions. Most other classical backtesting methodologies become invalid when evaluated before the end of the pre-specified time period set for testing; see Table \ref{tab:comp}. %
\end{remark}

\begin{remark}
    \label{rem:R1-1}
As explained in Section \ref{sec:2}, our e-processes are designed to test underestimation of risk measures.
If one is interested in testing overestimation, for the case of VaR, one can simply use $-L_t$ instead of $L_t$, and build an e-process.
This leads to a two-sided test by averaging  two e-processes, one for each side. On the other hand, testing overestimation of ES seems to be very challenging, and our current method does not apply. This asymmetry between the two sides is due to the nature of ES as a one-sided tail risk measure.
\end{remark}

\section{Choosing the betting process}\label{sec:betting} 
 
One of the essential steps in the testing procedure is choosing a betting process  $\bm{\lambda}=(\lambda_t)_{t\in [T]}$ in \eqref{eq:mtg}. Throughout this section,   $e$ is a $\mathcal{P}$-one-sided e-statistic for $\psi=(\rho,\phi):\M\to\R\times \R^{d-1}$ and $\mathcal{P}\subseteq\M$. We omit $\mathcal{P}$ when $\mathcal{P}$ is the domain of $\psi$. Any predictable process $\bm{\lambda}$ with values in $[0,1]$ yields a supermartingale in \eqref{eq:mtg} under $H_0$, and thus the testing procedure is valid at all stopping times by Theorem \ref{lem:Ville}. However, the statistical power of the tests, and the growth of the process $(M_t(\bm{\lambda}))_{t\in\{0,\dots,T\}}$ if the null hypothesis $H_0$ is false, heavily depends on a good choice of  $\bm{\lambda}$. 
The betting process is chosen by the tester, e.g., a regulator or an internal model risk examiner.

\subsection{GRO, GREE, GREL and GREM methods}  
Our methods are related to maximizing the expected log-capital originally proposed by \cite{Kelly1956}, adopted by \cite{GrunwaldHeideETAL2020} in their GRO (growth-rate optimal) criterion, and studied by \cite{Shafer2021} and \cite{WaudbySmithRamdas2020} for testing by betting. For an e-variable $E$ and a probability measure $Q$ representing an alternative hypothesis, the key quantity to consider is $\E^Q[\log E]$, which is called the e-power of $E$ under  $Q$ by \cite{vovk2024nonparametric}. 
 
Let $T$ be the time horizon of interest, which can be a finite integer or $\infty$.
Let $Q_t$, $t\in [T]$, be 
the distribution of $L_t$ given the information contained in $\mathcal F_{t-1}$.
When choosing the betting process $(\lambda_t)_{t\in [T]}$, we fix an upper bound $\gamma\in(0,1)$ and restrict $\lambda_t\in[0,\gamma]$ for all $t\in [T]$. 
The upper bound $\gamma$ is not restrictive and only prevents some ill-behaving cases. We can set $\gamma=1/2$ in our context (see Remark \ref{rem:gamma} below).
Below we formally introduce a few methods to determine the betting process. 

1. \textbf{GRO} (growth-rate optimal): We compute $\bm{\lambda}^{\mathrm{GRO}}=(\lambda_t^{\mathrm{GRO}})_{t\in [T]}$ by
\begin{equation}\label{eq:GROW1}
\lambda^{\mathrm{GRO}}_t = \lambda^{\mathrm{GRO}}_t(r,z) =\argmax_{\lambda\in [0,\gamma]}\E^{Q_t}[\log(1-\lambda+\lambda e(L_t,r,z))],
\end{equation} 
and we plug in $(r,z)=(r_t,z_t)$. 
The optimal $\lambda_t^\mathrm{GRO}$ in \eqref{eq:GROW1} can be calculated through a convex program as the function $\lambda \mapsto \log(1-\lambda+\lambda e(L_t,r,z))$ is concave.
This requires the knowledge of the conditional distribution $Q_t$  of $L_t$ given $\mathcal{F}_{t-1}$.
In practice, $Q_t$ is unknown to the tester, and one may need to choose a model to approximate $Q_t$. 
If the probability measure $Q_t$ is unknown but from a certain family, one can use the method of mixtures or mixture martingales \citep[see e.g.,][]{delaPenaETAL2004, delaPenaETAL2009} of alternative scenarios to obtain an e-process close to that based on the unknown true model. 
The optimizer $\lambda_t^{\rm GRO}$ may not be unique in some special cases, e.g., $e(L_t,r,z) $ is the constant  $1$ or the expectation in \eqref{eq:GROW1} is $\infty$, but it is unique in most practical cases.

Without specifying a particular model for the alternative hypothesis, we propose the following three methods to choose $\bm \lambda$, all trying to approximate the  distribution $Q_t$ under the alternative hypothesis using the sample until time point $t-1$. 

2. \textbf{GREE} (growth-rate for empirical e-statistics):
Let $E$ follow  the empirical distribution of the sample $e(L_s,r_s,z_s)_{s\le t-1}$. We compute $\bm{\lambda}^{\mathrm{GREE}}=(\lambda_t^{\mathrm{GREE}})_{t\in [T]}$  by 
\begin{equation}
\label{eq:GREE}
\begin{aligned}
\lambda^{\mathrm{GREE}}_t&=\argmax_{\lambda\in [0,\gamma]}\E[\log(1-\lambda+\lambda E)]  
 =\argmax_{\lambda\in [0,\gamma]}\frac{1}{t-1}\sum^{t-1}_{s=1}\log(1-\lambda+\lambda e(L_s,r_s,z_s)).
\end{aligned}
\end{equation} 

3. \textbf{GREL} (growth-rate for empirical losses): 
Let $L$ follow   the empirical distribution of the sample $(L_s)_{s\le t-1}$. We compute $\bm{\lambda}^{\mathrm{GREL}}=(\lambda_t^{\mathrm{GREL}})_{t\in [T]}$ by
\begin{equation}
\label{eq:GREL}
\begin{aligned}
\lambda^{\mathrm{GREL}}_t= \lambda^{\mathrm{GREL}}_t(r,z)&=\argmax_{\lambda\in [0,\gamma]}\E[\log(1-\lambda+\lambda e(L,r,z))]\\&=\argmax_{\lambda\in [0,\gamma]}\frac{1}{t-1}\sum^{t-1}_{s=1}\log(1-\lambda+\lambda e(L_s,r,z)),
\end{aligned}
\end{equation}
and we plug in $(r,z)=(r_t,z_t)$.

Both \eqref{eq:GREE}  and \eqref{eq:GREL} are plug-in methods for the measure $Q_t$, one using sample e-statistics and the other using sample losses.
The problems \eqref{eq:GREE} and \eqref{eq:GREL} can be solved directly via convex programming.
The GREE and GREL methods are equivalent when the risk forecasts $r_t$ and $z_t$ are constant across $t\in [T]$.
 To explain the main difference between GREE and GREL, GREE decides $\lambda_t$ without using the current forecast $(r_t,z_t)$ as input but uses $(r_s,z_s)$ for $s<t$,
whereas GREL uses $(r_t,z_t)$ to compute $\lambda_t$, but discards  $(r_s,z_s)$ for $s<t$. The forecast $(r_t,z_t)$ supplied by the bank could be informative, noisy, or even misleading, and there is no uniformly superior choice. 
The GREE and GREL methods are asymptotically optimal in different practical situations, which we study in Section \ref{sec:R2-52}. For both methods, one may use a moving window to compute the empirical distributions instead of using all previous data points; see Section \ref{sec:emp} for financial data analysis.

To take advantage of both methods, we propose an e-process by taking the average.

4. \textbf{GREM} (growth-rate for empirical mixture): We choose $\bm{\lambda}^{\mathrm{GREM}}=(\lambda_t^{\mathrm{GREM}})_{t\in [T]}$ such that 
$$M_{t}(\bm{\lambda}^{\mathrm{GREM}})=\frac{M_{t}(\bm{\lambda}^{\mathrm{GREE}})}{2}+\frac{M_{t}(\bm{\lambda}^{\mathrm{GREL}})}{2},$$
where $M$ is defined in \eqref{eq:mtg}. 
By Lemma 1 of \cite{VovkWang2020}, there exists a betting process for the GREM method. More precisely,
$$\lambda^{\mathrm{GREM}}_t=\frac{M_{t-1}(\bm{\lambda}^{\mathrm{GREE}})\lambda^{\mathrm{GREE}}_t+M_{t-1}(\bm{\lambda}^{\mathrm{GREL}})\lambda^{\mathrm{GREL}}_t}{M_{t-1}(\bm{\lambda}^{\mathrm{GREE}})+M_{t-1}(\bm{\lambda}^{\mathrm{GREL}})}.$$
The GREM method is asymptotically optimal for the practical cases when either the GREE or the GREL method is optimal (see Theorem \ref{th:opt}). Other ways of averaging than using equal weights is possible, if the tester has side information on which method likely works better.

Our construction of the e-processes is based on the same idea that underlines the GRAPA method introduced in \citet[Section B.2]{WaudbySmithRamdas2020}. Two main differences are that we build e-variables from backtest e-statistics and we use risk forecasts as input variables for the betting strategy. Constructing a betting process depending on predictions has previously been explored by \citet{HenziZiegel2021}.

In all simulation and empirical results, we use the monotone backtest e-statistics 
$e=e_p^Q$ for $\VaR_p$ and $e=e_p^\ES$ for $(\ES_p,\VaR_p)$. These choices are justified by results in Section \ref{sec:techresults}, which we briefly explain below.

\begin{remark}\label{rem:GRO}
Usually, growth-rate optimality (GRO) refers to the optimality among the set of \emph{all} e-variables for the null $r \ge \rho(L_t|\mathcal{F}_{t-1})$, $z = \phi(L_t |\mathcal{F}_{t-1})$, conditional on $\mathcal F_{t-1}$. \emph{A priori}, GRO as defined at \eqref{eq:GROW1} is just a restricted GRO, that is, GRO amongst the subset of e-variables of the form $1-\lambda+\lambda e\left(L_t, r, z\right)$. However, as we will see from results in Section \ref{sec:techresults}, for the important examples of the mean, the variance, and VaR, the seemingly restricted GRO is actually the usual unrestricted GRO, which implies essentially that the presented approach cannot be improved without imposing specific model assumptions. The same holds true for the case of ES under an additional assumption.
\end{remark}

A simple way to get an approximation of \eqref{eq:GREE} and \eqref{eq:GREL} is to use a Taylor expansion $\log(1+y)\approx y-y^2/2$ at $y=0$  and the first-order condition. 
This leads to  
\begin{equation}\label{eq:GREE_em}
\lambda^{\mathrm{GREL}}_t\approx 0\vee\frac{\sum^{t-1}_{s=1}e(L_s,r,z)-t+1}{\sum^{t-1}_{s=1}(e(L_s,r,z)-1)^2}\wedge \gamma
\end{equation}
for  the GREL method  \eqref{eq:GREL},
where    $a \vee b$ represents   the maximum of $a,b$ and $a \wedge b $ represents the minimum.
We replace $(r,z)$ in \eqref{eq:GREE_em} by $(r_s,z_s)$ for the GREE method \eqref{eq:GREE}.  The special cases of \eqref{eq:GREE_em} for VaR and ES are given in Section \ref{app:vares}.

\begin{remark}\label{rem:gamma}
    We restrict the betting process by the upper bound $\gamma < 1$ to avoid the e-process collapsing to $0$. For illustration, suppose that for each $t$, $X_t=e(L_t,r,z)$ given $\mathcal{F}_{t-1}$ takes   value $0$ with small probability and   value $2$ with large probability, so its expected value is larger than $1$ and the null hypothesis is not true. 
    As long as we do not observe $0$ up to time $t$, the empirical distribution is concentrated at $2$, leading to an optimal strategy $\lambda_t^{\rm GREE}=1$. This betting process yields an e-process that becomes $0$ as soon as we observe a $0$ from $X_t$ and therefore should be avoided.  
In all our numerical and data experiments, the optimal $\lambda_t$ from each method is typically quite small ($<0.1$) for tail risk measures like VaR and ES. Hence, it is harmless to set $\gamma=1/2$ by default in our context.
A different approach to obtain the betting process is to put a distribution on $\lambda$ and estimate by a ``posterior"; see \cite{agrawal2021optimal} and \cite{jang2023tighter}.
\end{remark}

\begin{remark}\label{rem:R1-under}
In our setting, the financial institution can arbitrarily report risk forecasts. This allows the possibility for the bank to ``game" the regulator by intentionally over-reporting the risk forecasts for a period of time before it starts under-reporting in another period of time. Such a gaming problem also happens for most traditional backtesting methods. Our method provides an efficient solution to such an issue by using a rolling window (e.g., 250 or 500 days) to calculate the betting process instead of using all past data. During the over-prediction period, the e-values $e(L_t,r_t,z_t)$ tend to be small on average with mean smaller than $1$. By \eqref{eq:GREE_em} and Proposition \ref{lem:lambda01} below, the resulting betting process with the GREE or GREL method will be equal or close to $0$, yielding an e-process being almost a constant $1$. After the bank starts to under-predict, the past data will exit the rolling window gradually. The GREE or GREL method will pick up the large e-values and start to choose a positive betting process. Moreover, how  the betting process will be chosen is not known to the bank.\footnote{Under the null hypothesis,  the e-process $M(\boldsymbol \lambda)$ is a supermartingale, no matter how $\boldsymbol \lambda$ is chosen. Therefore, the regulator can announce $\lambda_t$ 
after seeing the forecast $(r_t,z_t)$ and 
before seeing the next loss data point $L_t$.} This also increases the cost of gaming by the bank since the regulator can actively choose a positive betting process to detect evidence against the under-reports if she has sufficient reason to realize the bank under-predicts for a certain period. 
More details and simulation studies can be found in Section \ref{app:R1-EC-5}. 

\end{remark}

\subsection{Optimality of betting processes}\label{sec:R2-52}

Next, we discuss the optimality of the betting process. 
We first define an intuitive notion of asymptotic optimality.
For asymptotic results discussed in this section, we will assume an infinite time horizon; that is, we consider $t\in \N$. 
All statements on probability and convergence are with respect to the true probability generating the data.
\begin{definition}
For $(L_{t-1},r_t,z_t)_{t\in\N}$ adapted to $(\mathcal{F}_{t-1})_{t\in\N}$ and a given function $e:\R^{d+1}\to[0,\infty]$,
\begin{enumerate}[(i)]
\item two betting processes $\boldsymbol \lambda=(\lambda_t)_{t\in\N}$  and $\boldsymbol \lambda'=(\lambda'_t)_{t\in\N}$ are \emph{asymptotically equivalent}, denoted by $\boldsymbol \lambda \simeq \boldsymbol  \lambda'$, if
$$\frac{1}{T} (\log M_{T}(\boldsymbol \lambda ) - \log M_{T}(\boldsymbol \lambda' ) )\xrightarrow{L^1}0 \mbox{~~~
as $T\to\infty$},$$
where $M$ is defined in \eqref{eq:mtg};\footnote{The $L^1$-convergence $Z_T\xrightarrow{L^1} Z $ means $\E[|Z_T-Z|] \to 0$ as $T\to \infty$.}
\item a betting process $\boldsymbol \lambda $ is \emph{asymptotically optimal} if
 {$\boldsymbol \lambda \simeq (\lambda^{\mathrm{GRO}}_t(r_t,z_t))_{t\in\N}$}.
\end{enumerate}
\end{definition}
Intuitively, the asymptotic equivalence between two betting processes means that the long-term growth rates of the two resulting e-processes are the same. The asymptotic optimality of a betting process is defined by the asymptotic equivalence using the GRO method as a benchmark because GRO is the best-performing method if we know the full distributional information of the losses.

The following proposition characterizes the situations where the betting processes in the GRO method do not reach $0$ and $1$. In our formulation, $\bm\lambda$ is not allowed to reach $1$ due to the upper bound $\gamma<1$, but we nevertheless give a theoretical condition that the unconstrained optimizer is less than $1$.

\begin{proposition}\label{lem:lambda01}
For   $(r,z)\in\R\times\R^{d-1}$ and $t\in\N$ and any optimizer $\lambda^{\mathrm{GRO}}_t$ of \eqref{eq:GROW1}, the following statements hold.
\begin{enumerate}[(i)]
    \item $\lambda^{\mathrm{GRO}}_t(r,z)>0$ if and only if $\E^{Q_t}[e(L_t,r,z)|\mathcal{F}_{t-1}]>1$.
    \item With $\gamma=1$ in \eqref{eq:GROW1},   $\lambda^{\mathrm{GRO}}_t(r,z)<1$ if and only if $\E^{Q_t}[1/e(L_t,r,z)|\mathcal{F}_{t-1}]> 1$.
\end{enumerate}
\end{proposition}
The observation in Proposition \ref{lem:lambda01} also holds true for $Q_t$ replaced by an empirical measure as in  \eqref{eq:GREE} and \eqref{eq:GREL}. Hence, GREE or GREL will choose $\lambda_t>0$ only when the empirical mean of the e-statistic (specified differently in the two methods) is larger than $1$.

 Below we present an assumption for  the asymptotic analysis. The condition is very weak because the interesting case in backtesting is when $\E^{Q_t}[\log (e(L_t,r,z))]$ is small. 
Denote by $\psi^*(\mathcal P) \subseteq\psi(\mathcal{P})$ as the set of all values $(r,z)\in \psi(\mathcal{P})$ such that $e(x,r,z)<\infty$ for all $x\in\R$.  
\begin{assumption}\label{ass:finite}
For all $(r,z)\in\psi^*(\mathcal P)$, {$\sup_{t\in \N }\E^{Q_t}[|\log(e(L_t,r,z))|]<\infty$}.
\end{assumption}

The following theorem addresses the asymptotic optimality of the GREE, GREL and GREM methods in different situations, as well as their consistency. We say that a method with e-process $M$  has asymptotic full power if $Q(\sup_{t\in \{0,\dots,T\}} M_t\ge 1/\alpha)\to 1$ as $T\to\infty$ for all $\alpha \in (0,1)$, where $Q$ is the data generating probability.

\begin{theorem}\label{th:opt}
For $(L_{t-1},r_t,z_t)_{t\in\N}$ adapted to $(\mathcal{F}_{t-1})_{t\in\N}$ such that $(r_t,z_t)$ takes values in $\psi^*(\mathcal P)$ and $e:\R^{d+1}\to[0,\infty]$, under Assumption \ref{ass:finite} {and using $\gamma=1$ in the betting strategies}, the following statements hold.
\begin{enumerate}[(i)]
\item  $(\lambda^{\mathrm{GREE}}_t)_{t\in\N} $ is asymptotically optimal  if   $(e(L_t,r_t,z_t))_{t\in\N}$ is iid and $(r_t,z_t)_{t\in\N}$ is deterministic.
\item   $(\lambda^{\mathrm{GREL}}_t(r_t,z_t))_{t\in\N} $  is asymptotically optimal  if $(L_t)_{t\in\N}$ is iid and either:
\begin{enumerate}[(a)]
\item  $(r_t,z_t)_{t\in\N}$ takes finitely many possible values in $\R^d$.
\item $(r_t,z_t)$, $t\in\N$, are in a common compact set, $e(x,r,z)$ is continuous in $(r,z)$, and $(r_t,z_t)\xrightarrow{\mathrm{p}}(r_0,z_0)$ as $t\to\infty$ for some $(r_0,z_0)\in\R^d$.
\end{enumerate}
\item $(\lambda^{\mathrm{GREM}}_t)_{t\in\N}$ is asymptotically optimal if either $(\lambda^{\mathrm{GREE}}_t)_{t\in\N} $ or $(\lambda^{\mathrm{GREL}}_t(r_t,z_t))_{t\in\N} $ is asymptotically optimal.
\end{enumerate}
Moreover,  if $r_t<\rho(L_t|\mathcal F_{t-1})-\epsilon$ for some $\epsilon>0$ and each $t \in \N$, then GREE in  (i), GREL in  (ii), and GREM in  (i) and (ii) have asymptotic power one, using any backtest e-statistic $e$ for $\psi$.
\end{theorem}

Any choice of the betting strategy $\boldsymbol \lambda$ yields a valid and non-asymptotic test, and here  $T\to \infty$ is taken only to discuss optimality of the power. 
The asymptotic optimality results in Theorem \ref{th:opt} are based on strong, and perhaps unrealistic, assumptions; they are imposed for technical reasons. 
Nevertheless, we obtain some useful insight on the comparison between GREE and GREL.
Intuitively, the GREE method should outperform the GREL method when the backtest e-statistics $e(L_t,r_t,z_t)$, $t\in\N$, are iid and $(r_t,z_t)$ is not informative about how to choose $\lambda_t$ (i.e., they are noise), while the GREL method 
 should outperform  the GREE method when the losses $L_t$, $t\in\N$, are iid and  $(r_t,z_t)$ is informative on how to choose $\lambda_t$;\footnote{In this case,
we can statistically infer the value of $\rho(L_{t})$ from past data, and thus the values of $(r_t,z_t)$ are informative about whether it is likely an over-prediction (which means we should choose a large $\lambda_t$) or an under-prediction (which means we should choose a small or zero $\lambda_t$).}  recall that GREL uses  $(r_t,z_t)$ whereas GREE does not.
We illustrate the insights for the comparison between the GREE and GREL methods through Example \ref{rem:diff} below. In most practical cases, we  do not know clear patterns of the losses and forecasts as they arrive sequentially over time. In this sense, the GREM method is recommended. %

\begin{example}\label{rem:diff}
Let the size of training data be $l=10$, the sample size for testing be $n=1,000$, and $Z_1,\dots,Z_{n+l}$ be iid samples simulated from the standard normal distribution. We report the average performance of backtesting methods over $1,000$ simulations.
\begin{enumerate}[(a)]
\item The iid condition of the whole backtest e-statistics implies that the GREE method works better than the GREL method when losses and risk forecasts exhibit co-movements over time. Such situations are common in the financial market; for instance, risk forecasts will increase over time when a company is extending its business.
Assume that $L_t=(1+t/(n+l))Z_t$ for $t\in [n+l]$. This model represents the case where the financial institution's investment generates iid cash flow but the institution increases the investment amount over time. Following the increasing trend of the investment, the risk forecaster announces the under-estimated forecasts of $\VaR_{0.95}(L_t|\mathcal{F}_{t-1})$ and $\ES_{0.95}(L_t|\mathcal{F}_{t-1})$ as  $z_t=1.48(1+t/(n+l))$ and $r_t=1.86(1+t/(n+l))$, respectively, for $t\in [n+l]$. Figure \ref{fig:GREE} plots the realized losses $L_t$, ES forecasts $r_t$, and the e-processes obtained by the GRO, GREE, GREL and GREM methods for $t=l+1,\dots,n+l$.
\begin{figure}[t]
    \centering
    \caption{Realized losses and ES forecasts with a linear extending business (left panel); average log-transformed e-processes obtained by different methods over $1,000$ simulations (right panel)}
    \includegraphics[width=0.48\textwidth]{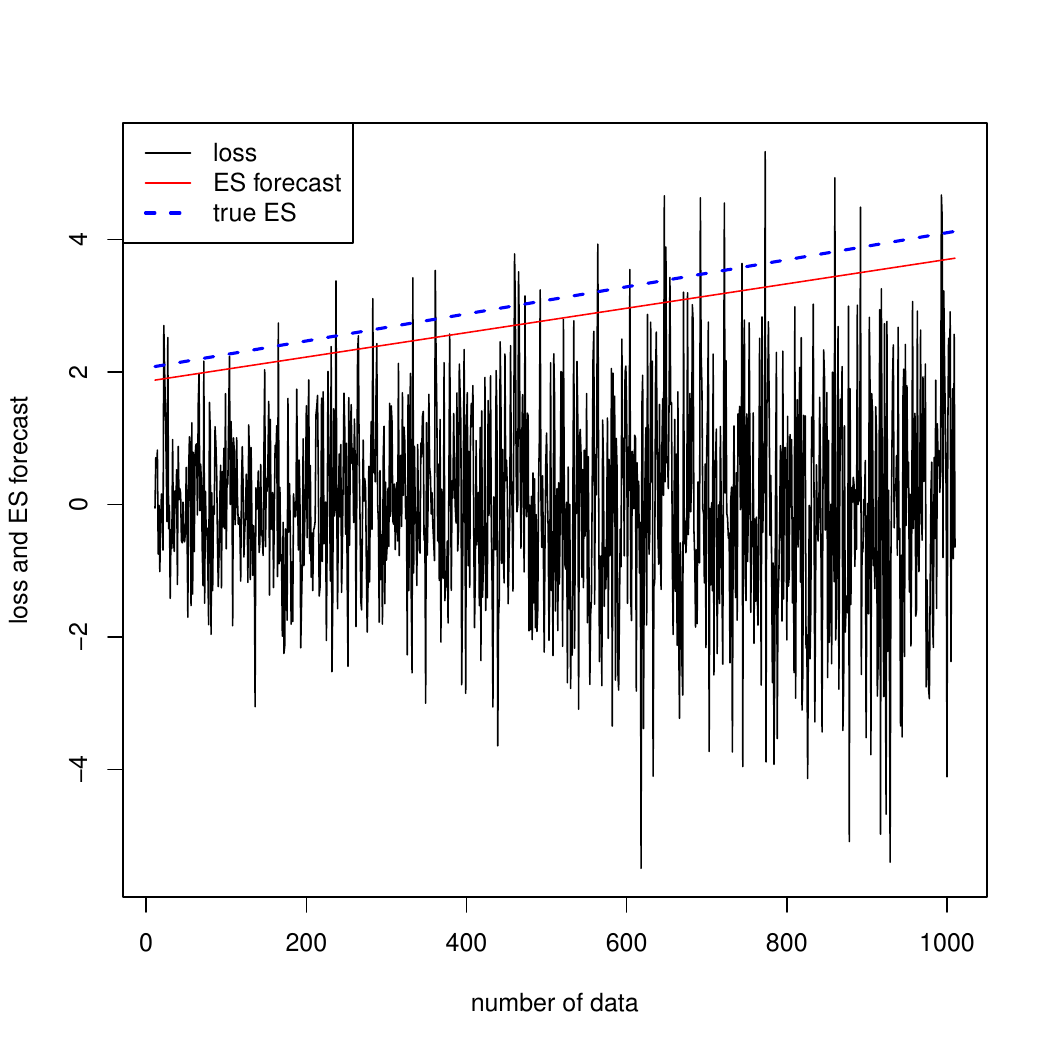}
    \includegraphics[width=0.48\textwidth]{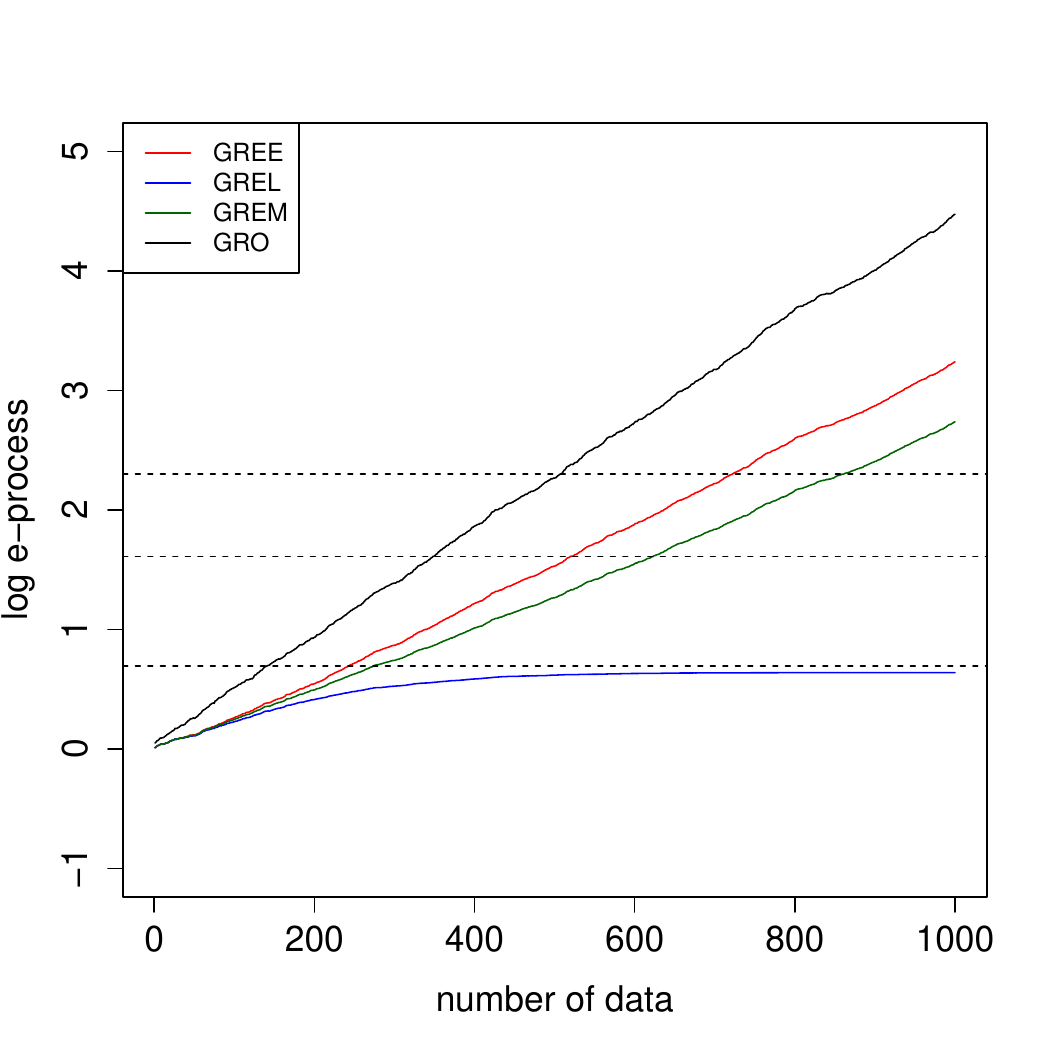}
    \label{fig:GREE}
\end{figure}
We observe from Figure \ref{fig:GREE} that the GREE e-process  dominates  the GREL e-process. This is consistent with the result of Theorem \ref{th:opt} by noting the co-movements of the losses and the VaR and ES forecasts which makes the backtest e-statistics $(e_p(L_t,r_t,z_t))_{t\in [n+l]}$ iid.

\item We consider a non-linear business cycle. Take the random losses to be $L_t=Z_t(1+\sin(\theta t))$ for $t\in [n+l]$, where $\theta=0.01$. The risk forecasts of $\VaR_{0.95}(L_t|\mathcal{F}_{t-1})$ and $\ES_{0.95}(L_t|\mathcal{F}_{t-1})$ also have a similar trend but are under-estimated. Namely, we have $z_t=1.48(1+\sin(\theta t))$ for VaR and $r_t=1.86(1+\sin(\theta t))$ for ES. The losses and forecasts, and the average log e-processes for different methods are plotted in Figure \ref{fig:GREE_nonlinear}.
\begin{figure}[t]
    \centering
    \caption{Realized losses and ES forecasts with a non-linear business cycle (left panel); average log-transformed e-processes obtained by different methods over $1,000$ simulations (right panel)}
    \includegraphics[width=0.48\textwidth]{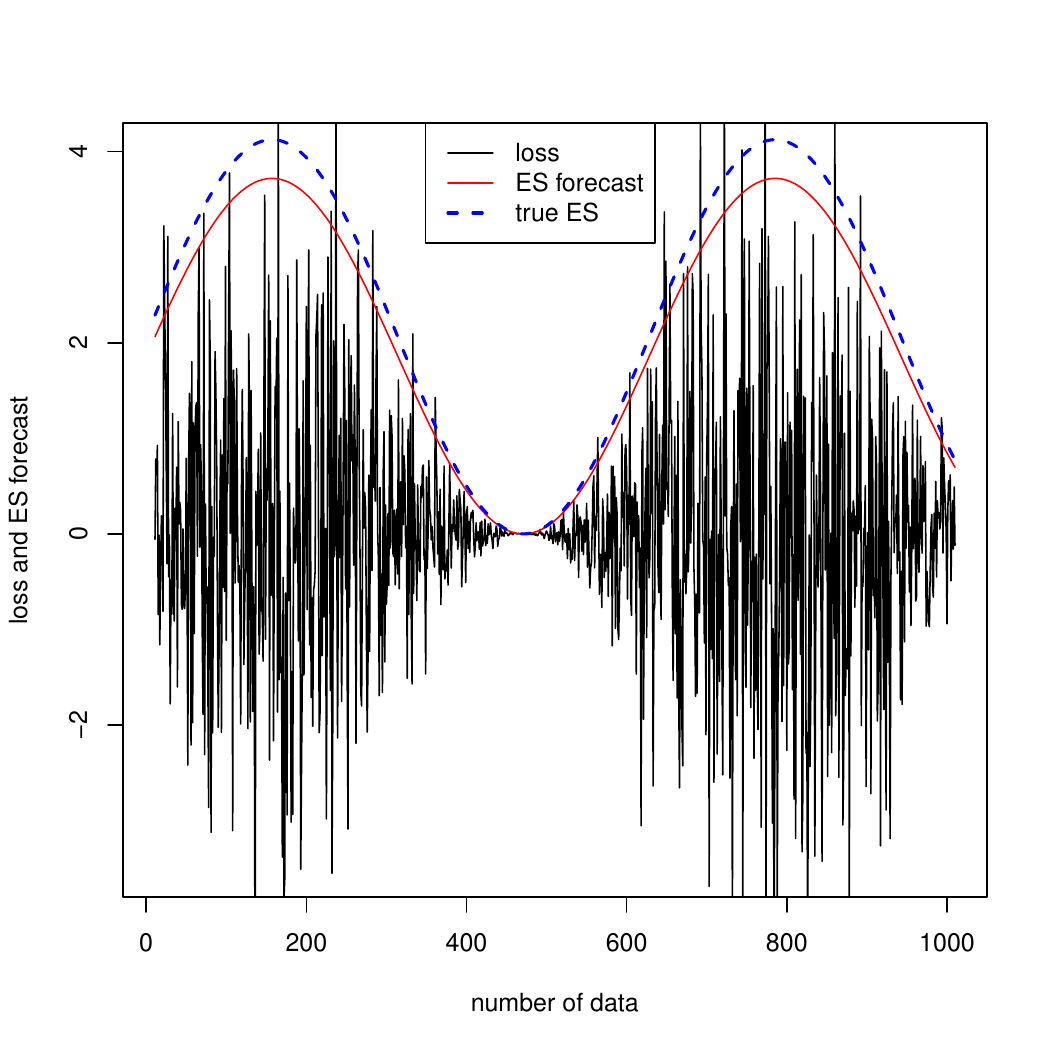}
    \includegraphics[width=0.48\textwidth]{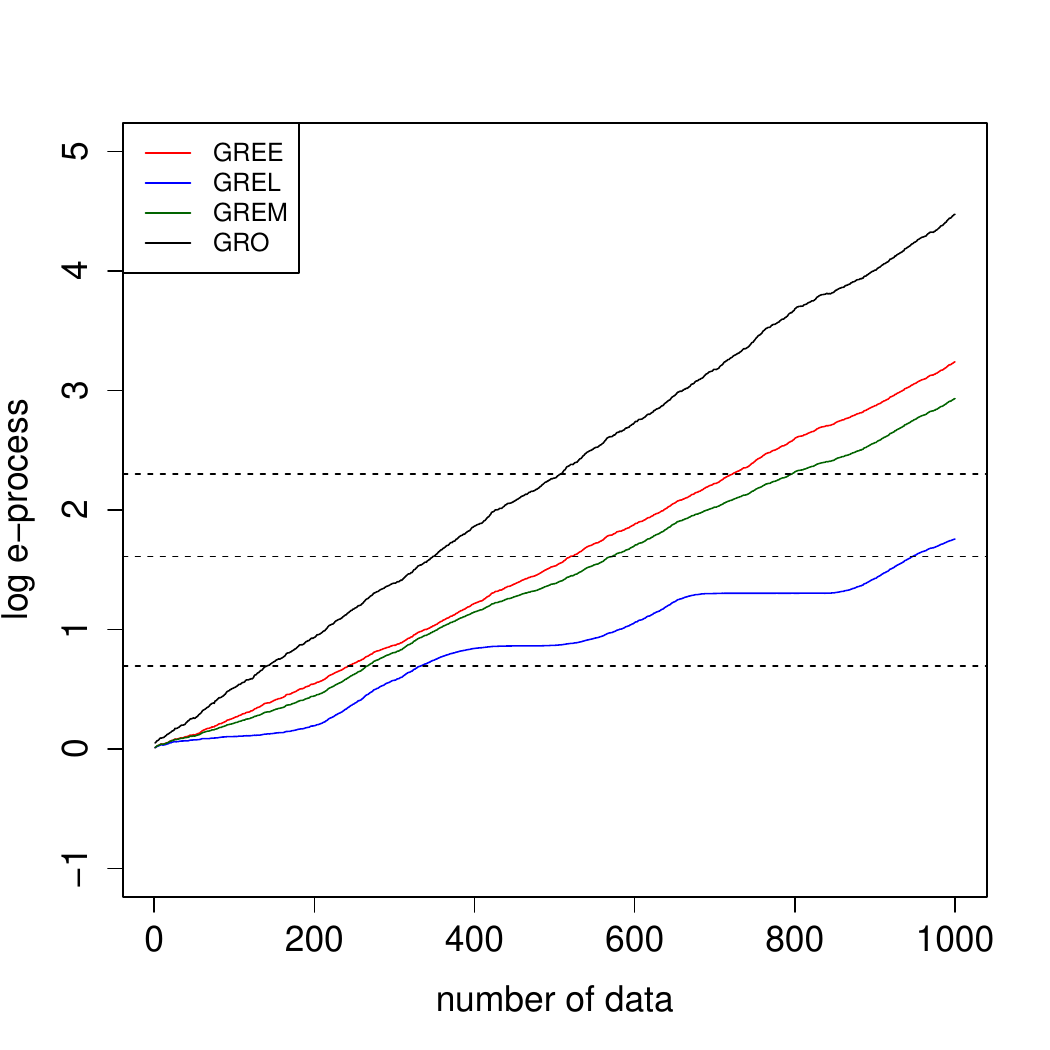}
    \label{fig:GREE_nonlinear}
\end{figure}
Similarly to (a), we also observe better performance of the GREE method than GREL because of the overall iid pattern of the whole e-statistics.

\item It is expected from Theorem \ref{th:opt} that the GREL method will dominate the GREE method when the losses exhibit an iid pattern and there is no clear evidence of co-movements between losses and risk forecasts. Let the random losses be $Z_1,\dots,Z_{n+l}$; thus, they are iid. Suppose that the risk forecaster announces the forecasts of $\VaR_{0.95}(Z_t|\mathcal{F}_{t-1})$ and $\ES_{0.95}(Z_t|\mathcal{F}_{t-1})$ to be $z_t=1.64 + \epsilon_t$ and $r_t=2.06+\epsilon_t$, respectively, for $t\in [n+l]$, where $\epsilon_1,\dots,\epsilon_{n+l}$ are iid samples uniformly distributed on the support $\{\pm i/10:i=0,\dots,5\}$. In this case, the forecaster is able to obtain risk forecasts close to the true values but is subject to a forecasting error $(\epsilon_t)_{t\in [n+l]}$. Figure \ref{fig:GREL} plots the realized losses $Z_t$, ES forecasts $r_t$, and the corresponding e-processes obtained by the GRO, GREE, GREL and GREM methods for $t=l+1,\dots,n+l$.
\begin{figure}[t]
    \centering
    \caption{Realized losses and ES forecasts with iid losses (left panel); average log-transformed e-processes obtained by different methods over $1,000$ simulations (right panel)}
    \includegraphics[width=0.48\textwidth]{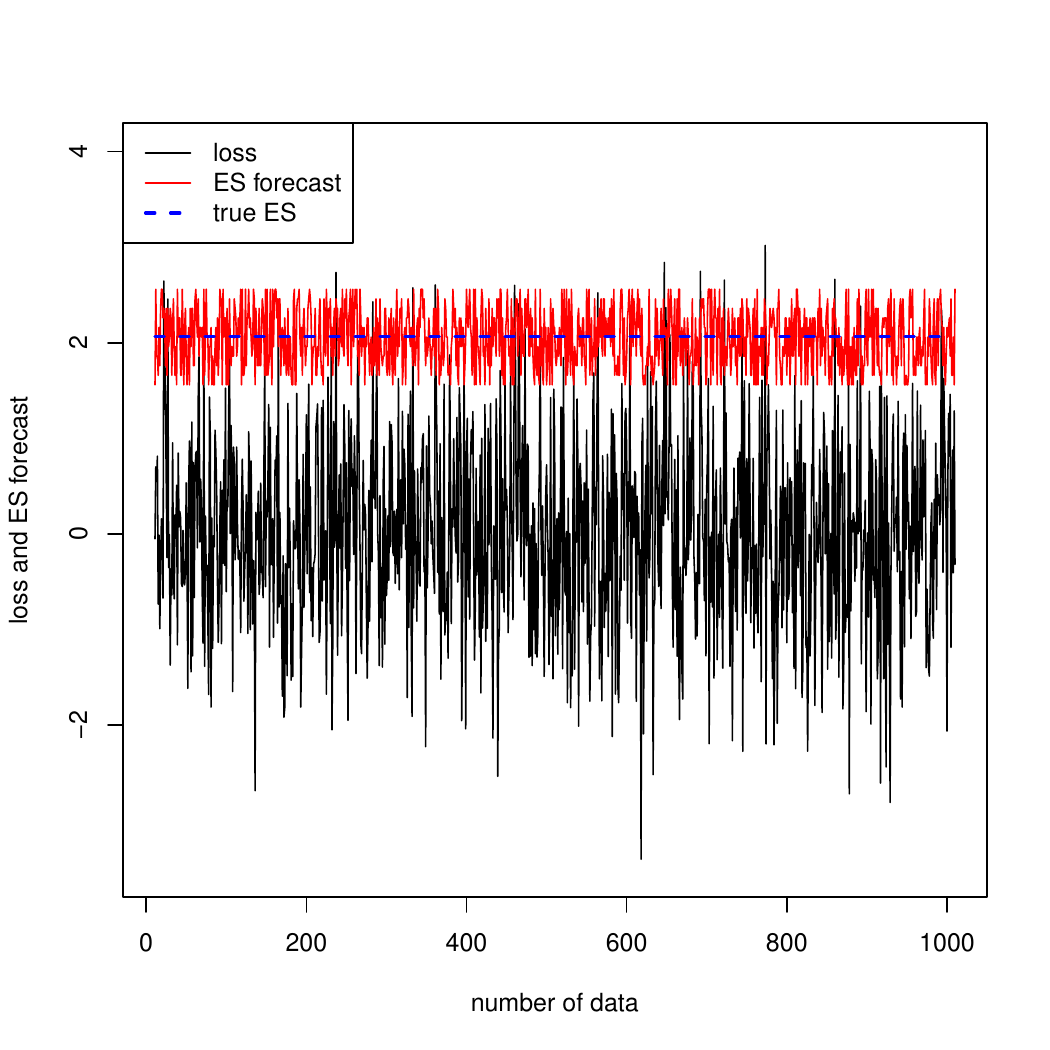}
    \includegraphics[width=0.48\textwidth]{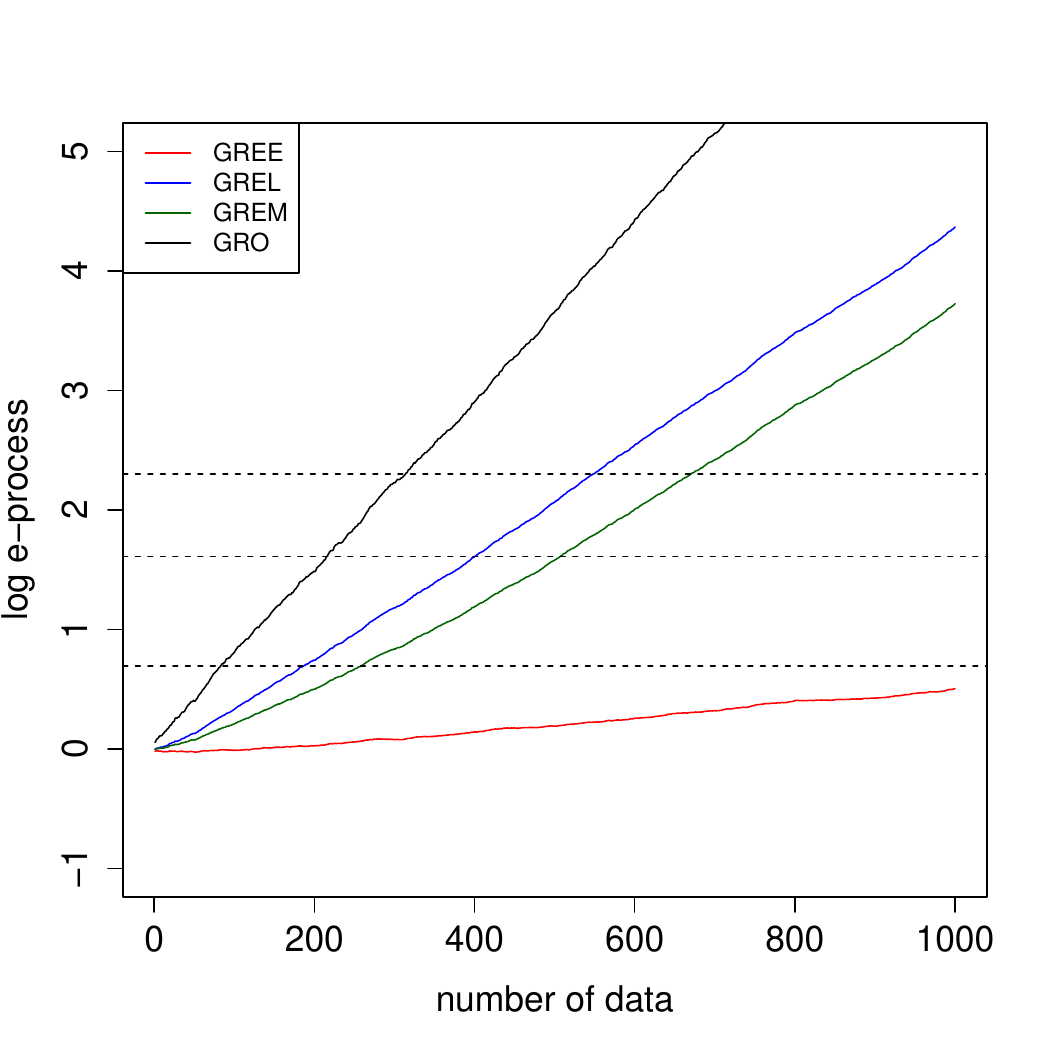}
    \label{fig:GREL}
\end{figure}
We observe from Figure \ref{fig:GREL} that the GREL method outperforms the GREE method. This example shows that the GREL method is able to detect evidence against risk forecasts due to downward fluctuations of the forecasts, while GREE does not perform well in this case because it only uses historical forecasts whose average is close to the true value.
\end{enumerate}
\end{example}

\section{Characterizing backtest e-statistics}\label{sec:techresults}

In this section, we present several  results on the characterization of backtest e-statistics. 
These results  justify the unique roles of the e-statistics we introduced in Section \ref{sec:2}, and their growth-rate optimality among all e-variables as discussed in Section \ref{sec:betting}.
A main practical message is that   $e_p^Q$ and $e_p^\ES$  are essentially the only useful choices for VaR and (ES,VaR), respectively, in building up e-processes in Section \ref{sec:etest}.\footnote{These results are  not needed for the statistical validity of our methodology.}

\subsection{Necessary conditions for the existence of backtest e-statistics}
\label{sec:32}

Not all functionals $\rho$ on $\M$ admit backtest e-statistics that are solely based on the information of $\rho$. 
Below we give a necessary condition for a backtest e-statistic to exist. 
A functional $\rho:\M\to \R$ is \emph{monotone} if $\rho(F)\le \rho(G)$ for all $F\le_1 G$, where $\le_1$ is the usual stochastic order; namely, $F\le_1 G$ if and only if $F\ge G$ pointwise on $\R$.
We also say that $\rho$ is \emph{uncapped} if for each $F\in \M$ and $r>\rho(F)$, there exists $\bar F \in \M$ such that 
$\bar F \ge_1 F$ and $\rho(\bar F)=r$. 
All monetary risk measures \citep{FollmerSchied2016} are monotone and uncapped. 
A functional $\rho:\M\to\R$ is \emph{quasi-convex} if $\rho(\lambda F+(1-\lambda)G)\le\max\{\rho(F),\rho(G)\}$ for all $\lambda\in[0,1]$ and $F,G\in\M$. Similarly,  $\rho$ is \emph{quasi-concave} if $-\rho$ is quasi-convex, and $\rho$ is \emph{quasi-linear} if it is both quasi-convex and quasi-concave.

\begin{proposition}\label{prop:1}
Suppose that $\rho:\M\to \R$ is 
 monotone and uncapped.  
If there exists a backtest e-statistic, then $\rho$ is quasi-convex.
\end{proposition}

When $\M$ is convex, quasi-convexity of $\rho$ is equivalent to the condition that the  set $\{F\in \M: \rho(F) \le r\}$ is convex for each $r\in \R$.
The requirement in Proposition \ref{prop:1} rules out a large class of coherent risk measures including ES.\footnote{In particular,  all comonotonic-additive coherent risk measures except for the mean are monotone and uncapped but not quasi-convex (see e.g., \citealt[Theorem 3]{WangWeiETAL2020}).  A risk measure is coherent (\citealp{ArtznerETAL1999}) if it is subadditive, cash additive, monotone, and positively homogeneous as a mapping from a set of random variables to real numbers.}  
As is shown in the following proposition, if the backtest e-statistic for $\rho$ is monotone, then $\rho$ is necessarily quasi-linear, which is stronger than the quasi-convexity  in Proposition \ref{prop:1}, and this result does not require that $\rho$ is monotone and uncapped. 

\begin{proposition}\label{prop:2} If there exists a monotone backtest e-statistic $e:\R^2\to [0,\infty]$  for $\rho:\M\to \R$, then $\rho$ is quasi-linear.
\end{proposition}

We say a functional $\rho:\M\to\R$ has \emph{convex level sets (CxLS)} if the set $\{F\in \M: \rho(F) = r\}$ is convex for each $r\in\R$. Quasi-linearity of $\rho$ is stronger than the condition that $\rho$ has CxLS, and they are equivalent when $\M$ is convex and $\rho$ is monotone.
Functionals with CxLS have been studied extensively in the recent literature due to their connection to elicitability and backtesting \citep{Gneiting2011, Ziegel2016}. For a recent summary of related results, see \cite{WangWei2020}.

\subsection{Characterizing backtest e-statistics for common risk measures}
\label{sec:tech}

First, we characterize all backtest e-statistics for the mean and for $(\var,\E)$; see also Examples \ref{ex:1}--\ref{ex:2}.

\begin{proposition}[Backtest e-statistics for the mean]\label{prop:unimean}
Let $a \in \R$ and $\PP \subseteq \M_1$ be the set of distributions with support in $[a,\infty)$. Any $\PP$-one-sided e-statistic for the mean is bounded above by 
\begin{equation}\label{eq:up_mean}
e'(x,r) = 1 + h(r)\frac{x-r}{r-a}, \quad x \ge a,\; r> a, 
\end{equation}
for some measurable function $h$ on $(a,\infty)$ with $0 \le h \le 1$. The upper bound $e'$ is a one-sided e-statistic for the mean. Moreover, $e'$ is a backtest e-statistic for the mean, if and only if $h > 0$. The functions $h$ and $r \mapsto (r-a)/h(r)$ are increasing if and only if $e'$ is a monotone backtest e-statistic for the mean.
\end{proposition}

Proposition \ref{prop:unimean} shows in particular that any e-variable for the null hypothesis $\E[L] \le r$, $L \ge a$ that is a function of $L$ must be dominated by an e-variable $e'(L,r)$ as given in \eqref{eq:up_mean}.

\begin{proposition}[Backtest e-statistics for the variance]\label{prop:variance}
Any one-sided e-statistic on $\M_2$ for $\psi=(\mathrm{var},\mathbb{E})$ is bounded above by
\begin{equation}\label{eq:variance}
e^{\prime}(x, r, z)=1+h(r, z) \frac{(z-x)^{2}-r}{r}+k(r,z)(x-z), \quad x, z \in \mathbb{R},~r > 0,
\end{equation}
for some measurable functions $h,k$ on $(0, \infty) \times \mathbb{R}$ with $0\le h\le 1$ and $rk^2\le 4h(1-h)$. The upper bound $e'$ is a one-sided e-statistic for $\psi$. Moreover, $e'$ is a backtest e-statistic for $\psi$ if and only if $h > 0$ and $k=0$; in this case, $e^{\prime}$ is further a monotone backtest e-statistic for $\psi$ if and only if  the functions $r\mapsto h(r,z)$ and $r \mapsto$
$r / h(r,z)$ are increasing for all $z\in\R$.\footnote{The statement and proof of this result have been corrected in Nov 2026 compared to the published version.}
\end{proposition}

Proposition \ref{prop:variance} shows in particular that any e-variable for the null hypothesis $\var[L] \le r$ and $\E[L] = z$ that is a function of $L$ must be dominated by an e-variable $e'(L,r,z)$ as given in \eqref{eq:variance}.

  The simplest backtest e-statistic for $\VaR$ is given  in \eqref{eq:eqp} in Example \ref{ex:5}.
The following theorem shows that all backtest e-statistics for $\VaR$ are essentially dominated by this simple choice.

\begin{theorem}[Backtest e-statistics for VaR]\label{prop:uniquantile}
Let $p \in (0,1)$. Any one-sided e-statistic  for $\VaR_p$ is bounded above by
\begin{equation}\label{eq:VaR_up}
e'(x,r) = 1 + h(r)\frac{p-\id_{\{x\le r\}}}{1-p},
\end{equation}
for some measurable function $h$ on $\R$ with $0 \le h \le 1$. The upper bound $e'$ is a one-sided e-statisic for $\VaR_p$. Moreover, $e'$ is a backtest e-statistic for $\VaR_p$, if and only if $h > 0$. The function $h$ is constant if and only if $e'$ is a monotone backtest e-statistic for $\VaR_p$.
\end{theorem}

Theorem \ref{prop:uniquantile} shows in particular that any e-variable for the null hypothesis $\VaR_p(L) \le r$ that is a function of $L$ must be dominated by an e-variable $e'(L,r)$ as given in \eqref{eq:VaR_up}.

Next, we consider backtest e-statistics for $(\ES,\VaR)$.
It is straightforward that $\ES_p$ is monotone, uncapped, and 
$\{F\in \M: \ES_p(F) \le r\}$ is not convex.\footnote{It might be interesting to note that  $\ES_p$ is concave on $\M$, implying that  the set $\{F\in \M: \ES_p(F) \ge r\}$ is convex for each $r$; see Theorem 3 of \cite{WangWeiETAL2020}.}
Hence, Proposition \ref{prop:1} implies that there is no backtest e-statistic for $\ES_p$ using solely the information of $\ES_p$. A similar point was made in \cite{AcerbiSzekely2017} that $\ES_p$ is not backtestable in some specific sense.  
By Theorem \ref{th:ep}, there exists a  backtest e-statistic $e^\ES_p$ for $(\ES_p,\VaR_p)$. 
In particular, $e^\ES_p(L,r,z)$ is an e-variable for the hypothesis
$ 
H_0: \ES_p(L)\le r \mbox{~and~} \VaR_p(L)=z.
$  
The following theorem characterizes one-sided and backtest e-statistics for $(\ES_p,\VaR_p)$ under the assumption that they are increasing in the loss $L$.

\begin{theorem}[Backtest e-statistics for ES]
\label{th:es}
Let $p \in (0,1)$. Any one-sided e-statistic on $\M_1$  for $\psi=\left(\mathrm{ES}_{p}, \mathrm{VaR}_{p}\right)$ that is increasing in its first argument (i.e., realized loss) is bounded above by 
$$
e^{\prime}(x, r, z)=1+h(r, z)\left(\frac{(x-z)_{+}}{(1-p)(r-z)}-1\right) + k(r,z) \frac{p-\id_{\{x\le z\}}}{1-p}, \quad x \in \mathbb{R},~z < r,
$$
for some measurable functions $h,k$ with $0\le h,k\le 1$ and $h+k \le 1$. The upper bound $e'$ is a one-sided e-statistic for $\psi$. Moreover, $e'$ is a backtest e-statistic for $\psi$, if and only if $h > 0$ and $k=0$. The functions $r\mapsto h(r,z)$ and $r \mapsto$ $(r-z) / h(r,z)$ are increasing for all $z < r$ if and only if $e^{\prime}$ is a monotone backtest e-statistic for $\psi$.
\end{theorem}

It is clear that in Theoroem  \ref{th:es}, if $e'(x,r,z)$ is continuous in $x$, then $k=0$ and $e'=1-h+ h e^\ES_p$.

Theorems \ref{prop:uniquantile} and \ref{th:es} illustrate the essential roles of $e^Q_p$ and $e^{\ES}_p$ among all possible choices of e-statistics for VaR and the pair (ES,VaR). All choices of one-sided e-statistics for $\VaR_p$ are dominated by an e-statistic of the form 
$
1- \lambda + \lambda e_p^Q (x,r),
$
and all those for $(\ES_p,\VaR_p)$ that are relevant for risk management are upper bounded by  
$
1- \lambda + \lambda e_p^\ES(x,r,z),
$ where $\lambda$ is a function taking values in $[0,1]$.
Therefore, in view of \eqref{eq:mtg}, the e-statistics $e$ can be without loss of generality chosen as $e^Q_p$ for $\VaR_p$ and $e^\ES_p$
for $(\ES_p,\VaR_p)$, and $\lambda$ can be chosen separately depending on the risk forecasts.

 Characterization results of backtest e-statistics for risk measures can be obtained by using the link between backtest e-statistics and identification functions that is presented in Section \ref{app:iden}, or by direct arguments. The first approach has the advantage of being more general but in the special cases we considered here, we would only obtain the results under slightly stronger conditions.

Results in this section have implications on the GRO e-variables discussed in Remark \ref{rem:GRO}. 
A GRO e-variable for the considered risk measures must have the form $1-\lambda+\lambda e\left(L_t, r, z\right)$ for suitable  $e$ in Section \ref{sec:2}; in case of ES, the optimality is restricted to e-variables increasing in the loss $L_t$.

\section{Simulation studies}
\label{sec:sim_stu}

In this section, we provide simulation studies on backtesting VaR and ES. This illustrates the details of our backtesting methodology numerically. Furthermore, we examine how different factors affect the quality of the backtesting procedure, especially the impact of the choice of the betting process in \eqref{eq:mtg}. We evaluate the backtesting performance when the risk measures are under-reported, over-reported, or reported exactly by the risk forecaster.

For all e-tests, we report evidence against the forecasts when the e-process exceeds thresholds $2$, $5$, or $10$. %
We call such evidence a detection. From the practical viewpoint, the three thresholds we choose form four zones for levels of alerts to financial institutions. This is in a similar sense to the standard three-zone approach for backtesting VaR in the financial industry.

\begin{remark}
In classical statistical terminology, what we call a \emph{detection} is a \emph{rejection} of the null hypothesis based on our e-test with thresholds $2$, $5$, and $10$, respectively. 
Our choice of using ``detection"  is  to emphasize that having detected evidence of moderate size such as $2$ with the e-test is a useful early warning that risk predictions might not be prudent enough. Recall that Jeffrey's threshold  of e-values for ``substantial" evidence is 3.2 and for  ``decisive" evidence is 10; see \cite{Shafer2021} and   \cite{VovkWang2021} for more discussions on observing moderately large e-values.
\end{remark}

The simulation and data analysis in Sections \ref{sec:sim_stu} and \ref{sec:emp}, together with those in Section \ref{app:num}, illustrate our main methodology.

\subsection{Backtests via stationary time series}
\label{sec:sim}

We apply our e-backtesting procedure to a setting with time series to test the hypotheses \eqref{eq:R2-H0VaR} and \eqref{eq:H0t}.
We simulate data from an AR$(1)$--GARCH$(1, 1)$ process:
$$L_t=\mu_t+\sigma_t Z_t, ~~\mu_t=-0.05+0.3 L_{t-1},~~\sigma^2_t=0.01+0.1\sigma^2_{t-1}Z^2_{t-1}+0.85\sigma^2_{t-1},~~t\in\N$$
where $\{Z_t\}_{t\in \N}$ is a sequence of iid innovations following a skewed-t distribution with shape parameter $\nu=5$ and skewness parameter $\gamma=1.5$.
These parameter values are from \cite{NoldeZiegel2017} and we use them for a direct comparison.
In total, $1,000$ independent simulations are produced, each of which includes a sample of size $500$ used for backtesting. A rolling window of size $500$ is applied for risk estimation at each time spot $t$.

For forecasting, we assume that the data follow an AR$(1)$--GARCH$(1, 1)$ process $\{L_t\}_{t\in \N}$ with $L_t=\mu_t+\sigma_tZ_t$, where $\{Z_t\}_{t\in \N}$ is assumed to be a sequence of iid innovations with mean $0$ and variance $1$, following a normal, t-, or skewed-t distribution. Thus, the forecaster has a correct time-series structure with possibly incorrect innovation. Here, $\{\mu_t\}_{t\in\N}$ and $\{\sigma_t\}_{t\in\N}$ are adapted to $(\mathcal{F}_{t-1})_{t\in \N}$. In addition, as a benchmark (column ``true" in Tables \ref{tab:var_perc} and \ref{tab:var}), we produce a series of forecasts with skewed-t innovations and the true $\{\mu_t\}_{t\in\N}$ and $\{\sigma_t\}_{t\in\N}$ as the data generating process. The details of the forecasting procedure are described in Section \ref{app:forecast}.
The risk forecaster deliberately under-reports, over-reports, or reports the exact point forecasts of ${\VaR}_p$ or $(\ES_p,\VaR_p)$ she obtains. 
For backtesting, the e-processes in \eqref{eq:mtg} are calculated with the betting process $(\lambda_t)_{t\in [T]}$ chosen  by the GREM method using Taylor approximation via \eqref{eq:GREE_em}. The results for the GREE and GREL methods and their comparison are demonstrated in Section \ref{app:comp}. We detect evidence against the forecasts when the e-processes exceed thresholds $2$, $5$, or $10$. We first present results for backtesting $\VaR_{0.99}$. The percentage of detections, the average number of days taken to detect evidence against the forecasts (conditional on detection occurring), and the average final log-transformed e-values are shown in Tables \ref{tab:var_perc} and \ref{tab:var}.

\begin{table}[t]
\centering
\small
\begin{center}
\caption{Percentage of detections (in $\%$) for $\VaR_{0.99}$ forecasts based on $1,000$ simulations of time series, $500$ trading days, and the GREM method: we use an  AR($1$)--GARCH($1,1$) model with four different innovations for the prediction;
$-10\%$ and $+10\%$ mean that the prediction has been manually under-reported or over-reported by $10\%$; a detection means that the  e-process reaches the corresponding threshold
}
\vspace{.1in}
\begin{tabular}{c c c c | c c c | c c c | c c c}
\toprule
	& \multicolumn{3}{c}{normal} &  \multicolumn{3}{c}{t} & \multicolumn{3}{c}{skewed-t} & \multicolumn{3}{c}{true} \\\midrule
	threshold & $2$ & $5$ & $10$ & $2$ & $5$ & $10$ & $2$ & $5$ & $10$ & $2$ & $5$ & $10$\\\midrule
	$-10\%$ & $99.6$ & $98.0$ & $94.3$ & $97.8$ & $88.8$ & $76.1$ & $46.0$ & $14.4$ & $5.9$ & $38.3$ & $10.7$ & $4.5$\\\midrule
	exact & $97.0$ & $87.9$ & $75.0$ & $86.8$ & $60.9$ & $40.2$ & $17.8$ & $2.5$ & $0.5$ & $15.0$ & $1.7$ & $0.2$\\\midrule
	$+10\%$ & $86.1$ & $62.4$ & $41.4$ & $62.5$ & $26.3$ & $11.9$ & $6.3$ & $0.4$ & $0$ & $3.9$ & $0.3$ & $0$\\\bottomrule
\end{tabular}
\label{tab:var_perc}
\end{center}
\end{table}

\begin{table}[t]
\centering
\footnotesize
\setlength{\tabcolsep}{3pt}
\caption{The average number of days taken to detect evidence against $\VaR_{0.99}$ forecasts conditional on detection, based on $1,000$ simulations of time series, $500$ trading days, and the GREM method: the models are the same as in Table \ref{tab:var_perc}; numbers in brackets are average final log-transformed e-values}
\label{tab:var}
\vspace{.1in}
\begin{tabular}{c c c c c | c c c c | c c c c | c c c c}
\toprule
	& \multicolumn{4}{c}{normal} & \multicolumn{4}{c}{t} & \multicolumn{4}{c}{skewed-t} & \multicolumn{4}{c}{true}\\
\midrule
	threshold & $2$ & $5$ & $10$ & & $2$ & $5$ & $10$ & & $2$ & $5$ & $10$ & & $2$ & $5$ & $10$ & \\
\midrule
	$-10\%$ & $116$ & $185$ & $228$ & $(5.489)$ & $156$ & $238$ & $285$ & $(3.390)$ & $230$ & $284$ & $332$ & $(0.3707)$ & $220$ & $312$ & $347$ & $(0.2200)$\\
\midrule
	exact & $158$ & $239$ & $287$ & $(3.311)$ & $200$ & $284$ & $322$ & $(1.759)$ & $219$ & $244$ & $217$ & $(-0.07467)$ & $234$ & $310$ & $441$ & $(-0.1466)$\\
\midrule
	$+10\%$ & $196$ & $277$ & $316$ & $(1.858)$ & $224$ & $305$ & $351$ & $(0.7341)$ & $183$ & $227$ & -- & $(-0.2135)$ & $209$ & $315$ & -- & $(-0.2428)$\\
\bottomrule
\end{tabular}
\end{table}

As expected, the results show that evidence against normal and t-innovations is more likely to be detected than against skewed-t innovations which is the true model. The percentage of detections for exact true forecasts, or the type-I error, is $0.2\%$ for threshold $10$.
Under-reporting VaR leads to earlier detections than reporting the exact VaR forecasts and the converse holds true for over-reporting. For the exact and over-reporting true forecasts, rejections are essentially type-I errors.  
If such an error is made, it tends to be early, and this is confirmed by the simulation results.

Results on backtests of ES
are reported  in Tables \ref{tab:es_perc} and \ref{tab:es}.
The results in Table \ref{tab:es_perc} confirm our intuition that 
under-reporting or using a wrong innovation can be detected  with a large probability, whereas forecasts from the true model and their more conservative versions appear the opposite.
Moreover, under-reporting (resp.~over-reporting) both of ES and VaR and under-reporting (resp.~over-reporting) only ES
have similar performance in terms of probability of detection and  time of detection.  
The average time to detection (Table \ref{tab:es}) is useful for risk management since early warnings (threshold 2) are often issued after about a fourth of the sampling time, and decisive warnings (threshold 10) after about half of the considered trading days.

A comparison with the traditional backtest method of  \cite{NoldeZiegel2017} is presented in Section \ref{app:type1}.

\begin{table}[t]
\centering
\small
\begin{center}
\caption{Percentage of detections ($\%$) for $\ES_{0.975}$ forecasts  based on simulations of time series, $500$ trading days, and the GREM method: we use an  AR($1$)--GARCH($1,1$) model with four different innovations for the prediction;
$-10\%$ ES (both) and $+10\%$ ES (both) mean that the prediction of ES (both VaR and ES) has been manually under-reported or over-reported by $10\%$; a detection means that the  e-process reaches the corresponding threshold}
\vspace{.1in}
\begin{tabular}{c c c c | c c c | c c c | c c c}
\toprule
	& \multicolumn{3}{c}{normal} & \multicolumn{3}{c}{t} & \multicolumn{3}{c}{skewed-t} & \multicolumn{3}{c}{true}\\\midrule
	threshold & $2$ & $5$ & $10$ & $2$ & $5$ & $10$ & $2$ & $5$ & $10$ & $2$ & $5$ & $10$ \\\midrule
	$-10\%$ ES & $99.8$ & $99.5$ & $98.5$ & $98.4$ & $88.8$ & $77.1$ & $47.6$ & $16.1$ & $6.2$ & $35.5$ & $9.2$ & $3.6$\\\midrule
	$-10\%$ both & $99.8$ & $99.5$ & $98.1$ & $98.5$ & $91.4$ & $82.0$ & $48.0$ & $15.7$ & $6.5$ & $36.1$ & $10.1$ & $4.2$\\\midrule
	exact & $99.3$ & $95.7$ & $88.3$ & $88.1$ & $63.9$ & $43.1$ & $18.8$ & $4.0$ & $0.8$ & $11.9$ & $1.7$ & $0.5$\\\midrule
	$+10\%$ both & $95.2$ & $80.4$ & $61.9$ & $64.9$ & $27.6$ & $9.9$ & $7.1$ & $1.0$ & $0$ & $4.2$ & $0.1$ & $0.1$\\\midrule
	$+10\%$ ES & $94.8$ & $79.8$ & $62.1$ & $70.0$ & $34.9$ & $15.6$ & $7.9$ & $1.1$ & $0.1$ & $4.6$ & $0.2$ & $0.1$\\\bottomrule
\end{tabular}
\label{tab:es_perc}
\end{center}
\end{table}

\begin{table}[t]
\centering
\footnotesize
\setlength{\tabcolsep}{3pt} %
\caption{Average number of days taken to detect evidence against $\ES_{0.975}$ forecasts conditional on detection. Simulations of time series with $500$ trading days using the GREM method. The models are the same as in Table~\ref{tab:es_perc}. ``--'' represents no detection; numbers in brackets are average final log-transformed e-values.}
\label{tab:es}
\vspace{.1in}
\begin{tabular}{c c c c c | c c c c | c c c c | c c c c}
\toprule
    & \multicolumn{4}{c}{normal} & \multicolumn{4}{c}{t} & \multicolumn{4}{c}{skewed-t} & \multicolumn{4}{c}{true}\\
\midrule
    threshold & $2$ & $5$ & $10$ & & $2$ & $5$ & $10$ & & $2$ & $5$ & $10$ & & $2$ & $5$ & $10$ & \\
\midrule
    $-10\%$ ES   & $89$  & $137$ & $176$ & $(6.671)$    & $151$ & $223$ & $277$ & $(3.428)$    & $224$ & $271$ & $264$ & $(0.5072)$   & $219$ & $303$ & $346$ & $(0.2544)$\\
\midrule
    $-10\%$ both & $94$  & $146$ & $189$ & $(6.347)$    & $141$ & $215$ & $271$ & $(3.720)$    & $218$ & $256$ & $251$ & $(0.4679)$   & $218$ & $282$ & $317$ & $(0.2292)$\\
\midrule
    exact        & $129$ & $201$ & $247$ & $(4.311)$    & $185$ & $267$ & $311$ & $(1.953)$    & $198$ & $195$ & $251$ & $(-0.04676)$ & $194$ & $252$ & $300$ & $(-0.1549)$\\
\midrule
    $+10\%$ both & $171$ & $250$ & $292$ & $(2.737)$    & $217$ & $283$ & $289$ & $(0.8275)$   & $141$ & $206$ & --    & $(-0.2072)$  & $148$ & $25$  & $57$  & $(-0.2522)$\\
\midrule
    $+10\%$ ES   & $168$ & $247$ & $297$ & $(2.702)$    & $207$ & $286$ & $298$ & $(0.9865)$   & $147$ & $158$ & $165$ & $(-0.2323)$  & $145$ & $186$ & $57$  & $(-0.2819)$\\
\bottomrule
\end{tabular}
\end{table}

\subsection{Monitoring structural change of time series}
\label{sec:structural}

We examine the power of our e-backtesting method to monitor the structural change of simulated time series data. We refer to \cite{ChuETAL1996} and \cite{BerkesETAL2004} for earlier work on monitoring the structural change of datasets. For a comparison with the results in \cite{HogaDemetrescu2022}, we use the same setup as described in their Section 6, and call their method the sequential monitoring method. We simulate the losses $\{L_t\}_{t\in\N}$ following the GARCH$(1,1)$ process:
$$L_t=-\sigma_tZ_t,~~\sigma^2_t=0.00001+0.04 L^2_{t-1}+\beta_t\sigma^2_{t-1},$$
where $\{Z_t\}_{t\in\N}$ is a sequence of iid innovations following a skewed-t distribution with shape parameter $\nu=5$ and skewness parameter $\gamma=0.95$, $\beta_t=0.7+0.25\id_{\{t>b^*\}}$ and $b^*\in[0,250]$ represents the time after which the model is subject to a structural change. We simulate $250$ presampled data for forecasting risk measures and another $250$ data for backtesting.

We choose the probability level for $\VaR_p$ and $\ES_p$ to be $p=0.95$. Via the presampled data, the forecaster obtains the forecasts of $\VaR_{0.95}(L_t|\mathcal{F}_{t-1})$ and $\ES_{0.95}(L_t|\mathcal{F}_{t-1})$ using empirical VaR and ES of the residuals and the estimated model parameters $\hat{\bm{\theta}}=(\hat\omega,\hat\alpha,\hat\beta)$. See Section \ref{app:structural} for details of the forecasting procedure.
Due to the model-free nature, we only use the losses and forecasts $(L_t,r_t,z_t)$ for our e-backtesting method, while the sequential monitoring method also uses the estimated volatility $\sigma_t(\hat{\bm{\theta}})$ by assuming the GARCH model of the losses. As suggested by \cite{HogaDemetrescu2022}, the Monte Carlo simulations detector with a rolling window performs the best among others for both VaR and ES monitoring. Therefore, we take this method for comparison with ours. We choose the size $m=50$ of the rolling window. The significance level of the sequential monitoring method is set to be $5\%$, while we choose the rejection threshold of our e-backtesting method to be $1/5\%=20$ (see Theorem \ref{lem:Ville}).

\begin{figure}[t]
    \centering
     \caption{Top panels: percentage of detections (\%) of $\VaR_{0.95}$ (left panels) and $\ES_{0.95}$ (right panels) forecasts over $10,000$ simulations of time series and $250$ trading days with structural changes at $b^*$; Bottom panels:  average run lengths (ARLs) of backtesting procedures; black lines (``monitor") represent results of the sequential monitoring method}
    \includegraphics[width=0.48\textwidth]{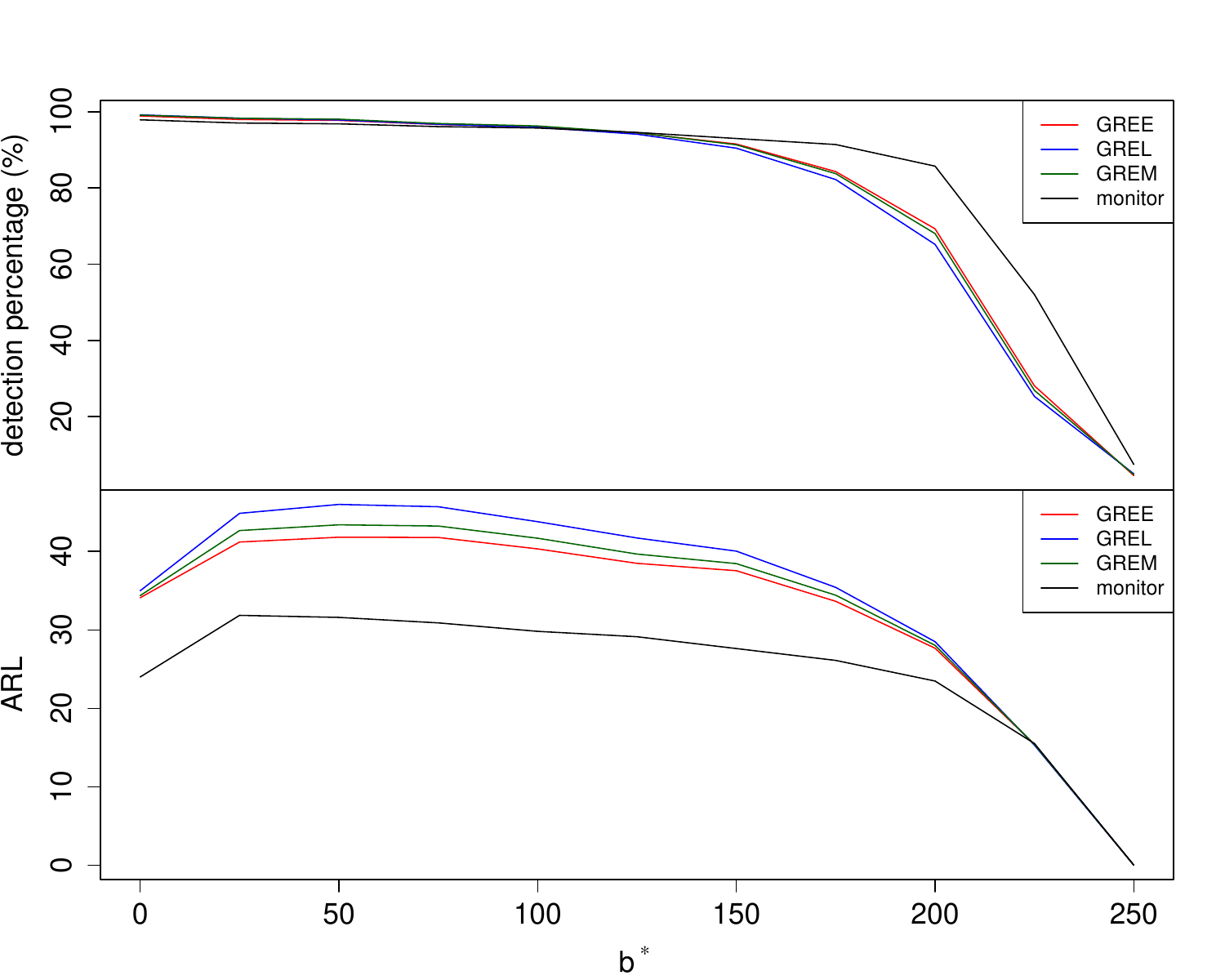}
    \includegraphics[width=0.48\textwidth]{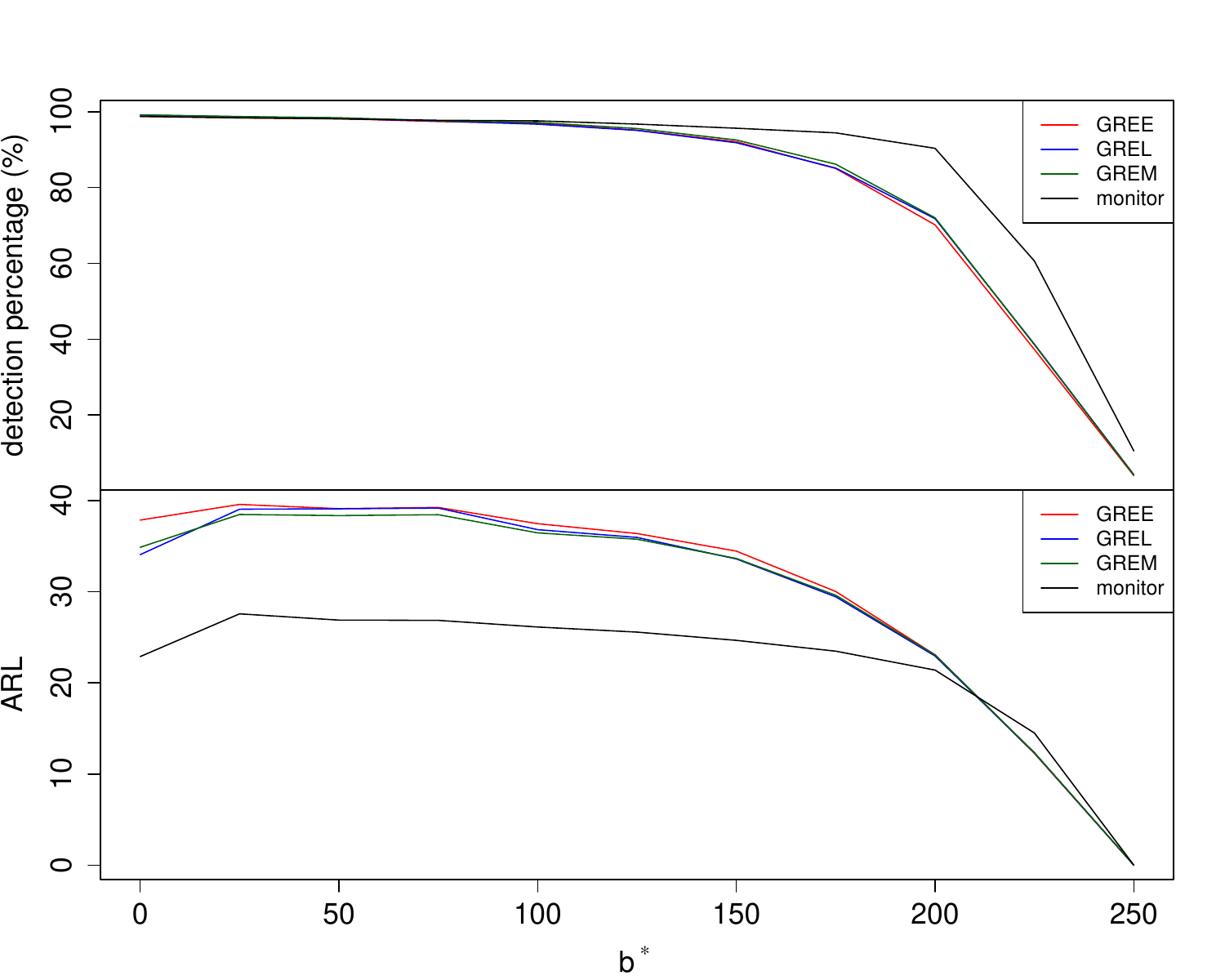}
    \label{fig:vares_compare}
\end{figure}

Figure \ref{fig:vares_compare} plots the average results we get based on $10,000$ simulations, where the betting processes of e-backtesting are chosen by the GREE, GREL or GREM method. The top panels plot the percentage of detections over the total $10,000$ simulations, including those before and after the structural changes at $t=b^*+1$, while the bottom panels show the average number of trading days from the structural changes at $t=b^*+1$ to detections through backtesting, given that detections occur after $t=b^*+1$. We call this quantity the average run length (ARL) as in \cite{HogaDemetrescu2022}.
We observe that the sequential monitoring method outperforms the GREE, GREL, and GREM methods in terms of the detection percentage and the detection speed (reflected by ARL).  This is not surprising, as our backtesting method is nonparametric and requires less model assumptions.\footnote{The backtesting methods based on cumulative violations in \cite{DuEscanciano2017} and \cite{HogaDemetrescu2022} require calculating the realized uniform variables known as probability integral transforms. For this, one needs to assume a parametric model (e.g., t-distribution) for the losses.}
On the other hand,  our test is one-sided while the sequential monitoring method is a two-sided test.
The GREE, GREL, and GREM methods exhibit reasonable performance for all values of $b^*$. From the ARL plots, the GREE, GREL and GREM methods detect evidence against the forecasts around $0$ to $30$ days later than the sequential monitoring method.

\section{Financial data analysis}
\label{sec:emp}

\subsection{The NASDAQ index}
\label{sec:NASDAQ}

\begin{figure}[t]
    \centering
     \caption{Negated percentage log-returns of the NASDAQ Composite index (left panel); $\ES_{0.875}$ and $\ES_{0.975}$ forecasts fitted by normal, t-, skewed-t distributions, and empirical risk forecasts (right panel) from Jan 3, 2000 to Dec 31, 2021}
    \includegraphics[width=0.48\textwidth]{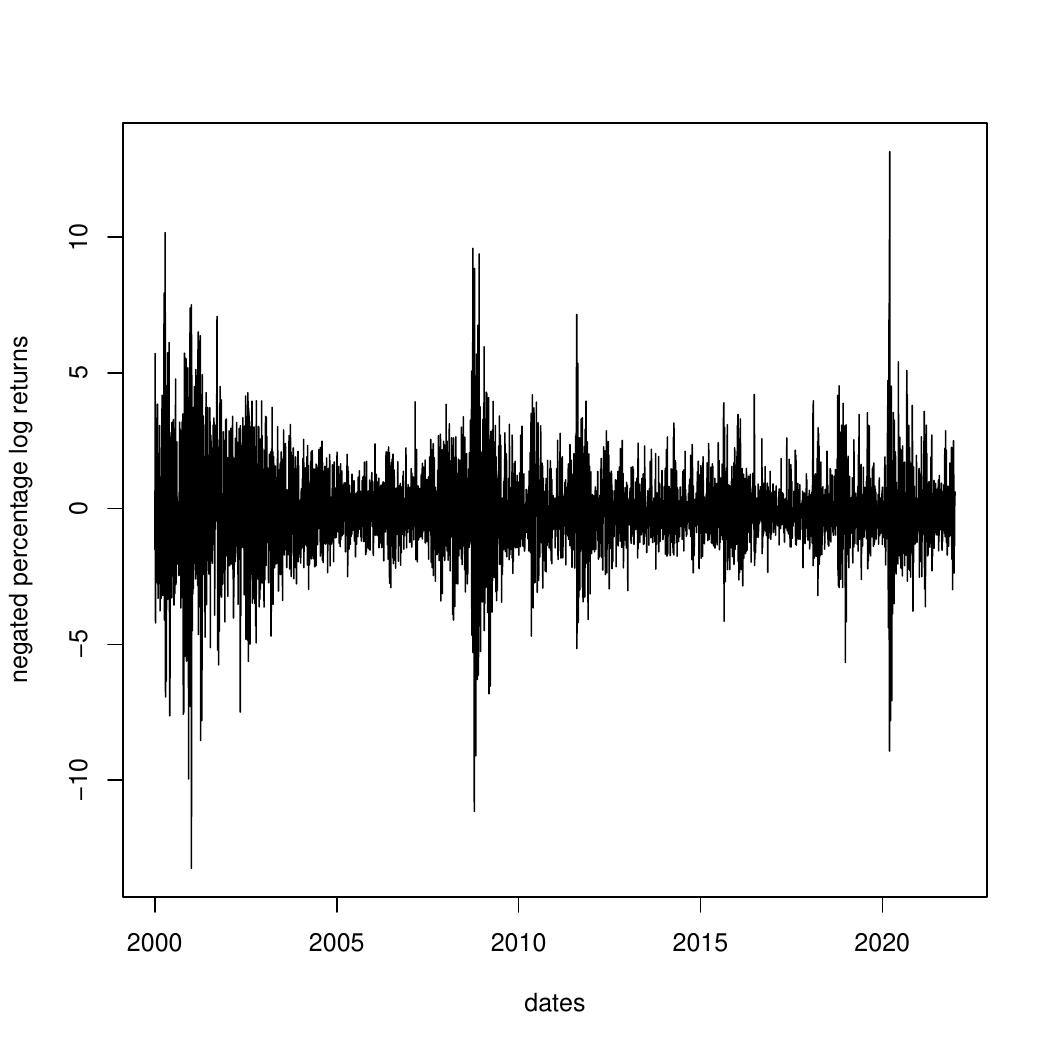}
    \includegraphics[width=0.48\textwidth]{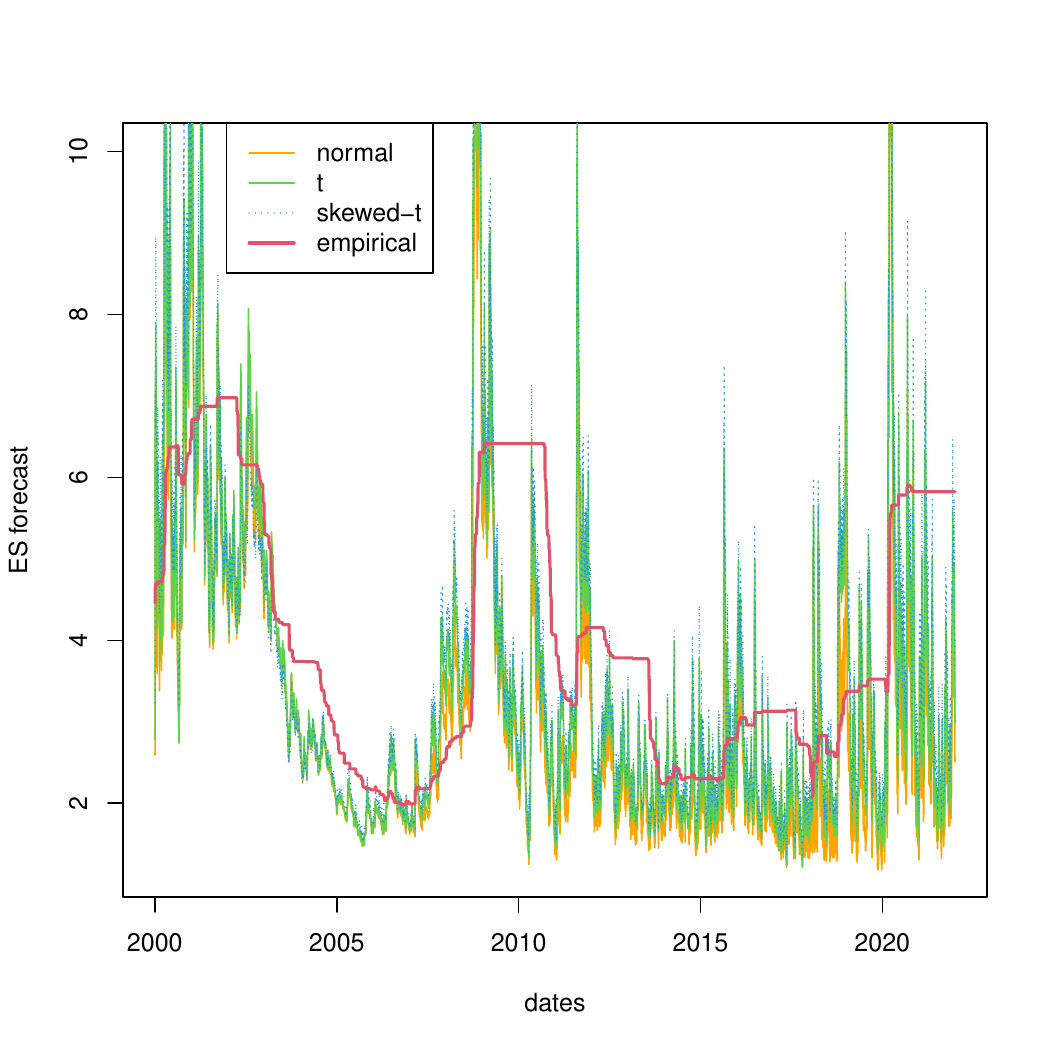}
    \label{fig:log_returns}
\end{figure}

We calculate the negated percentage log-returns using the data of the NASDAQ Composite index from Jan 16, 1996 to Dec 31, 2021. An AR$(1)$-GARCH$(1,1)$ model is fitted to the data with a moving estimation window of $500$ data points. The e-processes in \eqref{eq:mtg} are calculated with the betting process $(\lambda_t)_{t\in [T]}$ chosen by the GREE, GREL or GREM method.
Different from the backtesting methods used in Section \ref{sec:sim}, for each $t\in [T]$, the empirical mean in \eqref{eq:GREE_em} is calculated using a moving window of data in the past $500$ days. This choice  is made to reflect the practice of risk modeling where more recent data represent the current market and economic conditions better.
Therefore,  the first $500$ forecasts use $500$ data points each, and we start the backtesting procedure after the first $500$ forecasts are available, thus after the first $1,000$ data points. The sample size for backtesting is $5,536$, corresponding to forecasts of risk measures from Jan 3, 2000 to Dec 31, 2021. We plot the negated log-returns and the forecasts of $\ES_{0.975}$ fitted by normal, t-, and skewed-t distributions for the innovations over time in Figure \ref{fig:log_returns}. In addition to the parametric methods, we also plot the empirical risk forecasts with a nonparametric rolling window approach in Figure \ref{fig:log_returns}.

\begin{figure}[t]
    \centering
    \caption{Log-transformed e-processes testing $\ES_{0.975}$ with respect to the number of days for the NASDAQ index from Jan 3, 2005 to Dec 31, 2021; left panel: GREE method, middle panel: GREL method, right panel: GREM method}
    \includegraphics[width=0.32\textwidth]{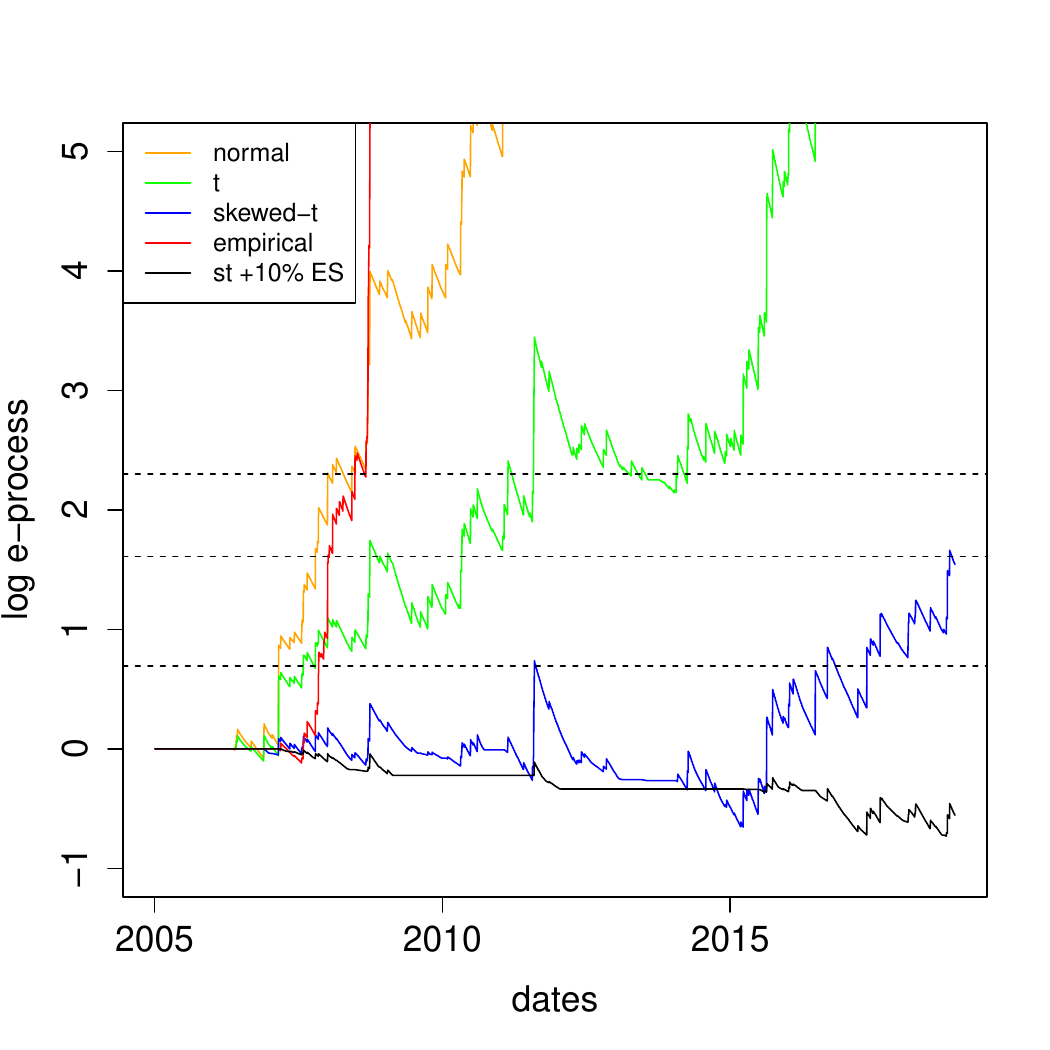}
    \includegraphics[width=0.32\textwidth]{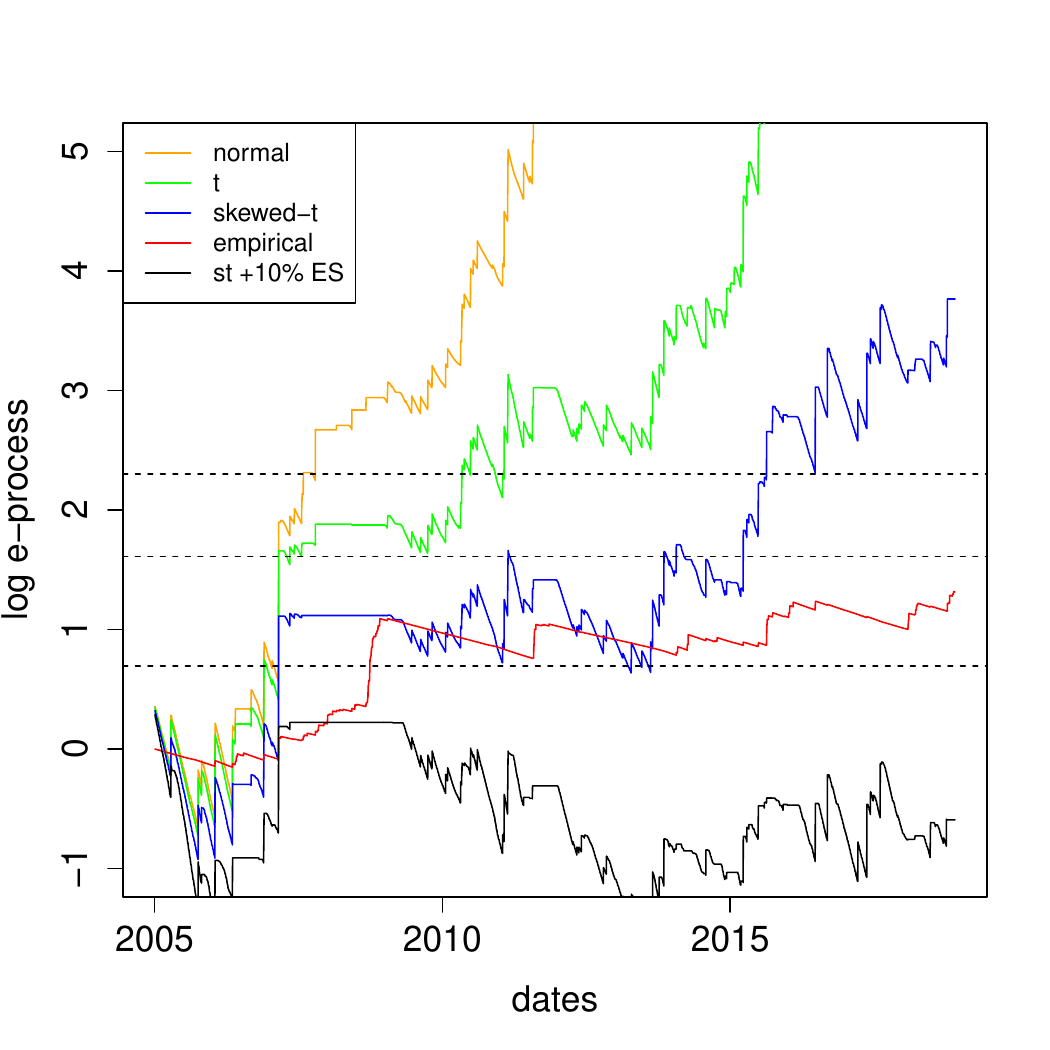}
    \includegraphics[width=0.32\textwidth]{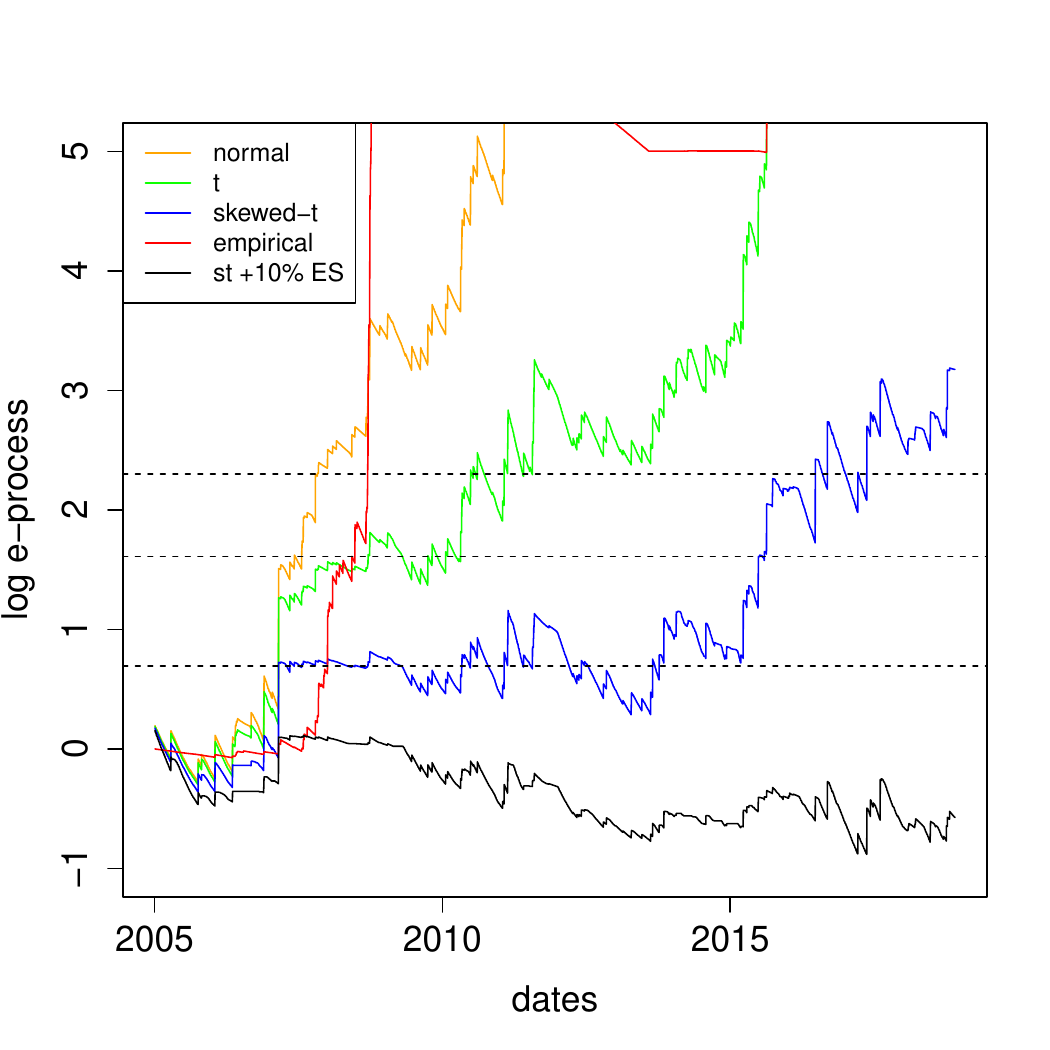}
    \label{fig:es_975_empirical2005}
\end{figure}

We present backtesting results using data from Jan 3, 2005 to Dec 31, 2021 to examine the impact of the 2007 -- 2008 financial crisis. Figure \ref{fig:es_975_empirical2005} shows the e-processes over time. Table \ref{tab:es_975_empirical2005} demonstrates the average $\ES_{0.975}$ forecasts and the number of days taken to detect evidence against  the forecasts, where the second last row contains the results for ES forecasts deliberately over-reported by $10\%$ assuming skewed-t innovations as a forecasting model that is prudent.

Most of the detections in Table \ref{tab:es_975_empirical2005} happen around $500$ -- $700$ trading days after Jan 3, 2005, where significant losses occurred during the financial crisis. Correspondingly, there are sharp jumps of the e-processes in Figure \ref{fig:es_975_empirical2005} at around $500$ -- $700$ trading days. In general, we observe that detections for lower thresholds $2$ and $5$ are significantly earlier than those for the final threshold $10$.
This allows regulators to get alerted much earlier than using the traditional p-tests when e-processes exceed the first threshold $2$ or further exceed $5$. The backtesting procedure may be stopped when an e-process exceeds $10$, which indicates a ``decisive" failure of the underlying model used by the financial institution.

\begin{table}[t]
\def\arraystretch{1.3}
\centering
\small
\begin{center}
\caption{Average $\ES_{0.975}$ forecasts (boldface in brackets) and the number of days taken to detect evidence against the forecasts for the NASDAQ index from Jan 3, 2005 to Dec 31, 2021; ``--" means no detection is detected till Dec 31, 2021}
\label{tab:es_975_empirical2005}
\vspace{.1in}
\begin{tabular}{c c c c c c c c c c c c c}
\toprule
& & \multicolumn{3}{c}{GREE} & & \multicolumn{3}{c}{GREL} & & \multicolumn{3}{c}{GREM}\\\cmidrule{3-5}\cmidrule{7-9}\cmidrule{11-13}
threshold & & $2$ & $5$ & $10$ & & $2$ & $5$ & $10$ & & $2$ & $5$ & $10$\\\midrule
normal & $\mathbf{(2.676)}$ & $540$ & $704$ & $756$ & & $479$ & $540$ & $650$ & & $540$ & $610$ & $713$\\
t & $\mathbf{(2.997)}$ & $650$ & $941$ & $1545$ & & $479$ & $540$ & $1344$ & & $540$ & $933$ & $1381$\\
skewed-t & $\mathbf{(3.202)}$ & $1661$ & $3477$ & -- & & $540$ & $1545$ & $2676$ & & $540$ & $2639$ & $2889$\\
st $+10\%$ $\ES$ & $\mathbf{(3.522)}$ & -- & -- & -- & & -- & -- & -- & & -- & -- & --\\
empirical & $\mathbf{(3.656)}$ & $719$ & $758$ & $876$ & & $941$ & $3823$ & -- & & $756$ & $862$ & $931$\\
\bottomrule
\end{tabular}
\end{center}
\end{table}

The GREL method generally yields larger e-processes than the GREE method except for the empirical forecasts. This may be because the sharp increase of losses upon the occurrence of the financial crisis violates the growth trend and co-movements of the losses and the risk forecasts, making the GREE method not favorable compared with the GREL method as discussed in Example \ref{rem:diff}. It seems from the result that the GREL method is more likely to detect evidence against the risk forecasts for extreme events (e.g., financial crisis) causing an abnormally sharp increase in losses. The GREL method does not perform well in detecting evidence against the empirical forecasts for both VaR and ES. This is expected because the empirical forecasts and the betting process of the GREL method are both obtained only by the information of the empirical distribution of losses, making GREL lack additional information to reject the empirical forecasts. Compared with the GREE and GREL methods, the performance of the GREM method is more stable in different cases where underestimation is detected. Therefore, the GREM method is recommended as a default choice for implementation. 

Although the two papers use different time frames for the real data analysis, our method is able to detect evidence against $\ES_{0.975}$ forecasts as efficiently as \cite{NoldeZiegel2017} (who tested forecasts with normal and skewed-t innovations). However, due to the nature of sequential and non-asymptotic tests, our method can tell the time when underestimation happens by observing the fast growth of e-processes. This feature is useful for practical financial regulation and cannot be achieved by traditional asymptotic tests based on p-values. A simple real data result for comparison to a p-value test is demonstrated in Section \ref{app:pvalue}.

\subsection{Optimized portfolios}
\label{sec:portfolio}

Apart from the NASDAQ index, we perform the e-backtesting procedure on data of a portfolio of $n=22$ stocks from Jan 5, 2001 to Dec 31, 2021.
Suppose that a bank invests in the above portfolio. After each trading day at time $t\in [T]$, the weights $$\mathbf{w}_t=(w^1_t,\dots,w^n_t)\in\Delta_n=\left\{(w_1,\dots,w_n)\in[0,1]^n: \sum^n_{i=1}w_i=1\right\}$$ are determined by a mean-variance criterion. Specifically, the bank solves the following optimization problem:\footnote{We use the mean-variance strategy to illustrate our method for its simplicity, despite its performance may not be empirically satisfactory; see e.g., \cite{DGU09}. Recall that our backtesting method does not require knowledge of the trading strategy or the statistical model, and can be applied to any trading strategy.
There are many other portfolio strategies that can be considered; see e.g., \cite{basak2001value} and \cite{mencia2009multivariate}.
}
$$\max_{\mathbf{w}_t\in\Delta_n}\E[-\mathbf{w}_t^\top \mathbf{L}_t]-\frac{\gamma}{2}\mathrm{var}(-\mathbf{w}_t^\top \mathbf{L}_t),$$
where $\mathbf{L}_t=(L^1_t,\dots,L^n_t)$ is the vector of negated percentage log-returns for all stocks in the portfolio with each $\{L^i_t\}_{t\in[T]}$ being modeled by an AR$(1)$--GARCH$(1,1)$ process for $i\in[n]$.
The bank reports $\VaR$ and   $\ES$ of the weighted portfolio by assuming $\mathbf{w}_t^\top\mathbf{L}_t$ to be normal, t-, or skewed-t distributed. Some of the assumptions in the estimation procedure are simplistic, and hence we do not expect to obtain precise risk forecasts. Suppose a financial institution reports its risk forecasts based on the naive approach described above. We are more likely to get detections if the simplistic assumptions lead to underestimation. The detailed setup and the list of stocks can be seen in Section \ref{app:portfolio}.

\begin{figure}[t]
    \centering
     \caption{Log-transformed e-processes testing $\ES_{0.975}$ with respect to the number of days for portfolio data from Jan 3, 2005 to Dec 31, 2021; left panel: GREE method, middle panel: GREL method, right panel: GREM method}
    \includegraphics[width=0.32\textwidth]{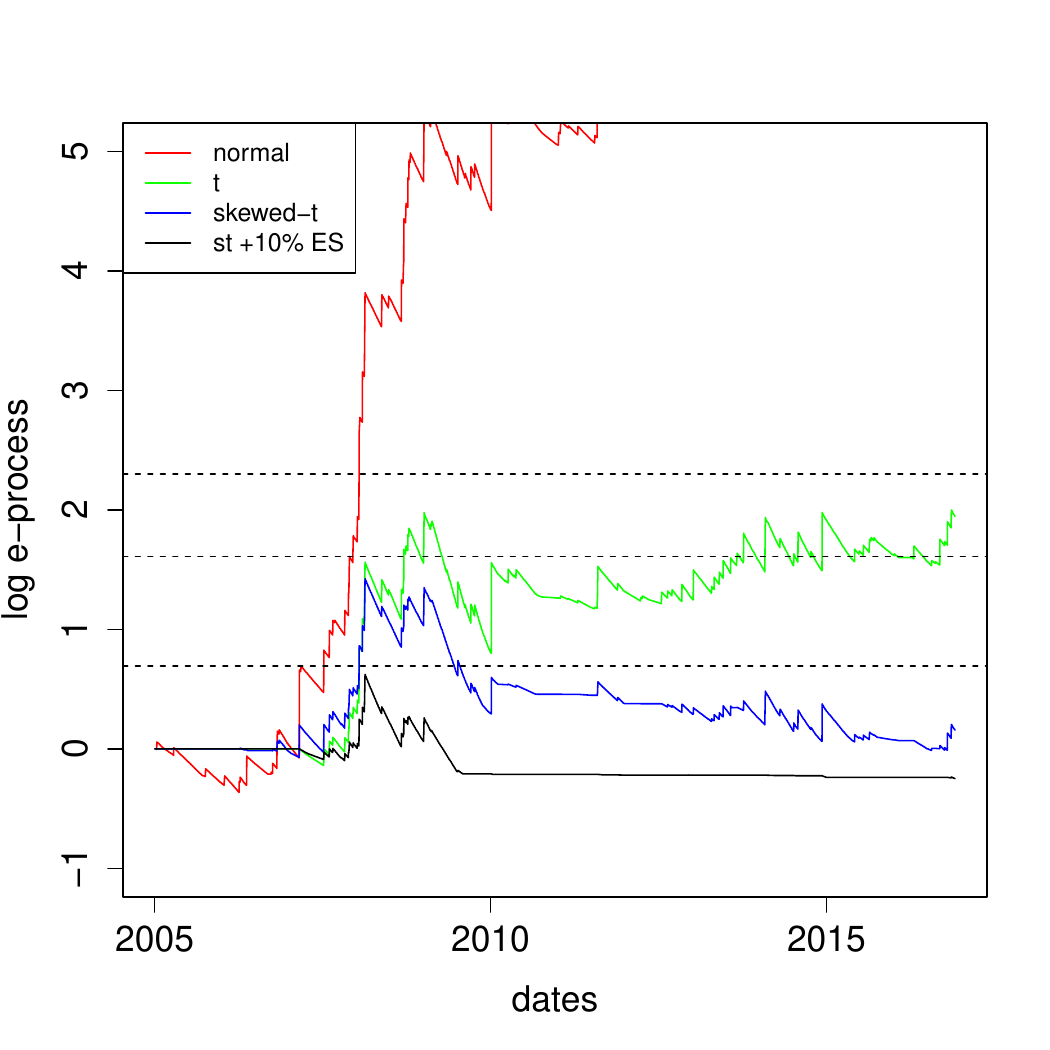}
    \includegraphics[width=0.32\textwidth]{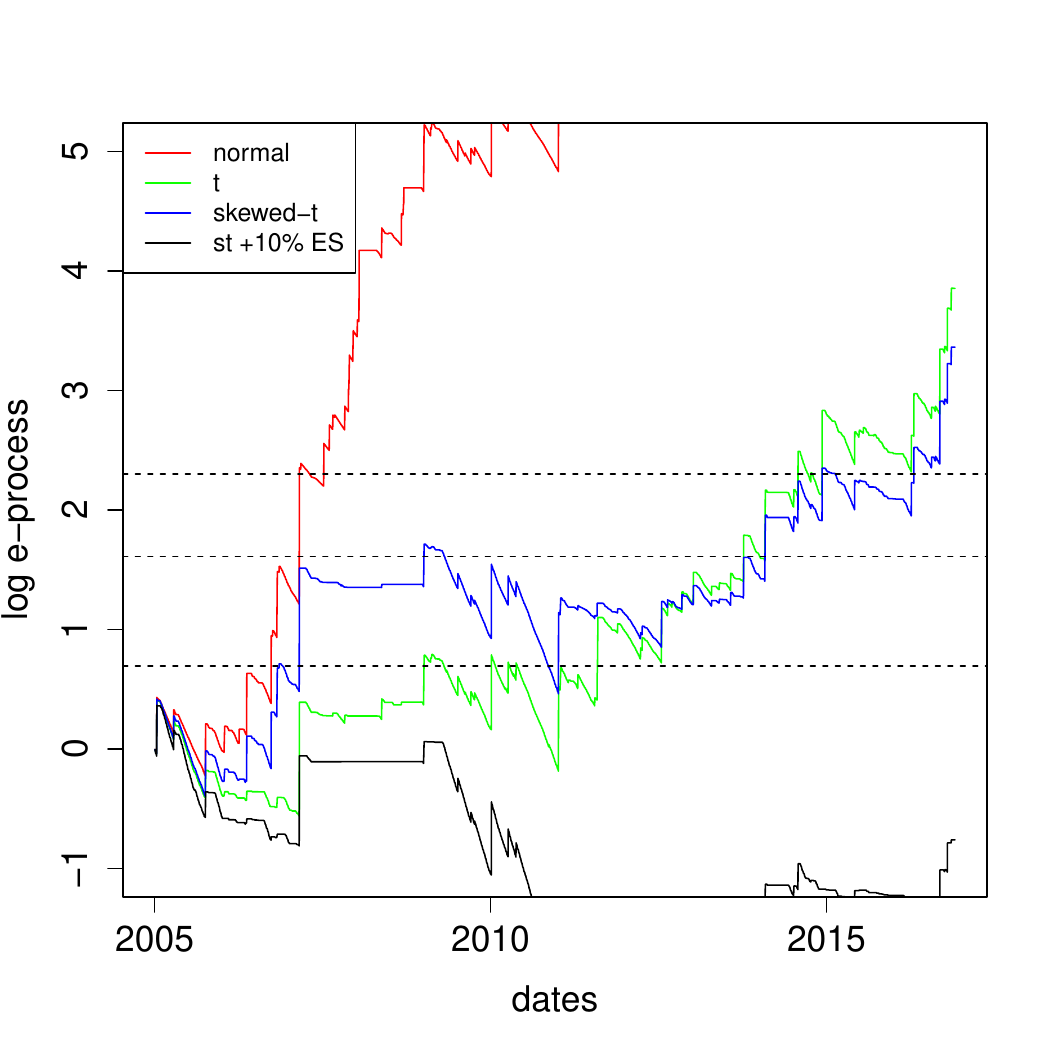}
    \includegraphics[width=0.32\textwidth]{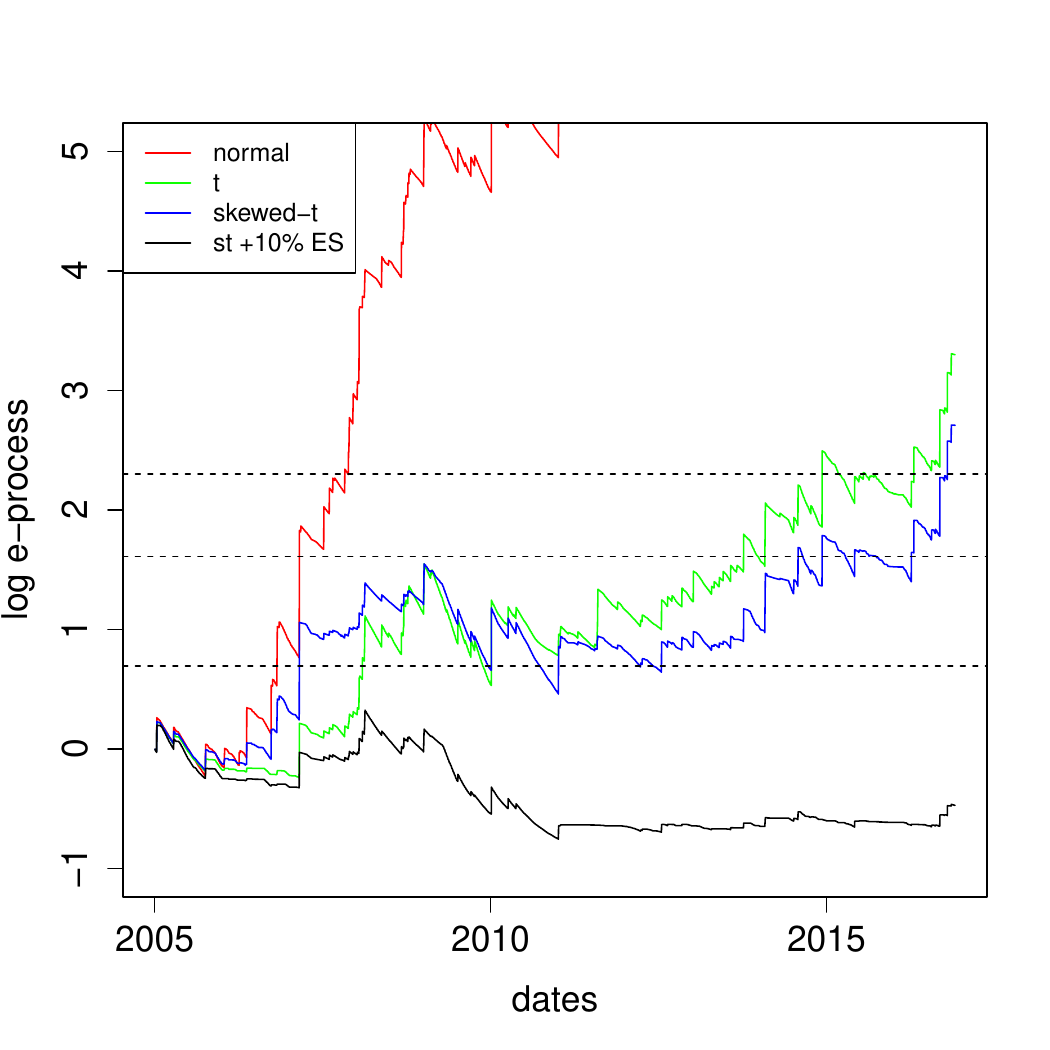}
    \label{fig:es_975_portfolio2005}
\end{figure}

Table \ref{tab:es_975_portfolio2005} shows the average forecasts of $\ES_{0.975}$ and backtesting results with different innovation distributions. The e-processes are plotted in Figure \ref{fig:es_975_portfolio2005}. The portfolio data differ from the simulated time series in the sense that the random losses and risk predictions exhibit much more complicated temporal dependence. Detections are obtained in most of the cases for thresholds $2$ and $5$ before large losses come in during the financial crisis in 2008. Due to the model-free nature, our e-backtesting method is able to detect evidence against risk forecasts when losses and risk forecasts exhibit complicated temporal dependence. This enables regulations for most real portfolio investments in financial markets.

\begin{table}[t]
\def\arraystretch{1.3}
\centering
\small
\begin{center}
\caption{Average $\ES_{0.975}$ forecasts (boldface in brackets) and the number of days taken to detect evidence against the forecasts for portfolio data from Jan 3, 2005 to Dec 31, 2021; ``--" means no detection is detected till Dec 31, 2021}
\vspace{.1in}
\begin{tabular}{c c c c c c c c c c c c c}
\toprule
& & \multicolumn{3}{c}{GREE} & & \multicolumn{3}{c}{GREL} & & \multicolumn{3}{c}{GREM}\\\cmidrule{3-5}\cmidrule{7-9}\cmidrule{11-13}
threshold & & $2$ & $5$ & $10$ & & $2$ & $5$ & $10$ & & $2$ & $5$ & $10$\\\midrule
normal & $\mathbf{(2.817)}$ & $547$ & $730$ & $767$ & & $438$ & $541$ & $541$ & & $461$ & $541$ & $714$\\
t & $\mathbf{(3.191)}$ & $767$ & $934$ & $3036$ & & $1009$ & $2207$ & $2411$ & & $778$ & $2207$ & $2502$\\
skewed-t & $\mathbf{(3.304)}$ & $767$ & -- & -- & & $469$ & $1009$ & $2502$ & & $541$ & $2411$ & $2972$\\
st $+10\%$ $\ES$ & $\mathbf{(3.635)}$ & -- & -- & -- & & -- & -- & -- & & -- & -- & --\\
\bottomrule
\end{tabular}
\label{tab:es_975_portfolio2005}
\end{center}
\end{table}

\section{Concluding remarks}

The e-backtesting method proposed in this paper is the first model-free and non-asymptotic backtest for ES, the most important risk measure in financial regulation implemented by \cite{BCBS2016}. Our methodology contributes to the backtesting issues of ES, which have been a central point of discussions in the risk management literature.  %
  Our methods are constructed using the recently developed notions of e-values and e-processes, which are shown to be promising in many application domains of statistics other than risk management. Some topics on which e-values become useful include sequential testing \citep{Shafer2021, GrunwaldHeideETAL2020},  multiple testing and false discovery control \citep{VovkWang2021, WangRamdas22},   probability forecast evaluation \citep{HenziZiegel2021}, meta-analysis in biomedical sciences \citep{terSchure2021}, and composite hypotheses \citep{WaudbySmithRamdas2020}. 
 Our paper connects two active areas of research through theoretical results and methodologies, and we expect more techniques from either world to be applicable to solve problems from the other. 
  
   Our e-test procedures feature advantages of e-values, including validity for all stopping times and feasibility for no assumptions on the underlying models. 
  Central to our proposed backtesting method, we use backtest e-statistics, which are useful also for  traditional testing problems, %
  although the main focus of the paper is backtesting. The characterization results in Section \ref{sec:techresults} give guidelines for choosing backtest e-statistics. Remarkably, for VaR and ES, essentially unique optimal choices of backtest e-statistics are identified, leaving little doubt on how to choose them in applications.  

If the sample size of a test is fixed, and accurate forecasts for the risk model  are available and to be tested together with forecasts for the risk measure, then traditional model-based methods may be recommended to use in practice, as they often have better power than our e-backtests. In the more realistic situations where the sample size is not fixed, or no models are to be tested along with the risk measure forecasts, our e-backtests are useful, and their multifaceted  attractive features are illustrated by our study. 

As for any other new statistical methodology, e-backtests have their own limitations, challenges, and 
undeveloped extensions. As the main limitation, since e-backtests require very little information on the underlying model, they could be less powerful than traditional model-based or p-value-based approaches. 
Therefore, there is a trade-off between flexibility and power that a risk practitioner has to keep in mind. 
For future directions, 
an important task is to obtain theoretically optimal betting processes using some data-driven procedures under practical assumptions. The methodology can be extended to more general risk measures and economic indices useful in different contexts, each demanding its own backtest e-statistics and backtesting procedure.

 \subsection*{Acknowledgments}
 We thank Zaichao Du, Chao Gao, Glenn Harrison, Marcin Pitera, Aaditya Ramdas and Neil Xu for helpful comments on the paper. RW acknowledges financial support from the Natural Sciences and Engineering Research Council (NSERC) of Canada (Grant Nos.~RGPIN-2024-03728, CRC-2022-00141). JZ acknowledge financial support from the Swiss National Science Foundation.

\bibliographystyle{plainnat}
\bibliography{biblio}

\newpage

\setcounter{page}{1}
\renewcommand{\thepage}{EC--\arabic{page}}

\setcounter{section}{0}
\renewcommand{\thesection}{\Alph{section}}

\setcounter{table}{0}
\renewcommand{\thetable}{EC.\arabic{table}}

\setcounter{lemma}{0}
\renewcommand{\thelemma}{EC.\arabic{lemma}}
\setcounter{proposition}{0}
\renewcommand{\theproposition}{EC.\arabic{proposition}}
\setcounter{theorem}{0}
\renewcommand{\thetheorem}{EC.\arabic{theorem}}
\setcounter{definition}{0}
\renewcommand{\thedefinition}{EC.\arabic{definition}}
\setcounter{corollary}{0}
\renewcommand{\thecorollary}{EC.\arabic{corollary}}
\setcounter{example}{0}
\renewcommand{\theexample}{EC.\arabic{example}}
\setcounter{equation}{0}
\renewcommand{\theequation}{EC.\arabic{equation}}

\appendix

\newpage

\begin{center}
    \Large \textbf {E-companion to ``E-backtesting"}
\end{center}

This e-companion contains four appendices. Section \ref{app:vares} contains the betting processes calculated via Taylor approximation for VaR and ES. Section \ref{app:iden} discusses the link between backtest e-statistics and identification functions.
 Section \ref{app:proofs} contains  proofs of all results. 
Section \ref{app:num} contains some   details of the simulation and data analysis.

\section{Taylor approximation formulas for GREE and GREL}
\label{app:vares}

We give formulas for the betting processes of the GREE and GREL methods for VaR and ES via Taylor approximation. For the GREL method, the special case of VaR, that is, taking $e=e^Q_p$ in \eqref{eq:GREE_em}, yields
$$\lambda^\mathrm{GREL}_t\approx 0\vee\frac{(1-p)\left((t-1)p-\sum^{t-1}_{s=1}\id_{\{L_s\le r\}}\right)}{(t-1)p^2+(1-2p)\sum^{t-1}_{s=1}\id_{\{L_s\le r\}}}\wedge \gamma.$$
For the special case of ES, taking $e=e^{\ES}_p$ in \eqref{eq:GREE_em}, the approximation is
\begin{align*} 
\lambda_t^\mathrm{GREL}\approx 0\vee\frac{(1-p)(r-z)\left(\sum^{t-1}_{s=1}(L_s-z)_+-(t-1)(1-p)(r-z)\right)}{\sum^{t-1}_{s=1}((L_s-z)_+-(1-p)(r-z))^2}\wedge \gamma.
\end{align*}
The corresponding formulas for the GREE method are obtained by replacing $r$ and $(r,z)$ by $r_s$ and $(r_s,z_s)$ in the $s$-th summand in above formulas, respectively.

\section{Link between backtest e-statistics and identification functions}
\label{app:iden}

The link between backtest e-statistics and identification functions is useful for deriving the characterization results of backtest e-statistics.
An integrable function $V:\R^{d+1}\to\R^d$ is said to be an \emph{$\M$-identification function} for a functional $\psi:\M\to\R^d$ if $\int_\R V(x,\psi(F))\d F(x)=\mathbf{0}$ for all $F\in\M$. Furthermore, $V$ is said to be \emph{strict} if
$$\int_\R V(x,y)\d F(x)=\mathbf{0}\iff y=\psi(F)$$
 for all $F\in\M$ and $y\in\R^d$ \citep{FisslerZiegel2016}. We say that $\psi$ is \emph{identifiable} if there exists a strict $\M$-identification function for $\psi$. %

There is a connection between backtest e-statistics and identification functions. Let $e:\R^2 \to \R$ be a backtest e-statistic  for $\rho:\M \to \R$. For $F \in \M$ and $ r \ge \rho(F) > r'$ it holds that
\[
\int e(x,r) \d F(x) \le 1  < \int e(x,r') \d F(x), 
\]
and hence, $1 - e(x,r)$ is often a strict identification function for $\rho$. Since identifiability of a functional coincides with eliciability under some assumptions detailed in \citet{SteinwartPasinETAL2014}, Proposition \ref{prop:2} is not surprising since elicitable functionals are known to have CxLS. 

A backtest e-statistic is called \emph{non-conservative} if $\int_\R e(x,\psi(F))\d F(x) = 1$ for all $P\in \PP$.

\begin{proposition}\label{prop:estatident}
Let $e:\R^{d+1} \to [0,\infty]$ be a non-conservative backtest e-statistic for $\psi=(\rho,\phi):\M \to \R^d$, and assume that $\phi$ has a $\PP$-identification function $v$. We have $V(x,r,z) = ( v(x,z),1-e(x,r,z))^{\top}$
is a $\PP$-identification function for $\psi$. 
\end{proposition}
\begin{proof}
Let $F \in \PP$. By assumption,
$$
\int V(x,\rho(F),\phi(F))\d F(x) = \left(\int v(x,\phi(F))\d F(x), 1 - \int e(x,\rho(F),\phi(F))\d F(x)\right)^\top = \mathbf{0}.  \qedhere
$$ 
\end{proof}

The connection of backtest e-statistics to identification functions is useful because under some regularity conditions there are characterization results for all possible identification functions for a functional \citep{Fissler2017,DimitriadisFisslerETAL2020}. Roughly speaking, given a monotone backtest e-statistic $e$ for $\psi$, then all other possible monotone backtest e-statistics $e'$ must be of the form
\begin{equation*}
e'(x,r) = 1 + h(r)(e(x,r) - 1)
\end{equation*}
for some non-negative function $h$. Clearly, $h$ must fulfill further criteria to ensure that $e'$ is a monotone backtest e-statistic for $\psi$. 

A further consequence of these considerations is that for a functional $\rho$ with monotone backtest e-statistic, there must be an %
identification function $V(x,r)$ that is bounded below by $-1$. This rules out a number of functionals including the expectation without further conditions on $\M$.

\section{Omitted proofs of all results}\label{app:proofs}

\begin{proof}[Proof of Lemma \ref{lem:R1}.]
The backward direction is a simple application of Jensen's inequality, which yields $$0<\E[\log (1-\lambda + \lambda E)]\le \log \E [ 1-\lambda + \lambda E] ~~\Longrightarrow ~~\E[E]>1.$$
To show the forward direction, 
it suffices to verify
$ \E [\log (1-\lambda +\lambda E)]>1$ for $\lambda>0$ small enough.
Note that  $\E [E]>1 $
implies $\E [E\wedge K]>1$ for some $K\ge 1$.
We denote by $Y=E\wedge K$ and let $x_-=(-x)_+$ for $x\in \R$.
Since $\E  [(Y-1)_+] - \E  [(Y-1)_-] = \E  [Y-1]>0$, there exists some $\epsilon\in (0,1)$ such that  
$$\frac{1}{1+\epsilon}\E [(Y - 1)_+ ] - \frac{1}{1-\epsilon}\E [(Y  - 1)_- ] >0.$$
Note that
$\log (1+x) \ge x/(1+\epsilon)$ for $x\in [0,\epsilon)$
and $\log (1+x) \ge x/(1-\epsilon)$ for $x\in (-\epsilon,0)$, that is,
$$
\log(1+x) \ge \frac{x_+}{1+\epsilon}
-\frac{x_-}{1-\epsilon} \mbox{~~~for $x\in (-\epsilon,\epsilon).$}
$$
Hence, for $\lambda\in (0,\epsilon/K)$, implying $\lambda (Y-1)\in (-\epsilon,\epsilon)$, we have  
\begin{align*}
\E [\log (1-\lambda +\lambda E)]&\ge \E [\log (1 + \lambda (Y-1) )] \\& 
\ge   \frac{1}{1+\epsilon}\E [ \lambda (Y-1)_+ ] 
- \frac{1}{1-\epsilon}\E [ \lambda (Y-1)_- ]  
>0,
\end{align*}  
thus showing the desired inequality.
\end{proof}

 \begin{proof}[Proof of Proposition \ref{prop:conservative}.]
 Suppose that $H_0$ in \eqref{eq:H0c} holds.  By the VaR-ES relation in \eqref{eq:var1} and \eqref{eq:es1},
 $$\E[e^{\mathrm{ES}}_p(L,r,z)]  =   \frac{\E\left[(L-z)_+\right]}{(1-p)(r-z)}  \le\frac{\E\left[   (L-\VaR_p(L))_+\right]}{(1-p)(\ES_p(L)-\VaR_p(L))}   =1.$$
Hence, $e^{\mathrm{ES}}_p(L,r,z)$ is an e-variable for \eqref{eq:H0c}.
 \end{proof}

\begin{proof}[Proof of Proposition \ref{lem:lambda01}.]
For all $(r,z)\in\R\times\R^{d-1}$ and $t\in\N$, write $X_t=e(L_t,r,z)$.

(i) Note that $\lambda^{\mathrm{GRO}}_t(r,z)>0$
is equivalent to $\E^{Q_t}[\log(1-\lambda+\lambda X_t)|\mathcal{F}_{t-1}]>0$
for some $\lambda$ that is $\mathcal F_{t-1}$-measurable.
Therefore, the equivalence statement follows from Lemma \ref{lem:R1}.

(ii) For the ``$\Leftarrow$" direction, suppose that $\lambda^{\mathrm{GRO}}_t(r,z)=1$. It is clear that $Q_t(X_t=0)=0$. It follows by continuity of $\lambda\mapsto\E^{Q_t}[\log(1-\lambda+\lambda X_t)|\mathcal{F}_{t-1}]$ and $\lambda\mapsto\E^{Q_t}[(X_t-1)/(1-\lambda+\lambda X_t)]$ that
$$0\le\left.\frac{\mathrm{d}}{\mathrm{d}\lambda}\E^{Q_t}[\log(1-\lambda+\lambda X_t)|\mathcal{F}_{t-1}]\right|_{\lambda=1}=\E^{Q_t}\left[\frac{X_t-1}{X_t}\middle|\mathcal{F}_{t-1}\right].$$
Hence, $\E^{Q_t}[1/X_t|\mathcal{F}_{t-1}]\le 1$.
For the ``$\Rightarrow$" direction, suppose that $\E^{Q_t}[1/X_t|\mathcal{F}_{t-1}]\le 1$. It follows that
$$\begin{aligned}
\E^{Q_t}[\log(1-\lambda+\lambda X_t)-\log(X_t)|\mathcal{F}_{t-1}]&=\E^{Q_t}\left[\log\left(\frac{1-\lambda}{X_t}+\lambda\right)\middle|\mathcal{F}_{t-1}\right]\\
&\le\E^{Q_t}\left[(1-\lambda)\left(\frac{1}{X_t}-1\right)\middle|\mathcal{F}_{t-1}\right]\le 0.
\end{aligned}$$
By strict concavity of $\lambda\mapsto\E^{Q_t}[\log(1-\lambda+\lambda X_t)|\mathcal{F}_{t-1}]$, we have $\lambda^{\mathrm{GRO}}_t(r,z)=1$, where the upper bound $\log(X_t)$ is obtained.
\end{proof}

Next we show Theorem \ref{th:opt}, which relies on Lemma \ref{lem:uni_conv} and Proposition \ref{lem:fixed} below.

\begin{lemma}\label{lem:uni_conv}
If $M:[0,1]\to\R$ and $M_t:[0,1]\to L^0$ are convex for all $t\in\N$, $M$ is continuous and $M_t(\lambda)\xrightarrow{\mathrm{a.s.}} M(\lambda)$ as $t\to\infty$ for all $\lambda\in[0,1]$, then $\sup_{\lambda\in[0,1]}|M_t(\lambda)-M(\lambda)|\xrightarrow{\mathrm{a.s.}}0$ as $t\to\infty$.\footnote{We slightly revised the statements of this result and Proposition \ref{lem:fixed}, and corrected the corresponding proofs in April 2026.}
\end{lemma}
\begin{proof}
For any $\eta>0$, take $\epsilon=\eta/8$. By continuity of $M$, there exists $\delta_0>0$, such that 
\begin{equation}\label{app:eq:cont1}
    |M(\lambda)-M(\lambda')|<\epsilon ~~\mbox{for all}~ |\lambda-\lambda'|<\delta_0.
\end{equation}
Take a partition $\lambda_1<\cdots<\lambda_{I-1}$ for $I\in\N\setminus\{1\}$, such that $\lambda_{i+1}-\lambda_i<\delta_0$ for all $i\in\{0,\dots,I-1\}$, where we write $\lambda_0=0$ and $\lambda_I=1$. For each $i\in\{0,\dots,I-1\}$, write $\tilde \lambda_i=(\lambda_i+\lambda_{i+1})/2$.
Because $M_t(\lambda)\xrightarrow{\mathrm{a.s.}} M(\lambda)$ as $t\to\infty$ for all $\lambda\in[0,1]$, there exists an event $A_\eta$ with $\mathbb{P}(A_\eta)=1$, such that for all $\omega\in A_\eta$, there exists $T(\omega)\in\N$, such that for all $t>T(\omega)$, we have 
\begin{equation}\label{app:eq:cont2}
    |M_t(\lambda_i)-M(\lambda_i)|<\epsilon ~\mbox{and}~ |M_t(\tilde\lambda_i)-M(\tilde\lambda_i)|<\epsilon~~\mbox{for all}~ i=0,\dots,I.
\end{equation}
Note that the statement in \eqref{app:eq:cont2} is made for each $\omega\in A_\eta$. 
For all $\lambda\in[0,1]$, there exists $i\in\{0,\dots,I\}$, such that $\lambda\in[\lambda_i,\lambda_{i+1}]$. Without loss of generality, we assume $\lambda\in[\lambda_i,\tilde\lambda_i]$. The case of $\lambda\in[\tilde\lambda_i,\lambda_{i+1}]$ can be shown analogously by symmetry.
We first claim that 
\begin{equation}\label{app:eq:lower}
    M_t(\lambda_i)-M_t(\lambda)\ge -|M_t(\lambda_i)-M_t(\tilde\lambda_i)|.
\end{equation}
Indeed, because $M_t$ is convex and $\lambda\in[\lambda_i,\tilde\lambda_i]$, we have
$M_t(\lambda)\le \max\{M_t(\lambda_i),M_t(\tilde\lambda_i)\}.$
It follows that
$$M_t(\lambda)-M_t(\lambda_i)\le \max\{M_t(\lambda_i),M_t(\tilde\lambda_i)\}-M_t(\lambda_i)\le |M_t(\lambda_i)-M_t(\tilde\lambda_i)|,$$
and thus \eqref{app:eq:lower} holds.
On the event $A_\eta$, we have by \eqref{app:eq:cont1} and \eqref{app:eq:cont2} that
\begin{equation}\label{app:eq:lower-2} |M_t(\lambda_i)-M_t(\tilde\lambda_i)|\le |M_t(\lambda_i)-M(\lambda_i)|+|M(\lambda_i)-M(\tilde\lambda_i)|+|M(\tilde\lambda_i)-M_t(\tilde\lambda_i)|\le 3\epsilon\end{equation}
for $t>T(\omega)$,
and similarly, $|M_t(\lambda_{i+1})-M_t(\tilde\lambda_i)|\le 3\epsilon$.
Because $M_t$ is convex, 
\begin{align}
    M_t(\lambda_i)-M_t(\lambda)&\le |M_t(\lambda_i)-M_t(\tilde\lambda_i)|+\frac{\tilde\lambda_i-\lambda}{\lambda_{i+1}-\tilde\lambda_i}|M_t(\lambda_{i+1})-M_t(\tilde\lambda_i)|\nonumber\\
    &\le|M_t(\lambda_i)-M_t(\tilde\lambda_i)|+|M_t(\lambda_{i+1})-M_t(\tilde\lambda_i)|\le 6\epsilon.\label{app:eq:upper}
\end{align}
By \eqref{app:eq:lower}, \eqref{app:eq:lower-2}, and \eqref{app:eq:upper},
\begin{equation}\label{app:eq:cont3}
    |M_t(\lambda_i)-M_t(\lambda)|\le 6\epsilon.
\end{equation}
Combining \eqref{app:eq:cont1}, \eqref{app:eq:cont2}, and \eqref{app:eq:cont3},
$$|M_t(\lambda)-M(\lambda)|\le |M_t(\lambda)-M_t(\lambda_i)|+|M_t(\lambda_i)-M(\lambda_i)|+|M(\lambda_i)-M(\lambda)|\le 8\epsilon=\eta$$
for $t>T(\omega)$. 
Now take a sequence $\eta_k\downarrow 0$ as $k\to\infty$ and the event $A=\bigcap_{k\in\N}A_{\eta_k}$. It follows that $\p(A)=1$. For all $\omega\in A$ and $k\in\N$, there exists $T_k(\omega)\in\N$, such that for all $t>T_k(\omega)$, $$\sup_{\lambda\in[0,1]}|M_t(\lambda)-M(\lambda)|\le \eta_k\downarrow 0~~\mbox{as}~k\to\infty.$$ Therefore, $\sup_{\lambda\in[0,1]}|M_t(\lambda)-M(\lambda)|\xrightarrow{\mathrm{a.s.}} 0$ as $t\to\infty$. 
\end{proof}

The next proposition says that under the iid assumption, the betting process computed from empirical distributions is asymptotically equivalent to that computed from the true distribution. 

\begin{proposition}\label{lem:fixed}
Let $X_1,X_2,\dots$ be non-negative iid random variables with $\E[|\log(X_1)|]<\infty$ and $\gamma\in[0,1]$. Let %
$$\lambda_t=\argmax_{\lambda\in[0,\gamma]}\frac{1}{t-1}\sum^{t-1}_{s=1}\log(1-\lambda+\lambda X_s);~~\lambda^*=\argmax_{\lambda\in[0,\gamma]}\E[\log(1-\lambda+\lambda X_t)],~~t\in\N.$$ 
We have
${T}^{-1}\sum^{T}_{t=1}(\log(1-\lambda_t+\lambda_t X_t))-\log(1-\lambda^*+\lambda^* X_t))\xrightarrow{L^1}0$ as $T\to\infty$.
\end{proposition}
\begin{proof}
Write $M(\lambda)=\E[\log(1-\lambda+\lambda X_t)]$ and
$$M_t(\lambda)=\frac{1}{t-1}\sum^{t-1}_{s=1}\log(1-\lambda+\lambda X_s)~~\mbox{for }\lambda\in[0,\gamma],~t\in\N.$$
By the strong law of large numbers, we have $M_t(\lambda)\xrightarrow{\mathrm{a.s.}} M(\lambda)$ as $t\to\infty$ for all $\lambda\in[0,\gamma]$.
Since the functions $M$ and $M_t$ are concave for all $t\in\N$, by Lemma \ref{lem:uni_conv}, we have $\sup_{\lambda\in[0,\gamma]}|M_t(\lambda)-M(\lambda)|\xrightarrow{\mathrm{p}}0$.
For all $\epsilon>0$, we have $$\sup_{\lambda:|\lambda-\lambda^{*}|\ge\epsilon}M(\lambda)\le M(\lambda^{*})$$ by the definition of $\lambda^{*}$ and the concavity of $M$. For all $t\in\N$, we have $M_{t}(\lambda_t)\ge M_{t}(\lambda^{*})$ by the definition of $\lambda_t$. Therefore, we have by Theorem 5.7 of \cite{vanderVaart1998} that $\lambda_t\xrightarrow{\mathrm{p}}\lambda^{*}$ as $t\to\infty$.
Because $\lambda_t$ is bounded for all $t\in\N$, $\{\lambda_t\}_{t\in\N}$ is uniformly integrable. It follows that $\lambda_t\to\lambda^{*}$ with respect to the $L^1$-norm as $t\to\infty$, denoted by $\lambda_t\xrightarrow{L^1}\lambda^{*}$; see e.g., \cite[Theorem 6.6.1]{Resnick2019}.

Next, we show that
\begin{equation}\label{eq:lim}
\frac{1}{T}\sum^T_{t=1}(\log(1-\lambda_t+\lambda_t X_t)-\log(1-\lambda^{*}+\lambda^{*} X_t))\xrightarrow{L^1}0
\end{equation}
as $T\to\infty$.
To simplify notation, write $Y_t(\lambda) = 1-\lambda +\lambda X_t = 1+(X_t-1)\lambda $, where $\lambda $ is a variable.
 
For all $\epsilon_1>0$ and $t\in\N$, by continuity and the monotone convergence theorem, there exists $\delta>0$, such that $\log(1-\delta)>-\epsilon_1$ and $\E[\log(1+(X_t-1)\delta)|X_t>1]<\epsilon_1$.
If $\lambda^*=0$, we have $\lambda_t\xrightarrow{\mathrm{p}}0$ as $t\to\infty$. Hence, there exists $N>0$, such that for all $t>N$, $\p(\lambda_t>\delta)<\epsilon_1$. We write $x_-=\min\{x,0\}$ and $x_+=\max\{x,0\}$ for $x\in\R$. It follows that
$$\begin{aligned}
    \E[ (\log Y_t (\lambda_t) )_-]\ge\E[\log(1-\lambda_t)]&\ge\log(1-\gamma)\p(\delta<\lambda_t\le\gamma)+\log(1-\delta)\p(0\le\lambda_t\le\delta)\\
    &\ge(\log(1-\gamma)-1)\epsilon_1,
\end{aligned}$$
and
$$\begin{aligned}
     \E[(\log Y_t (\lambda_t) )_+] 
    &=\E[\log Y_t (\lambda_t) (\id_{\{X_t>1,0\le\lambda_t\le\delta\}}+\id_{\{X_t>1,\delta<\lambda_t\le\gamma\}})]\\
    &\le\E[ \log Y_t (\delta)|X_t>1]\p(0\le\lambda_t\le\delta)+\E[\log Y_t (\gamma)|X_t>1]\p(\delta<\lambda_t\le\gamma)\\
    &\le(1+\E[\log Y_t (\gamma)|X_t>1])\epsilon_1.
\end{aligned}$$
It is clear that $\sup_{t\in\N}\E[\log Y_t (\gamma)|X_t>1]$ is bounded because $\sup_{t\in\N}\E[\log(X_t)]<\infty$. Hence, $\E[|\log Y_t (\lambda_t)|]<M_1\epsilon_1$ for some $M_1>0$ and for all $t>N$. Therefore, there exists $N_1>0$, such that for all $T>N_1$,
$$\begin{aligned}
    \E\left[\left|\frac{1}{T}\sum^T_{t=1} \log Y_t (\lambda_t) \right|\right]&\le\frac{1}{T}\sum^{N}_{t=1}\E[|\log Y_t (\lambda_t) |]+\frac{1}{T}\sum^{T}_{t=N+1}\E[|\log Y_t (\lambda_t) |] \le\epsilon_1+\frac{(T-N)M_1}{T}\epsilon_1.
\end{aligned}$$
Hence \eqref{eq:lim} holds for $\lambda^*=0$.

 If $\lambda^*=\gamma=1$, the argument for the case $\lambda^*=0$ and the fact $\E[\log(1/X_1)]<\infty$ imply 
$$\frac{1}{T}\sum^T_{t=1}(\log(1-\lambda_t+\lambda_t X_t)-\log(X_t))=\frac{1}{T}\sum^T_{t=1}\log(\lambda'_tX'_t+1-\lambda'_t)\xrightarrow{L^1}0,$$
where $\lambda'_t=1-\lambda_t$ and $X'_t=1/X_t$ for $t\in\N$. 

If $\gamma\in(0,1)$ and $\lambda^*\in(0,\gamma]$, we write $K=\max\{1/\lambda^*,1/(1-\lambda^*)\}<\infty$. Thus
$$\frac{X_{t}-1}{1+(X_{t}-1)\lambda^{*}}\in[-K,K]~~\text{for all } t\in\N.$$
For all $\epsilon_2>0$, it is clear by continuity that there exists $\delta>0$, such that $\log(1-\delta/K)>-\epsilon_2$. Because $\lambda_t\xrightarrow{L^1}\lambda^{*}$ as $t\to\infty$, there exists $N>0$, such that $\p(|\lambda_t-\lambda^*|>\delta)<\epsilon_2$ and $\E[|\lambda_t-\lambda^*|]<\epsilon_2$ for all $t>N$.
It follows that, for all $t>N$,  
$$\begin{aligned}
    &\E\left[(\log Y_t(\lambda_t) -\log Y_t(\lambda^*))_-\right]\\    &=\E\left[\log\left(1+\frac{X_{t}-1}{Y_t(\lambda^*)}(\lambda_t-\lambda^{*})\right)_-\right]\\
&=\E\left[\log\left(1+\frac{X_{t}-1}{Y_t(\lambda^*)}(\lambda_t-\lambda^{*})\right)\left(\id_{\{X_t\ge 1,0\le\lambda_t\le\lambda^*\}}+\id_{\{X_t<1,\lambda^*<\lambda_t\ge\gamma\}}\right)\right]\\
    &\ge\p(|\lambda_t-\lambda^*|\le\delta)\log(1-\delta/K)+\p(|\lambda_t-\lambda^*|>\delta)\Big(\E[\log Y_t(\lambda_t)|X_t\ge 1,0\le\lambda_t<\lambda^*-\delta]\\
    &\qquad +\E[\log Y_t(\lambda_t)|X_t< 1,\lambda^*+\delta\le\lambda_t<\gamma]-\E[\log Y_t(\lambda^*) ]\Big)\\
    &\ge(\log(1-\gamma)-\E[\log Y_t(\lambda^*) ]-1)\epsilon_2,
\end{aligned}$$
and
$$\begin{aligned}
    \E\left[(\log Y_t(\lambda_t) -\log Y_t(\lambda^*))_+\right]&=\E\left[\log\left(1+\frac{X_{t}-1}{Y_t(\lambda^*) }(\lambda_t-\lambda^{*})\right)_+\right] \le K\E[|\lambda_t-\lambda^*|]<K\epsilon_2.
\end{aligned}$$
Because $\E[\log Y_t(\lambda^*) ]$ is bounded, we have $E\left[|\log Y_t(\lambda_t) - \log Y_t(\lambda^*) |\right]<M_2\epsilon_2$ for some $M_2>0$ and for all $t>N$. Similar argument as the case of $\lambda^*=0$ leads to \eqref{eq:lim}.
\end{proof}

Proposition \ref{lem:fixed} gives a simplified illustration of the asymptotic optimality of the GREE method, which uses historical one-sided e-statistics as   iid input. This gives rise to a proof of Theorem \ref{th:opt}. 

\begin{proof}[Proof of Theorem \ref{th:opt}.]
Take $\gamma\in[0,1]$. For (i), because $(r_t,z_t)_{t\in\N}$ is deterministic and $(e(L_t,r_t,z_t))_{t\in\N}$ is iid,  we have for all $t\in\N$,
$$\lambda^{\mathrm{GRO}}_t(r_t,z_t)=\argmax_{\lambda\in[0,\gamma]}\E^{Q_t}[\log(1-\lambda+\lambda e(L_t,r_t,z_t))|\mathcal{F}_{t-1}]=\argmax_{\lambda\in[0,\gamma]}\E[\log(1-\lambda+\lambda X_t)|\mathcal{F}_{t-1}]=:\lambda^*,$$
where  $X_t=e(L_t,r_t,z_t)$.
The statement thus follows directly from Proposition \ref{lem:fixed}.

For (ii), we first show the result for any fixed $(r,z)\in\psi^*(\mathcal P)$. This follows directly from Proposition \ref{lem:fixed} by taking $X_t=e(L_t,r,z)$, $\lambda_t=\lambda^{\mathrm{GREL}}_t(r,z)$, and $\lambda^*=\lambda^{\mathrm{GRO}}_t(r,z)$ for $t\in\N$.

(a) Suppose that $(r_t,z_t)_{t\in\N}$ takes finitely many possible values in $\R^{\N}$. Let $(M_t)_{t\in\N}$ be defined in \eqref{eq:mtg}, $\bm{\lambda}^{\mathrm{GREL}}=(\lambda^{\mathrm{GREL}}_t(r_t,z_t))_{t\in\N}$ and $\bm{\lambda}^{\mathrm{GRO}}=(\lambda^{\mathrm{GRO}}(r_t,z_t))_{t\in\N}$. We have
$$\frac{1}{T}(\log(M_T(\bm{\lambda}^{\mathrm{GREL}}))-\log(M_T(\bm{\lambda}^{\mathrm{GRO}})))\xrightarrow{L^1}0$$
by taking mixtures of all possible values of $(r_t,z_t)_{t\in\N}$ that are finitely many.

(b) It suffices to show the result for $d=1$ and the general case holds similarly. Since $e(x,r)$ is continuous in $r$ and $r_t$, $t\in\N$, are in a common compact set, $e(x,r)$ is uniformly continuous with respect to $r$. Define $M(r,\lambda)=\E^{Q_t}[\log(1-\lambda+\lambda e(L_t,r))|\mathcal{F}_{t-1}]$ with $r\in\R$, $\lambda\in[0,\gamma]$ and $L_t\sim Q_t$ for $t\in\N$.

Let $Q_t$ be the empirical probability measure $\widehat{Q}_{t-1}$ for all $t\in\N$. We now prove that for all $t\in\N$ and $\epsilon>0$, there exists $\delta_1>0$, such that for all $|r-r'|<\delta_1$, $|\lambda^{\mathrm{GREL}}_t(r)-\lambda^{\mathrm{GREL}}_t(r')|\le \epsilon$. Suppose that the negated statement is true. Hence there exists $t\in\N$ and $\epsilon_0>0$, such that for all $\delta>0$, there exist $|r_\delta-r'_\delta|<\delta$, $|\lambda^{\mathrm{GREL}}_t(r_\delta)-\lambda^{\mathrm{GREL}}_t(r'_\delta)|>\epsilon_0$. Because $\lambda^{\mathrm{GREL}}_t(r)=\argmax_{\lambda\in[0,1]}M(r,\lambda)$ and $M(r,\lambda)$ is strictly concave in $\lambda$, we have $$\min\left\{M(r_\delta,\lambda^{\mathrm{GREL}}_t(r_\delta))-M(r_\delta,\lambda^{\mathrm{GREL}}_t(r'_\delta)),~M(r'_\delta,\lambda^{\mathrm{GREL}}_t(r'_\delta))-M(r'_\delta,\lambda^{\mathrm{GREL}}_t(r_\delta))\right\}>l$$
for some $l>0$.
By uniform continuity of $M(r,\lambda)$ with respect to $r$, there exists $\delta_0>0$, such that for all $|r-r'|<\delta_0$, $$\max\left\{\left|M(r,\lambda^{\mathrm{GREL}}_t(r))-M(r',\lambda^{\mathrm{GREL}}_t(r))\right|,\left|M(r,\lambda^{\mathrm{GREL}}_t(r'))-M(r',\lambda^{\mathrm{GREL}}_t(r'))\right|\right\}<l.$$
Therefore, 
$$\begin{aligned}
2l&<M(r_{\delta_0},\lambda^{\mathrm{GREL}}_t(r_{\delta_0}))-M(r_{\delta_0},\lambda^{\mathrm{GREL}}_t(r'_{\delta_0}))+M(r'_{\delta_0},\lambda^{\mathrm{GREL}}_t(r'_{\delta_0}))-M(r'_{\delta_0},\lambda^{\mathrm{GREL}}_t(r_{\delta_0}))\\
&\le \left|M(r_{\delta_0},\lambda^{\mathrm{GREL}}_t(r_{\delta_0}))-M(r'_{\delta_0},\lambda^{\mathrm{GREL}}_t(r_{\delta_0}))\right|+\left|M(r_{\delta_0},\lambda^{\mathrm{GREL}}_t(r'_{\delta_0}))-M(r'_{\delta_0},\lambda^{\mathrm{GREL}}_t(r'_{\delta_0}))\right|< 2l.
\end{aligned}$$
This leads to a contradiction.

Similarly, we can show that there exists $\delta_2>0$, such that for all $|r-r'|<\delta_2$, $|\lambda^{\mathrm{GRO}}(r)-\lambda^{\mathrm{GRO}}(r')|\le \epsilon$ by taking $Q_t$ to the probability measure $Q$ for the iid random variables $L_t$, $t\in\N$. Take $\hat\delta=\min\{\delta_1,\delta_2\}$. Because $r_t\xrightarrow{\mathrm{p}}r_0$, for all $\eta>0$, there exists $N\in\N$, such that $Q(|r_t-r_0|<\hat\delta)>1-\eta$ for all $t>N$. It follows that
$$\max\left\{Q(|\lambda^{\mathrm{GREL}}_t(r_t)-\lambda^{\mathrm{GREL}}_t(r_0)|> \epsilon),~Q(|\lambda^{\mathrm{GRO}}(r_t)-\lambda^{\mathrm{GRO}}(r_0)|> \epsilon)\right\}\le Q(|r_t-r_0|\ge \hat\delta)<\eta.$$
Since we also have $\lambda^{\mathrm{GREL}}_t(r_0)\xrightarrow{\mathrm{p}}\lambda^{\mathrm{GRO}}(r_0)$ as $t\to\infty$, it is clear that $\lambda^{\mathrm{GREL}}_t(r_t)\xrightarrow{\mathrm{p}}\lambda^{\mathrm{GRO}}(r_0)$ and $\lambda^{\mathrm{GRO}}(r_t)\xrightarrow{\mathrm{p}}\lambda^{\mathrm{GRO}}(r_0)$ as $t\to\infty$. By boundedness of the betting processes, we have $\lambda^{\mathrm{GREL}}_t(r_t)\xrightarrow{L^1}\lambda^{\mathrm{GRO}}(r_0)$ and $\lambda^{\mathrm{GRO}}(r_t)\xrightarrow{L^1}\lambda^{\mathrm{GRO}}(r_0)$ as $t\to\infty$. The result thus holds by \eqref{eq:lim}.

For (iii), write $\bm{\lambda}^{\mathrm{GREE}}=(\lambda^{\mathrm{GREE}}_t)_{t\in\N}$ and $\bm{\lambda}^{\mathrm{GREM}}=(\lambda^{\mathrm{GREM}}_t)_{t\in\N}$. It suffices to notice that 
$M_T(\bm{\lambda}^{\mathrm{GREM}})\ge \max \{ M_T(\bm{\lambda}^{\mathrm{GREL}}),M_T(\bm{\lambda}^{\mathrm{GREE}})\}/2$,
and by taking a limit as $T\to\infty$ we obtain the asymptotic optimality of $\bm{\lambda}^{\mathrm{GREM}}$ from that of 
$\bm{\lambda}^{\mathrm{GREE}}$
or 
$\bm{\lambda}^{\mathrm{GREL}}$.

To show the last statement on consistency, note that under the stated conditions, we can verify via Proposition \ref{lem:lambda01} that in each case, 
$$\lim_{T\to\infty} \frac{1}{T} \log(M_T(\bm{\lambda}^{\mathrm{GRO}}))>0.$$ 
This implies that the GRO method has 
asymptotic full power, and hence the same holds for the other methods, because they were shown above to be asymptotically optimal under the corresponding conditions.
\end{proof}

\begin{proof}[Proof of Proposition \ref{prop:1}.]
Let $e$ be a backtest e-statistic for $\rho$.  Write $\M_r(\rho)=\{F\in \M: \rho(F) \le r\}$.
Take $F,G\in \M_r(\rho)$ satisfying $\rho(F)=\rho(G)=r$. We have 
$\int e(x,r) \d F(x) \le 1$ and $\int e(x,r) \d G(x) \le 1$,
and hence 
$\int e(x,r) \d (\lambda F+ (1-\lambda) G) (x) \le 1$.
Thus, $\rho(\lambda F+(1-\lambda)G) \le r$.  
Next, for any $F,G\in \M_r(\rho)$, 
without loss of generality we assume $q:=\rho(F)\ge \rho(G)$. 
Take $\bar G\in \M$ such that $\bar G\ge_1 G$ 
and $\rho(\bar G) = q $. 
From the above analysis, we know that 
$\lambda F+(1-\lambda)\bar G\in \M_q(\rho)\subseteq  \M_r(\rho)$.
Since $\lambda F+(1-\lambda)  G \le_1 \lambda F+(1-\lambda)\bar G$,
we have $\lambda F+(1-\lambda)  G  \in  \M_r(\rho)$.
\end{proof}

\begin{proof}[Proof of Proposition \ref{prop:2}.]
Take $F,G\in \M $, $r\in \R$,  $\lambda \in[0,1]$, and write $H_\lambda = \lambda F+ (1-\lambda) G$.
First, suppose that 
$\rho(F),\rho(G)\le r$.
Since $(x,r)\mapsto e(x,r)$ is decreasing in $r$,  we have 
$\int e(x,r) \d F(x) \le 1$ and $\int e(x,r) \d G(x) \le 1$,
and hence 
$\int e(x,r) \d H_\lambda (x) \le 1$ for all $\lambda \in[0,1]$.
This implies $\rho(H_\lambda)\le r$.  
 Further, suppose that $\rho(F),\rho(G)\ge r$. Assume that $\rho(H_\lambda)<r$. Write $q=\rho(H_\lambda)$. There exists $\epsilon>0$, such that $q+\epsilon<r$. Since $(x,r)\mapsto e(x,r)$ is decreasing in $r$, $\int e(x,q+\epsilon)\d H_\lambda(x)\le 1$, $\int e(x,q+\epsilon)\d F(x)>1$, and $\int e(x,q+\epsilon)\d G(x)>1$. This leads to a contradiction. Therefore, $\rho(H_\lambda)\ge r$.
 Summarizing the above arguments, $\rho(F),\rho(G)\le r$ implies $\rho(H_\lambda)\le r$,
 and $\rho(F),\rho(G)\ge r$ implies $\rho(H_\lambda)\ge r$.
 This gives the quasi-linearity of $\rho$.
\end{proof}

The proof of Proposition \ref{prop:unimean} relies on the following lemma.
\begin{lemma}\label{lem:mean}
If a function $g:\R_+\to \R_+$ satisfies 
$\E[g(L)]\le 1 $ for all $L\ge 0$ with $\E[L]\le r$, then there exists $h \in [0,1]$ such that  $g(x)\le(1-h) + h x/r$ for $x\in \R$.
\end{lemma}
\begin{proof}
Without loss of generality we can assume $r=1$.
First, it is easy to note that $g(y)\le y$ for $y>1$; indeed, if $g(y)>y$ then taking a random variable $X$ with $\p(X=y)=1/y$ and $0$ otherwise  gives $\E[g(X)]> 1$ and breaks the assumption.
Moreover, $g(y)\le 1$ for $y\le 1$ is also clear, which in particular implies $g(1)\le 1$.

Suppose for the purpose of contradiction that the statement in the lemma does not hold. 
This means that for each $\lambda\in [0,1]$, either (a) $g(x) >(1-\lambda) + \lambda x$ for some $x<1$ or (b) $g(y) >(1-\lambda) + \lambda y$ for some $y>1$.
Since $g(y)\le y$ for $y>1$ and $g(x)\le 1$ for $x<1$, we know that
$\lambda =1$ implies (a)
and  $\lambda =0$ 
implies (b). 

We claim that there exists $\lambda_0\in (0,1)$ for which both (a) and (b) happen. 
To show this claim, let 
\begin{align*} 
\Lambda_0&= \{ \lambda \in [0,1]: g(y) >(1-\lambda) + \lambda y \mbox{ for some $y>1$}\};\\\Lambda_1&= \{ \lambda \in [0,1]: g(x) >(1-\lambda) + \lambda x \mbox{ for some $x<1$}\}.\end{align*}
Clearly, the above arguments show $\Lambda_0\cup \Lambda_1=[0,1]$, $0\in \Lambda_0$, and $1\in \Lambda_1$. Moreover, since the function $ (1-\lambda) + \lambda x$ is monotone in $\lambda$ for either $x<1$ or $x>1$, we know that both $\Lambda_0$ and $\Lambda_1$ are intervals. 
Let  
$ \lambda_*=\sup \Lambda_0$ and  
$ \lambda^*=\inf \Lambda_1.$
We will argue  $\lambda_*\not \in \Lambda_0$ 
and $ \lambda^* \not \in   \Lambda_1.$ 
If $\lambda_* \in \Lambda_0$, then there exists $y>1$ such that   $g(y) >(1-\lambda_*) + \lambda_* y $. By continuity, there exists $\hat \lambda_*>\lambda_*$ such that  $g(y) >(1-\hat \lambda_*) + \hat\lambda_* y $, showing that $\hat \lambda_*\in \Lambda_0$, contradicting the definition of $\lambda_*$. Therefore, $\lambda_* \not \in \Lambda_0$.
Similarly,  $\lambda^* \not \in \Lambda_1$, following the same argument.
If $\lambda_*  = \lambda^*$, then this point is not contained in $\Lambda_0\cup \Lambda_1$, a contradiction to $\Lambda_0\cup \Lambda_1=[0,1]$. Hence, it must be  $\lambda_* > \lambda^*$, which implies that $\Lambda_0\cap \Lambda_1 $ is not empty.

Let $x_0<1$ and $y_0>1$ be such that 
  $$ g(x_0)> 1-\lambda_0 +  \lambda_0 x_0  \mbox{~~~and~~~} g(y_0)> 1-\lambda_0 +  \lambda_0 y_0 .$$    
Let $X$ be distributed as 
$\p(X=y_0) = (1-x_0)/(y_0-x_0)$
and 
 $\p(X=x_0) = ( y_0-1)/(y_0-x_0)$, which clearly satisfies 
 $\E[X]=1$.  
Moreover, 
\begin{align*}
\E[g(X)] &=
\frac{ 1-x_0   }{y_0-x_0 }  g(y_0 ) + 
\frac{ y_0-1   }{y_0-x_0 } g(x_0)
\\&> 
\frac{ 1-x_0   }{y_0-x_0 } \left(1-\lambda_0 +  \lambda_0 y_0 \right) + 
\frac{ y_0-1   }{y_0-x_0 }\left(1-\lambda_0 +  \lambda_0 x_0\right)=1.
\end{align*}
This yields a contradiction.
\end{proof}

\begin{proof}[Proof of Proposition \ref{prop:unimean}.]
The first part of the proposition is immediate from Lemma \ref{lem:mean}. Example \ref{ex:1} implies that $e'$ is a one-sided e-statistic for the mean, and a backtest e-statistic if $h > 0$. Assume now that $e'$ is a backtest e-statistic for the mean. Considering any $L$ with $\E L = r' > r$, we find that $h > 0$. The condition that $e'$ is monotone implies the remaining conditions.
\end{proof}

The proof of Proposition \ref{prop:variance} relies on the following lemma.
\begin{lemma}\label{lem:var}
If a function $g:\R\to \R_+$ satisfies 
$\E[g(L)]\le 1 $ for all $L$ with $\E[L]=0$, $\E[L^2]\le r$, then there exists $h \in [0,1]$ and $k\in\R$ with $rk^2\le 4h(1-h)$, such that  $$g(x)\le(1-h) + h \frac{x^2}{r}+kx,~~x\in \R.$$
\end{lemma}

\begin{proof}
The proof is similar to that of Lemma \ref{lem:mean} with modification. 
Without loss of generality we assume $r=1$. Define
$$G(x)=\sup\left\{\E[g(Y)]:\E[Y]=0~\mbox{and}~\E[Y^2]=x^2\right\},~~x\ge 0.$$
First, it is easy to note that $G(y)\le y^2$ for $y>1$; indeed, for all random variables $Y$ with $\E[Y]=0$ and $\E[Y^2]=y^2$, take random variable $X$ such that $F_X=y^{-2}F_Y+(1-y^{-2})\delta_0$, where $\delta_0$ is the point mass at $0$. It follows that $\E[X]=0$, $\E[X^2]=1$, and thus
$$y^{-2}\E[g(Y)]\le y^{-2}\E[g(Y)]+(1-y^{-2})g(0)=\E[g(X)]\le 1.$$
Moreover, $G(y)\le 1$ for $0\le y\le 1$ is also clear, which in particular implies $G(1)\le 1$.

Suppose for the purpose of contradiction that there is no $h\in[0,1]$ such that $G(x)\le (1-h)+hx^2$ for all $x\ge 0$. 
This means that for each $\lambda\in [0,1]$, either (a) $G(x) >(1-\lambda) + \lambda x^2$ for some $0\le x<1$ or (b) $G(y) >(1-\lambda) + \lambda y^2$ for some $y>1$.
Since $G(y)\le y^2$ for $y>1$ and $G(x)\le 1$ for $0\le x<1$, we know that
$\lambda =1$ implies (a)
and  $\lambda =0$ 
implies (b). 
We see that there exists $\lambda_0\in (0,1)$ for which both (a) and (b) happen; this was shown in the proof of Lemma \ref{lem:mean}. 

Let $0\le x_0<1$ and $y_0>1$ be such that 
  $$ G(x_0)> 1-\lambda_0 +  \lambda_0 x_0^2  \mbox{~~~and~~~} G(y_0)> 1-\lambda_0 +  \lambda_0 y_0^2.$$  
There exist random variables $X_0,Y_0$ with $\E[X_0]=\E[Y_0]=0$, $\E[X^2_0]=x^2_0$, and $\E[Y^2_0]=y^2_0$, such that
$$\E[g(X_0)]>1-\lambda_0 +  \lambda_0 x_0^2,~~\E[g(Y_0)]>1-\lambda_0 +  \lambda_0 y_0^2.$$
Let random variable $X$ be such that $F_X=( y^2_0-1)/(y^2_0-x^2_0)F_{X_0}+(1-x^2_0)/(y^2_0-x^2_0)F_{Y_0}$, which clearly satisfies  $\E[X]=0$ and $\E[X^2]=1$.
Moreover, 
\begin{align*}
\E[g(X)] &=
\frac{ y^2_0-1   }{y^2_0-x^2_0 } \E[g(X_0)]+\frac{ 1-x^2_0   }{y^2_0-x^2_0 }  \E[g(Y_0)]
\\&> 
\frac{ y^2_0-1   }{y^2_0-x^2_0 }\left(1-\lambda_0 +  \lambda_0 x_0^2\right) + \frac{ 1-x^2_0   }{y^2_0-x^2_0 } \left(1-\lambda_0 +  \lambda_0 y_0^2 \right)
=1.
\end{align*}
This yields a contradiction. Therefore, there exists $h\in[0,1]$, such that $G(x)\le (1-h)+hx^2$ for all $x\ge 0$.

For fixed $x_1>0>x_2$, let random variable $Z$ be such that $\p(Z=x_1)=1-\p(Z=x_2)=-x_2/(x_1-x_2)$. We have $\E[Z]=0$ and $\E[Z^2]=-x_1x_2$. It follows that
$$\E[g(Z)]\le G(\sqrt{-x_1x_2})\le 1-h-hx_1x_2=\E[1-h+hZ^2],$$
and thus
\[
\frac{g(x_1)-1-h(x_1^2-1)}{x_1}
-
\frac{g(x_2)-1-h(x_2^2-1)}{x_2}
\le0.
\]
Hence, we can choose $k\in\R$ such that
$$\sup_{x>0}\frac{g(x)-1-h(x^2-1)}{x}\le k \le
\inf_{x<0}\frac{g(x)-1-h(x^2-1)}{x},$$
where both sides of the inequality are clearly finite. Together with the fact that $g(0)\le G(0)\le 1-h$, it follows that
$$g(x)\le 1-h+hx^2+kx,~~x\in\R.$$
Non-negativity of $g$ implies that $k^2\le 4h(1-h)$.
\end{proof}

\begin{proof}[Proof of Proposition \ref{prop:variance}.]
The first part of the proposition follows from Lemma \ref{lem:var}. Example \ref{ex:2} implies that $e'$ is a one-sided e-statistic for $(\var,\E)$, and it is a backtest e-statistic if $h > 0$ and $k=0$. Suppose now that $e'$ is a backtest e-statistic. Taking a random variable $L$ with mean $z$ finite variance $\var(L) > r$, it follows that $h > 0$. For contradition, if $k(r,z)\neq 0$ for some $r>0$ and $z\in\R$, then take random variable $X$ with $\E[X]=z-rk(r,z)/(2h(r,z))$ and $\var(X)>r$. It follows that
$$\E[e'(X,r,z)]=1+h(r,z)\left(\frac{\var(X)}{r}-1\right)-\frac{rk^2(r,z)}{4h(r,z)}<1,$$
for $\var(X)$ sufficiently close to $r$. This leads to a contradiction, thus $k=0$.
The final statement is directly obtained with the same argument as in the proof of Proposition \ref{prop:unimean}.
\end{proof}

For the proof of Theorem \ref{prop:uniquantile}, we show the following lemma.
\begin{lemma}\label{lem:VaR}
Fix $r\in \R$ and $p\in (0,1)$.
If a function $g:\R\to \R_+$ satisfies 
$\E[g(L)]\le 1 $ for all $L$ with $\VaR_p(L)\le r$, then there exists $h \in [0,1]$ such that  $$g(x)\le 1 + h\,   \frac{p -  \id_{\{x \le r\}}}{1-p},  \mbox{~~~~for $x\in \R$}.$$
\end{lemma}
\begin{proof}
Fix $r \in \R$ and consider the hypothesis
 $$
H_0: \VaR_p(L) \le r.
$$ 
The distribution $\delta_{x}$ for $x \le r$ is in $H_0$. Therefore, $g(x) \le 1$ for $x \le r$.

For any $q \in[p,1)$ and $\epsilon > 0$, $\epsilon' \ge 0$, the distribution $q\delta_{r-\epsilon'} + (1-q)\delta_{r+\epsilon}$ is in $H_0$. Therefore, $qY(r-\epsilon')+ (1-q)Y(r+\epsilon) \le 1$. Solving for $g(r+\epsilon)$ and taking the minimum over $q$, we find that
$$
g(r+\epsilon) \le \frac{1-g(r-\epsilon')p}{1-p},
$$
which implies
$$
g(r+\epsilon) \le \frac{1-\sup_{x\le r}g(x)p}{1-p}.
$$
Rearranging this inequality, we obtain $p\sup_{x\le r}g(x) \le 1 - (1-p) g(r + \epsilon)$, hence 
$$
p\sup_{x\le r}g(x) + (1-p)\sup_{x > r} g(x) \le 1.
$$
Denote by $h =1- \sup_{x\le r}g(x) \in [0,1]$ and $b = \sup_{x > r} g(x)$.  
Clearly, $g\le f$, where 
$$
f(x) = (1-h) \id_{\{x \le r\}} + b  \id_{\{x > r\}},~~~x\in \R.
$$ 
Taking a supremum over two-point distributions, it is easy to see that $\E[f(L)]\le 1$ from $\E[g(L)]\le1$.
Take any random variable $L$ with $\p(L>r)=1-p$.
Clearly $L$ satisfies $H_0$. 
It follows that 
$$
(1-h) p + b (1-p) = \E[f(L)] \le 1,
$$
leading to
$$
b  \le \frac{ 1 -p+hp }{1-p} = 1-h +\frac{h}{1-p}.
$$
Therefore, 
$$
g(x) \le f(x) \le 1-h + \frac{h}{1-p}  \id_{\{x > r\}},~~~x\in \R,
$$
as desired.
\end{proof}

\begin{proof}[Proof of Theorem \ref{prop:uniquantile}.]
The first part of the theorem follows from Lemma \ref{lem:VaR}. Example \ref{ex:3} shows that $e'$ is a one-sided e-statistic for $\VaR_p$. In order to obtain that $e'$ is a backtest e-statistic, it is necessary and sufficient that $h > 0$. Fix $x \in \R$. For $r \ge x$, the function $r \mapsto e'(x,r)$ is decreasing if and only if $h$ is increasing; for $r < x$, the same function is decreasing if and only if $h$ is decreasing. Since these considerations hold for any $x \in \R$, $h$ has to be constant if $e'$ is monotone.
\end{proof}

We use the following lemma for the proof of Theorem \ref{th:es}.
\begin{lemma}\label{lem:es}
Fix $(r,z)$ with $r>z$ and $p\in (0,1)$.
If an increasing function $g:\R\to \R_+$ satisfies 
$\E[g(L)]\le 1 $ for all $L$ with  $\ES_p(L)\le r$ and $\VaR_p(L)=z $,  then there exist $h,k \in [0,1]$ with $h+k\le 1$ such that  \begin{equation}\label{eq:R1-ES}
g(x)\le(1-h-k) + h   \frac{ (x-z)_+}{(1-p)(r-z)} + k \frac{\id_{\{x>z\}}}{1-p} ,  \mbox{~~~~for $x\in \R$}.
\end{equation}
\end{lemma}

\begin{proof}
Without loss of generality, assume $z=0$.
First note that the condition on $g$ implies $g(0)\le 1$.
Write $\theta = g(0)\in [0,1]$. Consider
$$
H_0: \ES_p(L)\le r \mbox{~and~} \VaR_p(L)=0.
$$

Let $X$ be any e-variable independent of a Bernoulli random variable $B$ with mean $p$.
Let $L=  rX(1-B)$. It is clear that $\VaR_p(L)= 0$ 
and $\ES_p(L) =r\E[X]\le r$.
Hence, $L$ satisfies $H_0$. 
It follows that $\E[g(L)]\le1$, leading to
$$
p \theta + (1-p)\E[g( rX)] \le 1,
$$ 
and thus 
$$
\E\left [ \frac{1-p}{1-p\theta }g(rX)\right] \le 1.
$$
Since $X$ is an arbitrary e-variable, by Lemma \ref{lem:mean}, there exists $\lambda \in [0,1]$ such that
$$
\frac{1-p}{1-p\theta } g(rx) \le 1-\lambda  +  \lambda x  \mbox{~~for all $x\ge 0$}.
$$
Equivalently, we can write
$$
 g(x) \le \frac{1-p\theta }{1-p} \left(1-\lambda  +  \lambda \frac x r\right)  \mbox{~~for all $x\ge 0$}.
$$
This implies $$\theta = g(0) \le  \frac{1-p\theta }{1-p} (1-\lambda ),$$
and thus
$$(1-\lambda ) (1- p\theta )  - (1-p)\theta = 1-\lambda - \theta + p\theta \lambda \ge 0.$$
Also, monotonicity of $g$ implies $g(x) \le g(0)= \theta $ for $x\le 0$. Putting these together, we get 
\begin{align*}
g(x) & \le \theta \id_{\{ x \le 0\} } + \frac{1-p\theta }{1-p}   \left(1-\lambda  +  \lambda \frac x r\right)  \id_{\{ x  > 0\} }
\\ &=\theta   + \left(\frac{1-p\theta }{1-p} (1-\lambda) -\theta\right)\id_{\{x>0\}} +  \frac{1-p\theta }{1-p} \lambda \frac{x_+}{r}   
\\ &=\theta   + \left( 1-\lambda -\theta + p\theta \lambda \right)\frac{\id_{\{x>0\}}}{{1-p}} + (1-p\theta)\lambda \frac{ x_+   }{(1-p)r}.
\end{align*}
Noting that $\theta + (1-\lambda -\theta + p\theta \lambda ) +  (1-p\theta)\lambda=1$,
and each of the three terms is in $[0,1]$, we obtain that \eqref{eq:R1-ES} holds.
\end{proof}

\begin{proof}[Proof of Theorem \ref{th:es}.]
The first part of the theorem follows directly from Lemma \ref{lem:es}. Suppose that $e'$ is a backtest e-statistic for $(\ES_p,\VaR_p)$. For all $r\in\R$, $z< r$, $\epsilon>0$, and for some $q\in(p,1)$, consider a random variable $$X=\left((r-z)(1-p)/(1-q)+\epsilon\right)\id_{A}+z,$$ where $\mathbb{P}(A)=1-q$. It follows that $\E[(X-z)_+]=(1-p)(r-z)+\epsilon(1-q)$ and $\mathrm{ES}_p(X)=r+\epsilon(1-q)/(1-p)>r$, since $\VaR_p(X) = z$.
Then,
\begin{equation*}\label{eq:es}
\begin{aligned}
1<\E[e'(X,r,z)]&=1+h(r, z)\left(\frac{\E[(X-z)_{+}]}{(1-p)(r-z)}-1\right) + k(r, z) \frac{p-\mathbb{P}(X\le z)}{1-p}\\
&=1+h(r, z)\frac{(1-q)\epsilon}{(1-p)(r-z)} + k(r, z) \frac{p-q}{1-p}.
\end{aligned}
\end{equation*}
Arbitrariness of $\epsilon$ implies that $k=0$ and $h > 0$.
The final statement is obtained by the same argument as in the proof of Proposition \ref{prop:unimean}.
\end{proof}

\section{Supplementary simulation and data analysis}
\label{app:num}

\subsection{Forecasting procedure for stationary time series data}
\label{app:forecast}

This section describes the details of the forecasting procedure for VaR and ES in Section \ref{sec:sim}. We assume that the data generated above follow an AR$(1)$--GARCH$(1, 1)$ process $\{L_t\}_{t\in\N}$ with $L_t=\mu_t+\sigma_tZ_t$, where $\{Z_t\}_{t\in\N}$ is assumed to be a sequence of iid innovations with mean $0$ and variance $1$, and $\{\mu_t\}_{t\in\N}$ and $\{\sigma_t\}_{t\in\N}$ are $\mathcal{F}_{t-1}$-measurable. Specifically, we have
\begin{equation}\label{app:eq:musigma}
\mu_t=c+\psi L_{t-1}~~\text{and}~~\sigma^2_t=\alpha_0+\alpha_1\sigma^2_{t-1}Z^2_{t-1}+\beta\sigma^2_{t-1},~~t\in\N.
\end{equation}
We first assume the innovations $Z_t$ to follow a normal, t-, or skewed-t distribution, and estimate $\{\hat{\mu}_t\}_{t\in\N}$ and $\{\hat{\sigma}^2_t\}_{t\in\N}$ through obtaining the maximum likelihood estimators of $(c,\psi,\alpha_0,\alpha_1,\beta)$ under the assumption on the distribution of $Z_t$. For t- and skewed-t distributions, parameters are estimated by the maximum likelihood method via the standardized residuals
$\{(L_t-\hat{\mu}_t)/\hat{\sigma}_t\}_{t\in\N}.$
For a risk measure $\rho$ ($\rho=\VaR_p$ or $\rho=\ES_p$), the value of $\rho(Z_t)$ can be calculated explicitly for the assumed parametric models (see e.g., \cite{McNeilETAL2015}, \citet{NoldeZiegel2017} and \citet{PattonZiegelETAL2019}). For estimation, the estimated parameters are plugged into these formulas resulting in estimates $\widehat{\rho(Z_t)}$. The final risk predictions are then $\hat{\mu}_t + \hat{\sigma}_t\widehat{\rho(Z_t)}$, where $\hat{\mu}_t$ and $\hat{\sigma}_t$ are computed from \eqref{app:eq:musigma} with the estimated parameters.
Table \ref{app:tab:forecast} shows the average of the forecasts of $\VaR$ and $\ES$ at different levels over all $1,000$ trials and all trading days, where the last line shows the average forecasts of VaR and ES using the true information of the data generating process.

\begin{table}[H]
\centering
\small
\begin{center}
\caption{Average point forecasts of $\VaR$ and $\ES$ at different levels over $1,000$ simulations of time series and $500$ trading days; values in boldface are underestimated by at least $10\%$ compared with values in the last line}
\vspace{.1in}
\begin{tabular}{c c c c c c c}
\toprule
	& $\overline{\VaR}_{0.95}$ & $\overline{\VaR}_{0.99}$ &  $\overline{\VaR}_{0.875}$ & $\overline{\ES}_{0.875}$ & $\overline{\VaR}_{0.975}$ & $\overline{\ES}_{0.975}$\\\midrule
	normal & $ 0.619$ & $\mathbf{0.906}$ & $0.411$ & $\mathbf{0.620}$ & $\mathbf{0.752}$ & $\mathbf{0.910}$\\\midrule
	t & $\mathbf{0.534}$ & $\mathbf{0.999}$ & $\mathbf{0.300}$ & $\mathbf{0.576}$ & $\mathbf{0.722}$ & $\mathbf{1.065}$\\\midrule
	skewed-t & $0.676$ & $1.281$ & $0.369$ & $0.727$ & $0.922$ & $1.358$\\\midrule
	true & $0.674$ & $1.271$ & $0.368$ & $0.723$ & $0.918$ & $1.343$\\\bottomrule
\end{tabular}
\label{app:tab:forecast}
\end{center}
\end{table}

\subsection{Comparing GREE and GREL methods for stationary time series}
\label{app:comp}

This section serves as a supplement to Section \ref{sec:sim} by demonstrating the results of backtesting VaR and ES using the GREE and GREL methods through Taylor approximation in \eqref{eq:GREE_em}. Meanwhile, we compare the performance of the GREE and GREL methods.
The results of VaR are shown in Tables \ref{app:tab:var_perc} and \ref{app:tab:var} and those for ES are shown in Tables \ref{app:tab:es_perc} and \ref{app:tab:es}. The GREL method is better than the GREE method in terms of percentage of detections in all cases of VaR and ES. This is consistent with the result in Theorem \ref{th:opt} because for the time series data, the losses used by the GREL method are relatively closer to an iid pattern compared to the whole e-statistics used by the GREE method. This is also not a contradiction to the slightly longer expected time to detection conditional on the detection of the GREL method, noting that the GREL method detects more often.

\begin{table}[H]
\centering
\small
\begin{center}
\caption{Percentage of detections ($\%$) for $\VaR_{0.99}$ forecasts over $1,000$ simulations of time series and $500$ trading days}
\vspace{.1in}
\begin{tabular}{c c c c | c c c | c c c}
\toprule
	& \multicolumn{9}{c}{GREE} \\\cmidrule{2-10}
	& \multicolumn{3}{c}{normal} &  \multicolumn{3}{c}{t} & \multicolumn{3}{c}{skewed-t} \\\midrule
	threshold & $2$ & $5$ & $10$ & $2$ & $5$ & $10$ & $2$ & $5$ & $10$\\\midrule
	$-10\%$ & $99.0$ & $96.5$ & $92.1$ & $95.5$ & $84.2$ & $72.4$ & $38.3$ & $13.7$ & $5.6$\\\midrule
	exact & $95.0$ & $84.4$ & $72.0$ & $79.5$ & $55.4$ & $37.2$ & $14.0$ & $2.8$ & $0.8$\\\midrule
	$+10\%$ & $80.5$ & $56.8$ & $38.2$ & $52.3$ & $22.0$ & $9.8$ & $5.3$ & $0.6$ & $0$\\\midrule
	& \multicolumn{9}{c}{GREL} \\\cmidrule{2-10}
	& \multicolumn{3}{c}{normal} &  \multicolumn{3}{c}{t} & \multicolumn{3}{c}{skewed-t} \\\midrule
	threshold & $2$ & $5$ & $10$ & $2$ & $5$ & $10$ & $2$ & $5$ & $10$\\\midrule
	$-10\%$ & $99.7$ & $98.3$ & $94.5$ & $98.2$ & $88.7$ & $76.7$ & $51.0$ & $15.1$ & $6.6$\\\midrule
	exact & $97.8$ & $88.6$ & $75.9$ & $87.4$ & $62.4$ & $39.8$ & $24.3$ & $3.2$ & $0.6$\\\midrule
	$+10\%$ & $87.7$ & $65.0$ & $43.9$ & $67.5$ & $28.9$ & $13.6$ & $10.8$ & $0.4$ & $0.1$\\\bottomrule
\end{tabular}
\label{app:tab:var_perc}
\end{center}
\end{table}

\begin{table}[H]
\centering
\footnotesize
\begin{center}
\caption{The average number of days taken to detect evidence against $\VaR_{0.99}$ forecasts conditional on detection over $1,000$ simulations of time series and $500$ trading days; numbers in brackets are average final log-transformed e-values}
\vspace{.1in}
\begin{tabular}{c c c c c | c c c c | c c c c}
\toprule
	& \multicolumn{12}{c}{GREE}\\\cmidrule{2-13}
	& \multicolumn{4}{c}{normal} &  \multicolumn{4}{c}{t} & \multicolumn{4}{c}{skewed-t}\\\midrule
	threshold & $2$ & $5$ & $10$ & & $2$ & $5$ & $10$ & & $2$ & $5$ & $10$ &\\\midrule
	$-10\%$ & $123$ & $186$ & $228$ & $(5.475)$ & $159$ & $236$ & $278$ & $(3.327)$ & $206$ & $260$ & $300$ & $(0.2856)$\\\midrule
	exact & $164$ & $239$ & $283$ & $(3.236)$ & $197$ & $272$ & $311$ & $(1.638)$ & $189$ & $229$ & $265$ & $(-0.1012)$ \\\midrule
	$+10\%$ & $197$ & $268$ & $300$ & $(1.734)$ & $217$ & $280$ & $318$ & $(0.5933)$ & $158$ & $224$ & -- & $(-0.1706)$\\\midrule
	& \multicolumn{12}{c}{GREL}\\\cmidrule{2-13}
	& \multicolumn{4}{c}{normal} &  \multicolumn{4}{c}{t} & \multicolumn{4}{c}{skewed-t}\\\midrule
	threshold & $2$ & $5$ & $10$ & & $2$ & $5$ & $10$ & & $2$ & $5$ & $10$ &\\\midrule
	$-10\%$ & $116$ & $185$ & $233$ & $(5.338)$ & $158$ & $241$ & $293$ & $(3.290)$ & $239$ & $281$ & $330$ & $(0.3492)$ \\\midrule
	exact & $160$ & $241$ & $295$ & $(3.240)$ & $196$ & $286$ & $332$ & $(1.736)$ & $233$ & $238$ & $289$ & $(-0.1463)$\\\midrule
	$+10\%$ & $189$ & $284$ & $330$ & $(1.849)$ & $226$ & $304$ & $358$ & $(0.7599)$ & $230$ & $211$ & $377$ & $(-0.3472)$\\\bottomrule
\end{tabular}
\label{app:tab:var}
\end{center}
\end{table}

\begin{table}[H]
\centering
\footnotesize
\begin{center}
\caption{Percentage of detections ($\%$) for $\ES_{0.975}$ forecasts over $1,000$ simulations of time series and $500$ trading days}
\vspace{.1in}
\begin{tabular}{c c c c | c c c | c c c}
\toprule
	& \multicolumn{9}{c}{GREE}\\\cmidrule{2-10}
	& \multicolumn{3}{c}{normal} & \multicolumn{3}{c}{t} & \multicolumn{3}{c}{skewed-t}\\\midrule
	threshold & $2$ & $5$ & $10$ & $2$ & $5$ & $10$ & $2$ & $5$ & $10$ \\\midrule
	$-10\%$ ES & $99.7$ & $98.9$ & $97.9$ & $96.7$ & $85.6$ & $73.8$ & $39.9$ & $14.9$ & $6.1$\\\midrule
	$-10\%$ both & $99.8$ & $98.9$ & $97.3$ & $97.9$ & $89.2$ & $79.3$ & $41.1$ & $13.6$ & $6.4$\\\midrule
	exact & $98.6$ & $93.2$ & $86.1$ & $83.0$ & $59.8$ & $41.1$ & $13.5$ & $3.2$ & $1.1$\\\midrule
	$+10\%$ both & $91.1$ & $75.7$ & $59.8$ & $54.2$ & $22.3$ & $10.2$ & $5.1$ & $0.8$ & $0$\\\midrule
	$+10\%$ ES &  $91.0$ & $76.1$ & $60.2$ & $60.6$ & $28.3$ & $13.8$ & $6.0$ & $1.1$ & $0.1$\\\midrule
\end{tabular}
\begin{tabular}{c c c c | c c c | c c c}
	& \multicolumn{9}{c}{GREL}\\\cmidrule{2-10}
	& \multicolumn{3}{c}{normal} & \multicolumn{3}{c}{t} & \multicolumn{3}{c}{skewed-t}\\\midrule
	threshold & $2$ & $5$ & $10$ & $2$ & $5$ & $10$ & $2$ & $5$ & $10$ \\\midrule
	$-10\%$ ES & $99.9$ & $99.2$ & $98.2$ & $97.5$ & $86.2$ & $74.1$ & $49.3$ & $16.6$ & $6.5$\\\midrule
	$-10\%$ both & $99.9$ & $99.0$ & $97.6$ & $98.1$ & $89.2$ & $78.8$ & $49.9$ & $17.4$ & $6.6$\\\midrule
	exact & $99.2$ & $94.6$ & $85.7$ & $87.6$ & $62.4$ & $41.9$ & $25.3$ & $5.6$ & $0.9$\\\midrule
	$+10\%$ both & $94.1$ & $77.3$ & $56.2$ & $66.0$ & $31.6$ & $12.5$ & $11.5$ & $1.5$ & $0.1$ \\\midrule
	$+10\%$ ES & $94.4$ & $76.7$ & $57.4$ & $72.7$ & $36.1$ & $16.9$ & $13.1$ & $1.7$ & $0.2$\\\bottomrule
\end{tabular}
\label{app:tab:es_perc}
\end{center}
\end{table}

\vspace{-.2in}

\begin{table}[H]
\centering
\footnotesize
\begin{center}
\caption{The average number of days taken to detect evidence against $\ES_{0.975}$ forecasts conditional on detection over $1,000$ simulations of time series and $500$ trading days; ``--" represents no detection; numbers in brackets are average final log-transformed e-values}
\vspace{.1in}
\begin{tabular}{c c c c c | c c c c | c c c c}
\toprule
	& \multicolumn{12}{c}{GREE}\\\cmidrule{2-13}
	& \multicolumn{4}{c}{normal} & \multicolumn{4}{c}{t} & \multicolumn{4}{c}{skewed-t}\\\midrule
	threshold & $2$ & $5$ & $10$ & & $2$ & $5$ & $10$ & & $2$ & $5$ & $10$ & \\\midrule
	$-10\%$ ES & $95$ & $141$ & $181$ & $(6.596)$ & $158$ & $226$ & $274$ & $(3.365)$ & $213$ & $250$ & $234$ & $(0.3991)$\\\midrule
	$-10\%$ both & $95$ & $152$ & $194$ & $(6.341)$ & $149$ & $220$ & $269$ & $(3.674)$ & $207$ & $242$ & $224$ & $(0.3594)$ \\\midrule
	exact & $139$ & $201$ & $250$ & $(4.278)$ & $193$ & $265$ & $307$ & $(1.822)$ & $208$ & $233$ & $220$ & $(-0.1011)$\\\midrule
	$+10\%$ both & $177$ & $249$ & $292$ & $(2.669)$ & $219$ & $267$ & $286$ & $(0.6485)$ & $106$ & $109$ & -- & $(-0.1843)$\\\midrule
	$+10\%$ ES & $174$ & $248$ & $295$ & $(2.625)$ & $210$ & $266$ & $288$ & $(0.8098)$ & $90$ & $96$ & $156$ & $(-0.2005)$\\\midrule
	& \multicolumn{12}{c}{GREL}\\\cmidrule{2-13}
	& \multicolumn{4}{c}{normal} & \multicolumn{4}{c}{t} & \multicolumn{4}{c}{skewed-t}\\\midrule
	threshold & $2$ & $5$ & $10$ & & $2$ & $5$ & $10$ & & $2$ & $5$ & $10$ & \\\midrule
	$-10\%$ ES & $97$ & $147$ & $189$ & $(6.344)$ & $155$ & $231$ & $282$ & $(3.238)$ & $235$ & $264$ & $277$ & $(0.4990)$\\\midrule
	$-10\%$ both & $99$ & $154$ & $201$ & $(5.963)$ & $146$ & $221$ & $271$ & $(3.511)$ & $223$ & $263$ & $278$ & $(0.4565)$\\\midrule
	exact & $134$ & $209$ & $258$ & $(4.027)$ & $191$ & $266$ & $318$ & $(1.892)$ & $208$ & $233$ & $220$ & $(-0.09266)$\\\midrule
	$+10\%$ both & $174$ & $257$ & $291$ & $(2.577)$ & $217$ & $289$ & $298$ & $(0.8661)$ & $186$ & $207$ & $70$ & $(-0.3171)$\\\midrule
	$+10\%$ ES & $173$ & $254$ & $296$ & $(2.557)$ & $215$ & $282$ & $301$ & $(1.007)$ & $189$ & $185$ & $271$ & $(-0.3653)$\\\bottomrule
\end{tabular}
\label{app:tab:es}
\end{center}
\end{table}

\subsection{Type-I errors for backtesting stationary time series}
\label{app:type1}

As shown by Remark \ref{rem:R1-threshold} and the results in Section \ref{sec:sim_stu}, the practical type-I errors are usually much smaller than the theoretical bound $\alpha$ in Theorem \ref{lem:Ville} when the detection threshold is $1/\alpha$ and the betting process is chosen properly. 
To see the type-I error specified by Theorem \ref{lem:Ville}, we simulate the same time series data as that in Section \ref{sec:sim} but with different sample sizes up to $10^5$ for backtesting. Under the true model, we report risk forecasts with skewed-t innovations and the true $\{\mu_t\}_{t\in\N}$ and $\{\sigma_t\}_{t\in\N}$ as the data generating process. Table \ref{app:tab:type1} shows the type-I errors for different detection thresholds and methods of choosing the betting processes. As a comparison, the bottom of Table \ref{app:tab:type1} also shows the type-I errors for the traditional backtest in \cite{McNeilFrey00} and \cite{NoldeZiegel2017} with significance levels $5\%$, $2\%$, and $1\%$, where the hypothesis is rejected if it is rejected at any sample size (this is of course invalid). We can see that the type-I errors increase as the sample size increases. For sample size $10^5$ (usually much smaller in practice), the type-I errors for the e-backtesting approaches are smaller than but close to the theoretical bounds indicated by Theorem \ref{lem:Ville}. The type-I error of the traditional backtest explodes for the large sample size $10^5$ due to its lack of optional validity, as expected.

\begin{table}[H]
\centering
\small
\begin{center}
\caption{Percentage of detections ($\%$) for true $\VaR_{0.99}$ forecasts over $1,000$ simulations of time series and different sample sizes}
\vspace{.1in}
\begin{tabular}{c c c c | c c c | c c c | c c c}
    \toprule
    sample size & \multicolumn{3}{c|}{$10^5$} & \multicolumn{3}{c|}{$10^4$} & \multicolumn{3}{c}{$10^3$} & \multicolumn{3}{c}{$500$}\\\midrule
    threshold ($1/\alpha$) & $2$ & $5$ & $10$ & $2$ & $5$ & $10$ & $2$ & $5$ & $10$ & $2$ & $5$ & $10$\\\midrule
    constant $\lambda_t=0.01$ & $38.9$ & $16.7$ & $8.6$ & $38.9$ & $16.7$ & $8.6$ & $37.4$ & $15.1$ & $7.1$ & $36.3$ & $10.9$ & $4.3$\\\midrule
    GREE & $32.5$ & $11.0$ & $5.7$ & $24.1$ & $7.7$ & $3.8$ & $14.9$ & $3.0$ & $0.9$ & $10.2$ & $1.1$ & $0.2$\\\midrule
    GREL & $41.8$ & $16.4$ & $8.7$ & $41.7$ & $16.0$ & $8.2$ & $26.8$ & $5.8$ & $2.0$ & $20.1$ & $3.0$ & $0.4$\\\midrule
    GREM & $38.8$ & $14.1$ & $7.1$ & $34.9$ & $12.0$ & $5.9$ & $20.4$ & $4.0$ & $1.2$ & $15.0$ & $1.7$ & $0.2$\\\bottomrule
    & \\\toprule
    Significance level & $5\%$ & $2\%$ & $1\%$ & $5\%$ & $2\%$ & $1\%$ & $5\%$ & $2\%$ & $1\%$ & $5\%$ & $2\%$ & $1\%$\\\midrule
    Traditional & $89.7$ & $82.4$ & $73.8$ & $14.6$ & $7.8$ & $4.5$ & $2.2$ & $0.6$ & $0.2$ & $1.7$ & $0.5$ & $0.3$ \\\bottomrule
    \end{tabular}
\label{app:tab:type1}
\end{center}
\end{table}

\subsection{Forecasting procedure for time series  with structural change}
\label{app:structural}

This section provides  details for the forecasting procedure for time series data with structural change in Section \ref{sec:structural}. After a burn-in period of length $1,000$, $500$ data points are simulated, within which $250$ presampled data $L_1,\dots,L_{250}$ are for forecasting risk measures and the rest $250$ data $L_{251},\dots,L_{500}$ are for backtesting. The forecaster obtains the estimates $\hat{\bm{\theta}}$ of the model parameters $\bm{\theta}=(\omega,\alpha,\beta_t)$ once using the standard Gaussian QML for the presampled $250$ losses. For $t\in\{251,\dots,500\}$, the forecasts of $\VaR_{0.95}(L_t|\mathcal{F}_{t-1})$ and $\ES_{0.95}(L_t|\mathcal{F}_{t-1})$ are obtained by $z_t=\sigma_t(\hat{\bm{\theta}})\widehat{\VaR}_{0.95}$ and $r_t=\sigma_t(\hat{\bm{\theta}})\widehat{\ES}_{0.95}$, respectively, where $\widehat{\VaR}_{0.95}$ and $\widehat{\ES}_{0.95}$ are empirical forecasts of VaR and ES using presampled residuals $L_1/\sigma_1(\hat{\bm{\theta}}),\dots,L_{250}/\sigma_{250}(\hat{\bm{\theta}})$. We choose the size $m=50$ of the rolling window and a leg $d=5$ of autocorrelations.\footnote{The leg $d=5$ is not necessary for the Monte Carlo simulations detector. We choose it to be consistent with the simulation setting of \cite{HogaDemetrescu2022} for comparison.}

\subsection{P-value trajectories for NASDAQ data analysis}
\label{app:pvalue}

\begin{figure}[H]
\caption{Left panel: p-value trajectories testing $(\VaR_{0.975},\ES_{0.975})$ with respect to number of days for the NASDAQ index from Jan 3, 2005 to Dec 31, 2021; right panel: the number of days taken for the traditional p-value test to detect evidence against the $(\VaR_{0.975},\ES_{0.975})$ forecasts}
\begin{minipage}{0.5\textwidth}
\centering
    \includegraphics[width=\textwidth]{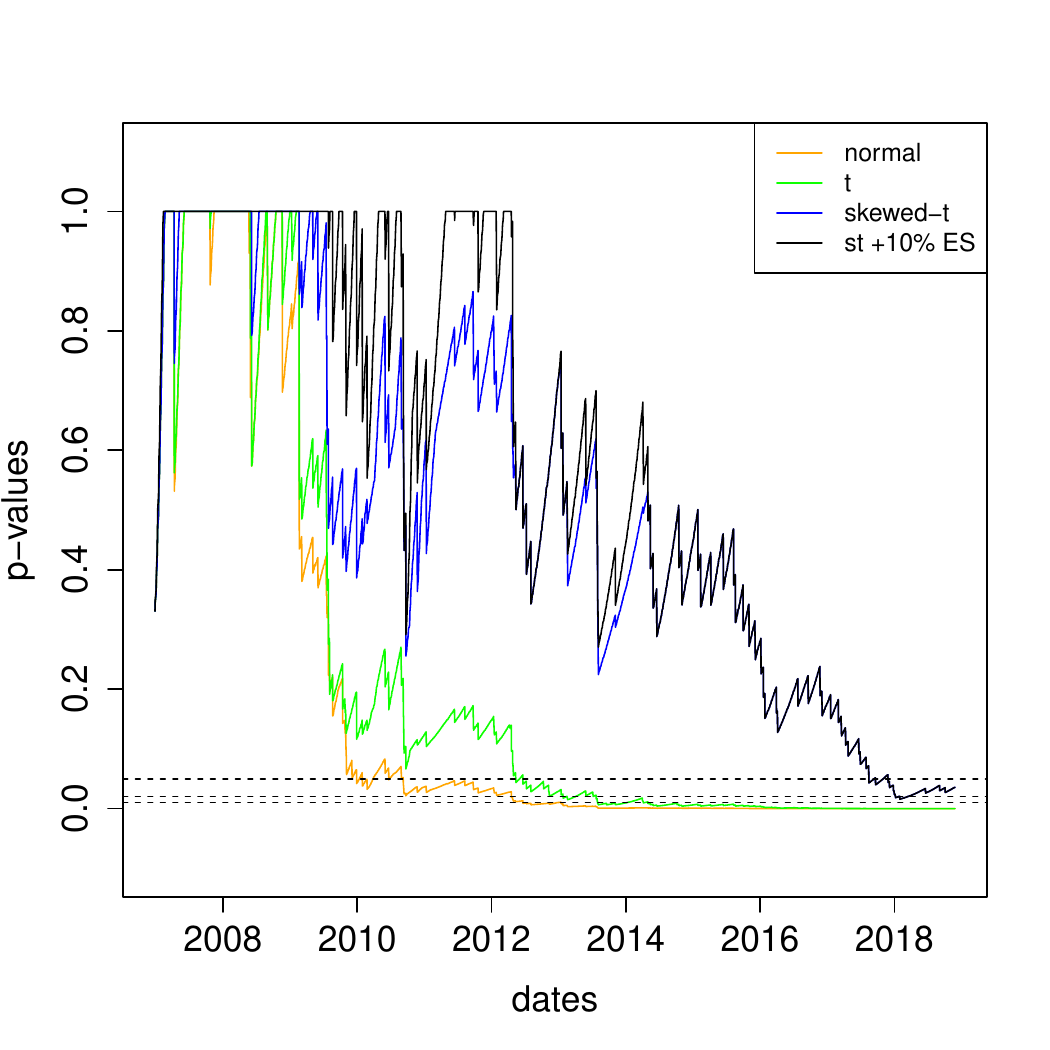}
\end{minipage}
\begin{minipage}{0.48\textwidth}
\centering
\begin{tabular}{c c c c}
\toprule
threshold & $0.05$ & $0.02$ & $0.01$\\\midrule
normal & $756$ & $1337$ & $1381$ \\
t & $1354$ & $1529$ & $1659$ \\
skewed-t & $2677$ & $2776$ & $3477$ \\
st $+10\%$ $\ES$ & $2677$ & $2776$ & $3477$ \\
\bottomrule
\end{tabular}
\end{minipage}
\label{app:fig:pvalue}
\end{figure}

We perform the traditional backtest in \cite{McNeilFrey00} and \cite{NoldeZiegel2017} for the same NASDAQ index used in Section \ref{sec:NASDAQ} for $(\VaR_{0.975},\ES_{0.975})$. For an ad hoc comparison to the e-processes in Figure \ref{fig:es_975_empirical2005}, we do the traditional backtest ``sequentially" and plot a process of p-values over the trading days in Figure \ref{app:fig:pvalue}. During the 2007-2008 financial crisis, we observe similar downward jumps for all p-value trajectories. However, all forecasts, no matter whether they are conservative or not, are eventually rejected in 2021. The result clearly shows that ``sequentially"  performing a traditional p-value test violates  validity.

\subsection{Detailed setup of data analysis for optimized portfolios}
\label{app:portfolio}

For the data of optimized portfolios we use in Section \ref{sec:portfolio}, the first $500$ data points are used for the initial forecast and another $500$ data points are for computing the first value of the betting process of the backtesting procedure. The final sample for backtesting contains $4280$ negated percentage log-returns from Jan 3, 2005 to Dec 31, 2021. The selected stocks are those with the largest market caps in the 11 S\&P 500 sectors divided by the GICS level 1 index as of Jan 3, 2005. The list of selected stocks is shown in Table \ref{tab:stock_list}.

\begin{table}[H]
    \centering
    \footnotesize
    \caption{22 selected stocks in S\&P 500 sectors divided by GICS level 1 as of Jan 3, 2005 for the portfolio}
    \vspace{.1in}
    \begin{tabular}{c c c}
 \toprule
         \textbf{Communication Services} & \textbf{Customer Discretionary} & \textbf{Consumer Staples}\\\midrule
         Verizon Communications Inc. & Time Warner Inc. & The Procter \& Gamble Co.\\
         AT\&T Inc. & The Home Depot, Inc. & Walmart Inc.\\\midrule
        \textbf{Energy} & \textbf{Financials} & \textbf{Health Care}\\\midrule
        Exxon Mobil Corp. & Citigroup Inc. & Johnson \& Johnson\\
        Chevron Corp. & Bank of America Corp. & Pfizer Inc.\\\midrule
        \textbf{Industrials} & \textbf{Information Technology} & \textbf{Materials}\\\midrule
        United Parcel Service Inc. & International Business Machines Corp. & EI du Pont de Nemours and Co.\\
        General Electric Co. & Microsoft Corp. & The Dow Chemical Co.\\\midrule
        \textbf{Real Estate} & \textbf{Utilities}\\\midrule
        Weyerhaeuser Co. & Exelon Corp.\\
        Simon Property Group Inc. & The Southern Co.\\\bottomrule
    \end{tabular}
    \label{tab:stock_list}
\end{table}

\begin{figure}[H]
    \centering
    \caption{Portfolio data fitted by different distribution from Jan 3, 2005 to Dec 31, 2021; left panel: negated percentage log-returns, right panel: $\ES_{0.975}$ forecasts}
    \includegraphics[width=0.48\textwidth]{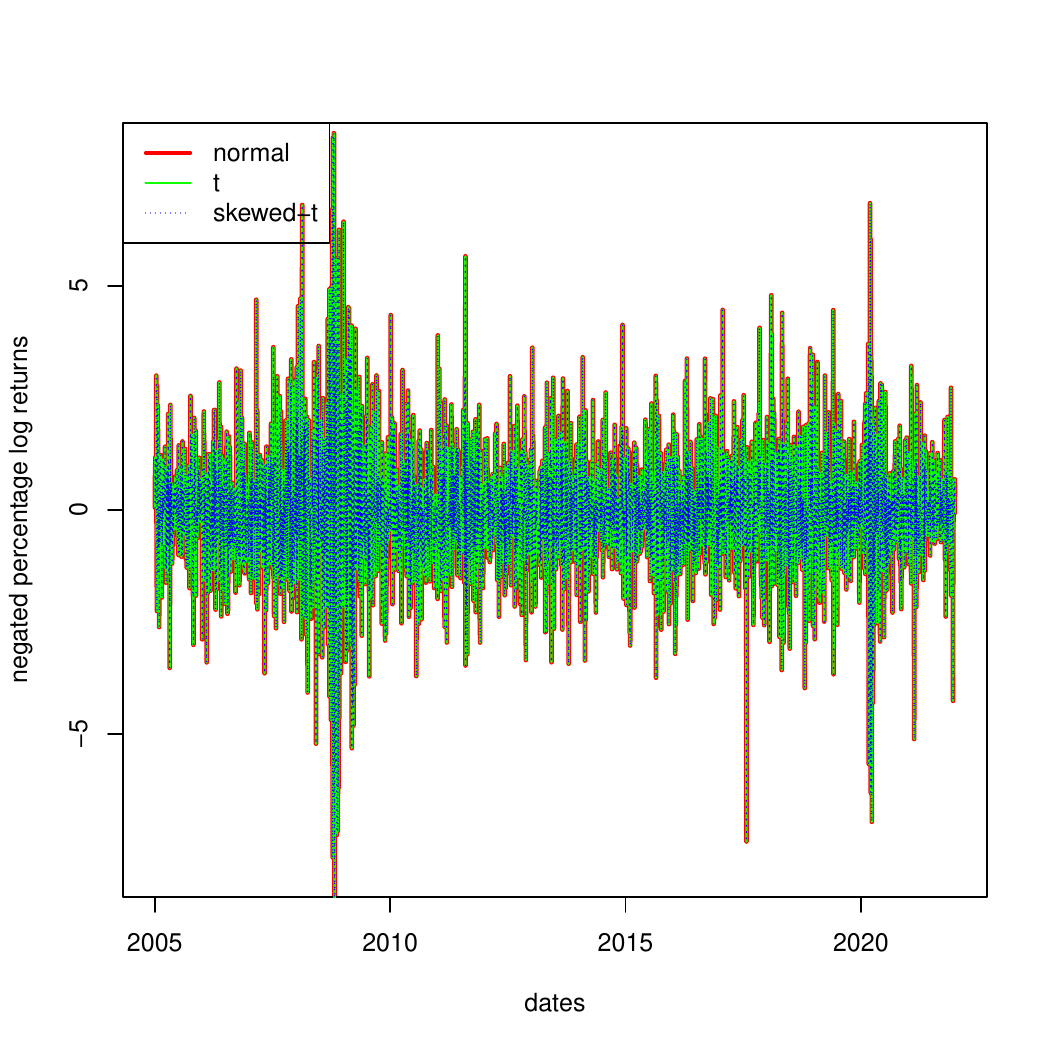}
    \includegraphics[width=0.48\textwidth]{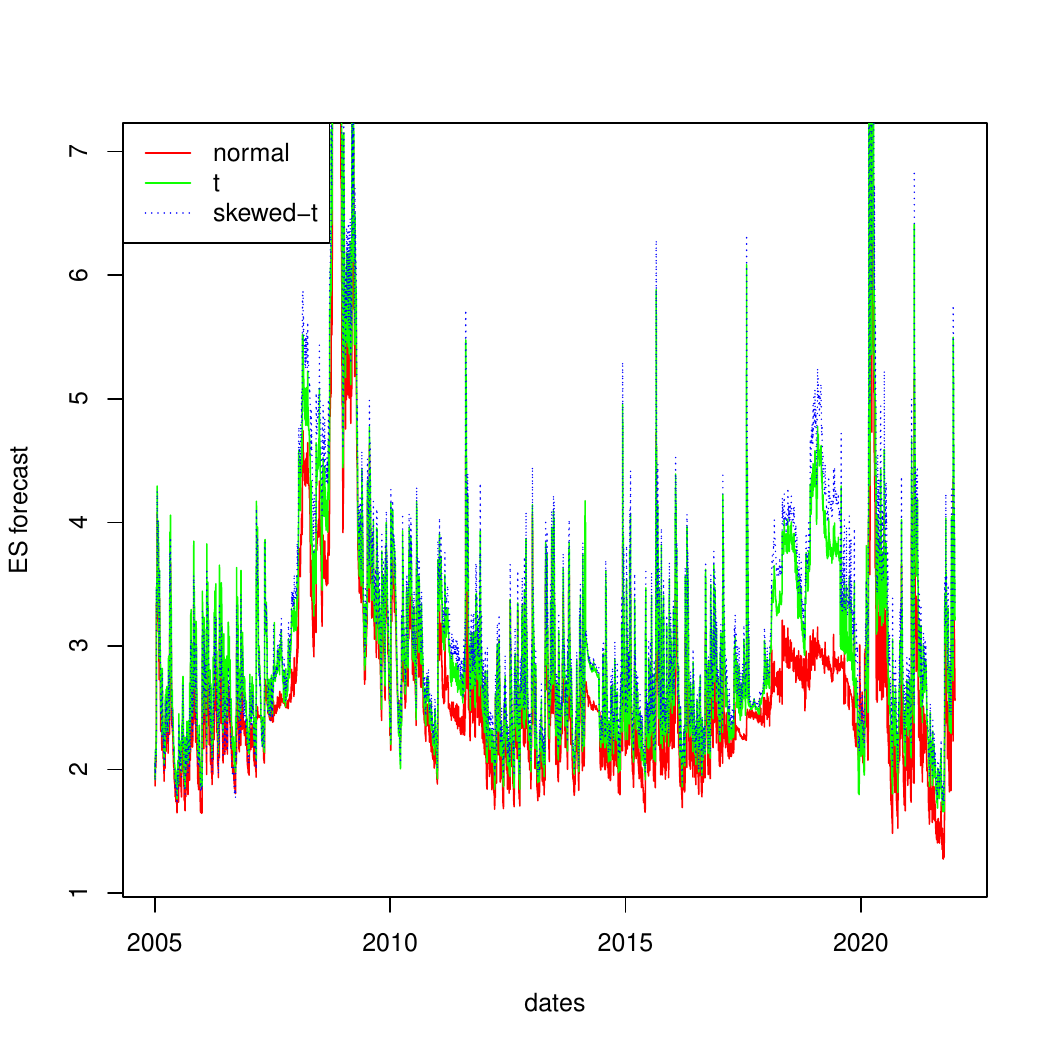}
    \label{app:fig:Portfolio}
\end{figure}

For forecasting, we assume
$L^i_t=\mu^i_t+\sigma^i_tZ^i_t$ with $\mu^i_t$ and $\sigma^i_t$ defined to be the same as \eqref{app:eq:musigma} for all $i\in[n]$. The innovations $\{Z^i_t\}_{t\in\N}$ are iid with respect to time with mean $0$ and variance $1$ for $i\in[n]$, assumed to be normal, t-, or skewed-t distributed.
If a stock delists from the S\&P 500 during the period from Jan 3, 2005 to Dec 31, 2021, it is  removed from the portfolio as soon as it delists with all of its weight redistributed to the other stocks in the portfolio.
The bank reports the $\VaR$ and the $\ES$ of the weighted portfolio by assuming $\mathbf{w}_t^\top\mathbf{L}_t$ to be normal, t-, or skewed-t distributed, respectively, with the mean $\sum^n_{i=1}w^i_t\mu^i_t$ and the variance $(w^1_t\sigma^1_t,\dots,w^n_t\sigma^n_t)^\top\Sigma_t(w^1_t\sigma^1_t,\dots,w^n_t\sigma^n_t)$, where $\Sigma_t$ is the covariance matrix of $(Z^1_t,\dots,Z^n_t)$. {The assumption is true when the innovations follow the normal distribution or follow the t-distribution with the same degree of freedom. We use this assumption to approximate the true distribution of the portfolio for t-distributions of different degrees of freedom and skewed-t distributions.} The parameters of the t- and skewed-t distributions of the weighted portfolio are estimated by the maximum likelihood method assuming the negated percentage log-return of the portfolio to be the AR$(1)$-GARCH$(1,1)$ process with innovations belonging to the same class of distribution. Figure \ref{app:fig:Portfolio} shows the negated log-returns of the portfolio and the forecasts of $\ES_{0.975}$ over time assuming different innovation distributions.

\section{Deliberate over-forecast strategy}
\label{app:R1-EC-5}

\subsection{Deliberate over-forecast strategy and a solution to this}

Due to the flexibility of our setting that allows for arbitrary reports of the risk forecasts, it is possible that the financial institution tries to``game" the regulator to escape the penalty from a detection. Specifically, the financial institution may first deliberately produce conservative risk forecasts to keep the e-process at a low level due to its multiplicative structure, and then start under-reporting risks to avoid the e-process exceeding the detection threshold quickly. Below we show that this issue can be addressed by our e-backtesting method with a properly calculated betting process.

By Proposition \ref{lem:lambda01} and \eqref{eq:GREE_em}, a positive betting process can only be chosen if the empirical mean of e-values is larger than $1$. Thus, when the realized e-values $e(L_t,r_t,z_t)$ are very small (due to over-forecasting), the GREE or GREL method will choose a small betting process (equal or close to $0$) during the over-reporting period. This keeps the e-process being almost a constant $1$. However, if we calculate the betting process using the empirical distribution of all past data, it may still take many data points for the GREE or GREL method to realize the under-reports after a long period of over-prediction. This problem can be solved by a slight modification of our method. Instead of using all past data points, we use a rolling window (e.g., 250 or 500 days) to determine the betting process in the GREE or GREL method. This guarantees the past over-report data to exit the window gradually, and thus prevents the GREE or GREL method from being dragged by the over-prediction for too long. Moreover, since the bank is not supposed to know the betting process $\boldsymbol \lambda$, the gaming cost for the bank increases to cover later under-prediction using intentional over-prediction. The regulator can actively choose a positive betting process to detect evidence against the null faster if she suspects that the bank under-predicts in a certain period.

Most of the existing backtesting approaches (such as the standard test for VaR by counting the number of breaches) are also faced with similar problems of being gamed by the bank through an intentionally long over-prediction period before the under-reports (leading to fewer breaches than theoretical). Our method provides a better solution to the issue than other methods mostly due to its nature of sequential testing and its strategic choices of the betting processes.

\subsection{A simulation of stationary time series with gamed forecasts}

We perform a simulation to demonstrate that our method can address this problem efficiently. We simulate the same AR$(1)$--GARCH$(1,1)$ process as that in Section \ref{sec:sim}, with a sample size $2,000$ and rolling window $500$. Assume that the financial institution first produces $1,000$ conservative risk forecasts, under AR$(1)$--GARCH$(1,1)$, by assuming skewed-t innovations and deliberately over-reporting $10\%$ of the exact forecasts; while it then produces $1,000$ aggressive risk forecasts by assuming normal innovations. When calculating the betting process for our e-backtesting method, different from Section \ref{sec:sim}, we use a moving window of size $500$ instead of using all historical data.\footnote{For the first $500$ data points, the moving window sizes are less than $500$.}

\begin{table}[H]
    \centering
    \caption{Percentage of detections (\%), the average number of days (after $1,000$ days) taken to detect evidence against risk forecasts, and average final log-transformed e-values for $\VaR_{0.99}$ and $\ES_{0.975}$ over $1,000$ simulations of time series and $2000$ trading days using the GREM method}
    \vspace{.1in}
    \begin{tabular}{c c c c | c c c}
    \toprule
	& \multicolumn{3}{c}{$\VaR_{0.99}$} & \multicolumn{3}{c}{$\ES_{0.975}$}\\\midrule
    threshold & $2$ & $5$ & $10$ & $2$ & $5$ & $10$\\\midrule
    percentage of detections & $99.7$ & $98.1$ & $94.8$ & $100$ & $100$ & $99.9$ \\\midrule
    days taken for detections & $302$ & $467$ & $541$ & $249$ & $409$ & $488$\\\midrule
    final log e-value & \multicolumn{3}{c|}{$6.118$} & \multicolumn{3}{c}{$7.332$} \\\bottomrule
    \end{tabular}
    \label{app:tab:detection}
\end{table}

\begin{figure}[H]
    \centering
    \caption{Average log-transformed e-processes testing $\VaR_{0.99}$ (left) and $\ES_{0.975}$ (right) with respect to the number of data points over $1,000$ simulations of time series and $2000$ trading days}
    \includegraphics[width=0.48\textwidth]{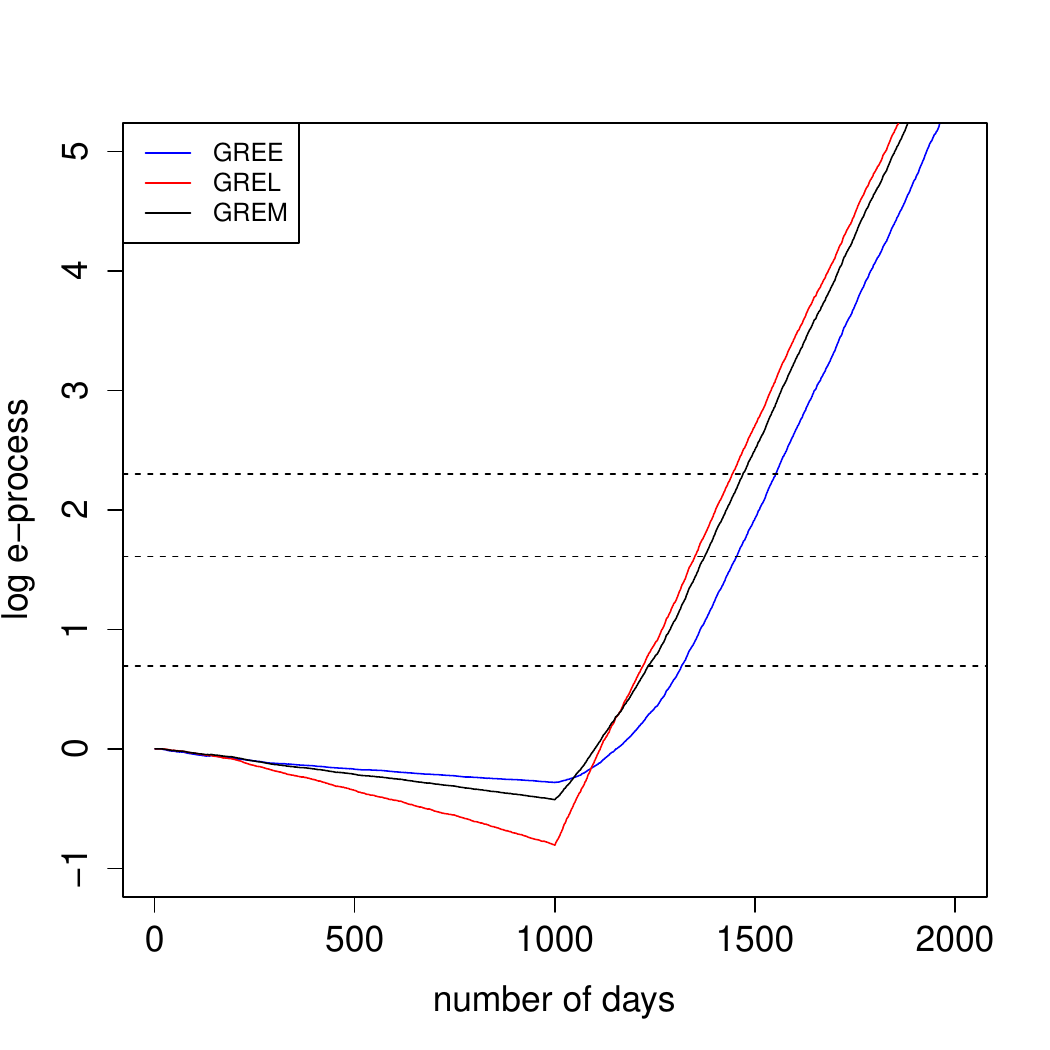}
    \includegraphics[width=0.48\textwidth]{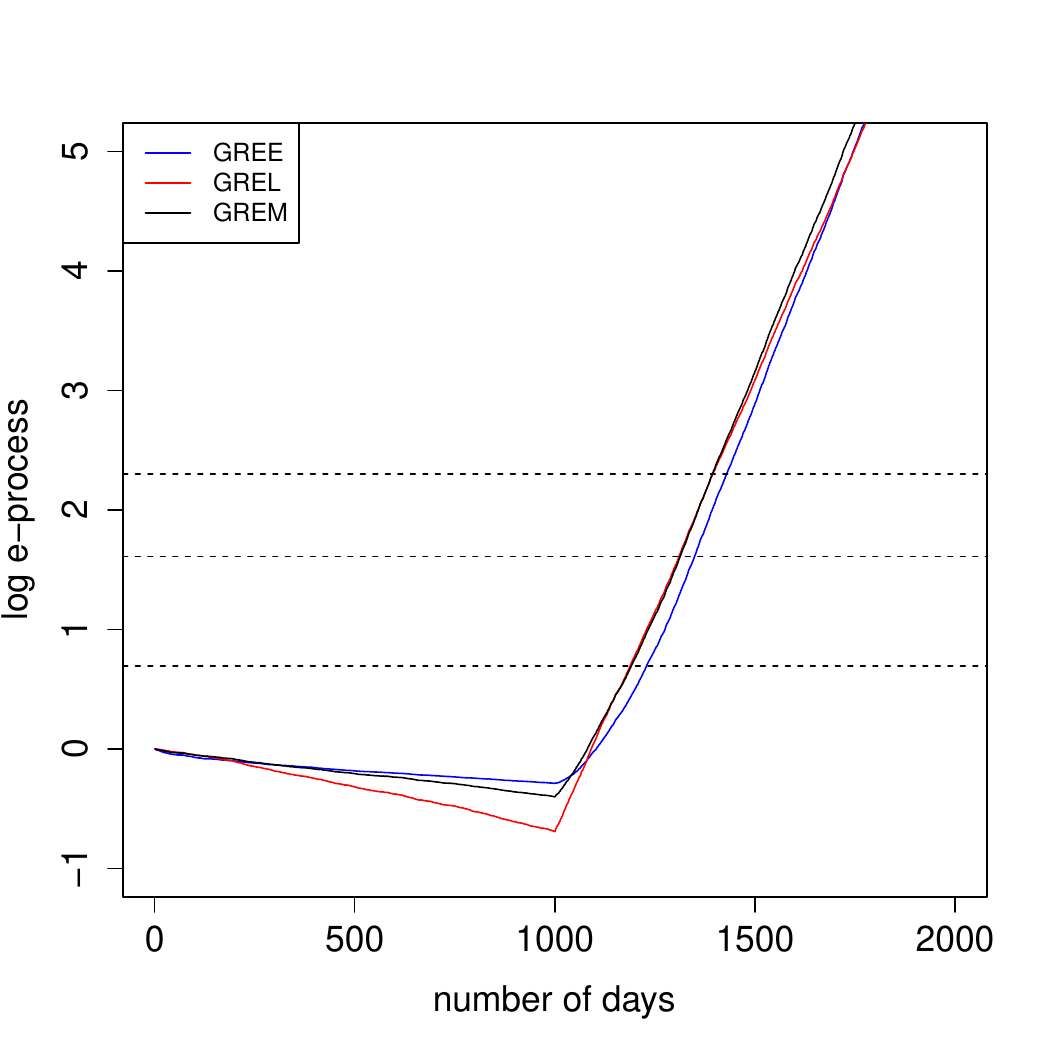}
    \label{app:fig:detection}
\end{figure}

Table \ref{app:tab:detection} shows the average result over $1,000$ simulations for the GREM method. After receiving $1,000$ conservative forecasts from the financial institution, the regulator detects evidence against $\VaR_{0.99}$ and $\ES_{0.975}$ forecasts at threshold $2$ after $250$ to $300$ days, which are $100$ to $150$ days later than those in Tables \ref{tab:var} and \ref{tab:es} but are still timely for the regulator to get alerted. Part of the delay is also caused by the fact that the moving window approach gives  less accurate estimates than using all historical data, assuming that the data are stationary. This suggests that the e-process is not seriously undermined by the over-reporting strategy of the financial institution. Figure \ref{app:fig:detection} shows that the log e-process
for the GREE method is always close to $0$ during the over-reporting period and grows steadily after the under-reports start to come in. On the other hand, the log e-process for the GREL method first drops a little lower to a point above $-1$, and then increases sharply as soon as the financial institution starts to under-report. The drop in the e-process for the GREL method does happen during the over-report period. However, the GREM method shares the advantages of both GREE and GREL methods, which guarantees that the e-process stays near $1$ (the GREE method will automatically choose the betting process near $0$) during the over-reporting period and grows quickly during the under-reporting period (the GREL method quickly starts to choose a positive betting process). In practice, we suggest the moving window approach and the GREM method to resolve similar gaming issues, although there may still be some inevitable delays.

\end{document}